\documentclass[final,11pt]{article}

\usepackage{eurosym,lscape,cancel,savesym,amsmath,ulem}
\def\em{\it}
\savesymbol{subequations}
\usepackage{cases}

\usepackage{ifthen,bm,multirow}
\usepackage{dsfont}
\usepackage{pgf}
\usepackage{latexsym, amsfonts, amssymb}
\usepackage{alltt,latexsym, verbatim}
\usepackage{float}
\usepackage{amsfonts}
\usepackage{xspace}
\usepackage{chicago} 

\newtheorem{theorem}{Theorem}[section]

\newtheorem{lem}{Lemma}[section]
\newtheorem{pro}{Proposition}[section]
\newtheorem{cor}{Corollary}[section]
\newtheorem{conj}{Conjecture}[section]

\newtheorem{rem}{Remark}[section]
\newtheorem{com}{Comments}[section]
\newtheorem{ex}{Example}[section]
\newtheorem{defi}{Definition}[section]
\newtheorem{hyp}{Assumption}[section]

\numberwithin{equation}{section}

\newcommand{\bt}{\begin{theorem}}\newcommand{\et}{\end{theorem}}
\newcommand{\bl}{\begin{lem}}\newcommand{\el}{\end{lem}}
\newcommand{\bp}{\begin{pro}}\newcommand{\ep}{\end{pro}}
\newcommand{\bcor}{\begin{cor}}\newcommand{\ecor}{\end{cor}}
\newcommand{\bconj}{\begin{conj}}\newcommand{\econj}{\end{conj}}

\newcommand{\bd}{\begin{defi} \rm }\newcommand{\ed}{\end{defi} }
\newcommand{\brem }{\begin{rem} \rm }\newcommand{\erem }{\end{rem}}
\newcommand{\bcom}{\begin{com} \rm }\newcommand{\ecom }{\end{com}}
\newcommand{\brems }{\begin{rem} \rm }\newcommand{\erems }{\end{rem}}
\newcommand{\bex}{\begin{ex} \rm }\newcommand{\eex}{\end{ex}}
\newcommand{\bhyp}{\begin{hyp} \rm }\newcommand{\ehyp}{\end{hyp}}

\def\proof{\noindent \textbf{\emph{\textbf{Proof}.$\qqq$}}}
\def\finproof {\hfill $\Box$ \vskip 5 pt }

\def \be{\begin{eqnarray}}
\def \ee{\end{eqnarray}}
\def \b*{\begin{eqnarray*}}
\def \e*{\end{eqnarray*}}

\def \[{[\,\!\![}
\def \]{]\,\!\!]}

\def \1{{\bf 1}}

\def \proof{{\noindent \bf Proof. }}
\def\ttt{t \in [0,\Ts]}
\newcommand{\bea}{\begin{eqnarray*}}
\newcommand{\eea}{\end{eqnarray*}}
\newcommand{\beqa}{\begin{eqnarray}}
\newcommand{\eeqa}{\end{eqnarray}}

\def\R{{\mathbb R}}
\def\proof{\noindent {\it Proof. $\, $}}
\def\finproof {\hfill $\Box$ \vskip 5 pt }
\def\I{\mathds{1}}
\def\sp{\,,\ \,}\def\sp{,\ \,}
\def\l{\label}

\def\bal{\begin{aligned}}
\def\eal{\end{aligned}}

\def\ttt{{t \in [0,\Ts]}}
\def\ttd{{t \in [0,\tb]}}
\def\hat{\widehat}

\def\myt{t}

\def\mym{m}

\def\xitau2{\xi_{(\thetau)}}\def\xitau2{\xi}
\def\xiitau2{\xi_i_{(\thetau)}}\def\xiitau2{\xi_i}
\def\xintau2{\tilde{\xi}_{(\thetau)}}\def\xintau2{\tilde{\xi}}

\def\Ltau2{\tilde{\xi}(\tau_{\mym},\tau_1)}
\def\chiitau2{P^i_{\tau}}

\newcommand{\beq}{\begin{eqnarray*}}
\newcommand{\eeq}{\end{eqnarray*}}

\def\mym{-}

\def\thelambda{\lambda^+}\def\thelambda{\lambda}\def\thelambda{c}
\def\thelambdam{\lambda}\def\thelambdam{\lambda^-}\def\thelambdam{\bar{\thelambda}}
\def\thelambdat{\tilde{\thelambda}}
\def\theg{g}

\def\mygb{\bar{g}}\def\mygb{\hat{g}}\def\mygb{\tilde{g}}

\def\tilde{\widetilde}

\def\emph{}

\def\gg{{\mathbb G}}\def\gg{{\mathbb F}}
\def\G{{\cal G}}\def\G{{\cal F}}

\def\gg{{\cal G}}\def\gg{\mathbb{G}}
\def\G{{\cal G}}

\def\E{\mathbb{E}}

\def\Rf{\theOm}\def\Rf{\mathfrak{r}}
\def\Rf{\theOm}\def\Rf{\mathfrak{r}}\def\Rf{\mathfrak{r}^b}

\def\thisR{R}\def\thisR{\rho}\def\thisR{R}\def\thisR{\rho^b}

\def\thetau{\vartheta}
\def\bthet a{\theta}
\def\tb{{\bar{\tau}}}\def\tb{\thetau}

\def\Rt{\mathbf{R}}\def\Rt{\mathsf{R}}\def\Rt{R}

\def\Ts{\bar{T}}\def\Ts{\mathcal{T}}\def\Ts{T}

\def\tb{\thetau}\def\tb{{\bar{\tau}}}

\def\Ts{\mathcal{T}}\def\Ts{\mathsf{T}}\def\Ts{\bar{T}}\def\Ts{T}

\def\ind{\mathds{1}}
\def\e{z}\def\e{e}

\def\iota{\mathbf{l}}

\def\myt{t}

\newcommand{\beql}[1]{\beqa\label{#1}\begin{aligned}}
\newcommand{\eeql}{\eal\eeqa}
\newcommand{\bel}{\bea\bal}
\newcommand{\eel}{\end{aligned}\eea}

\def\eee{\end{document}}
\def\ind {1\!\!1}\def\ind{\mathds{1}}
\def\I{\ind}

\def\G{{\cal G}}

\def\gg{{\mathbb G}}

 \def\Q{\mathbb Q}

\def\E {{\mathbb E} }

\def\R{{\mathbb R}}

\def\finproof {\hfill $\Box$ \vskip 5 pt }

\def\bal{\begin{aligned}}
\def\eal{\end{aligned}}

\def\finproof {\hfill $\Box$ \vskip 5 pt }

\def\Proba{\Proba}\def\Proba{\mathbb{Q}}

\def\sp{,\ \, }

\def\ttt{{t \in [0,T]}}

\def\xitau2{\xi_{(\thetau)}}
\def\xiitau2{\xi_i_{(\thetau)}}
\def\xintau2{\tilde{\xi}_{(\thetau)}}
\def\Ltau2{\tilde{\xi}(\tau_0,\tau_1)}
\def\chiitau2{P^i_{\tau}}

\def\Xt{\tilde{X}}\def\Xt{X}

\def\thetau{\tau}

\def\cG{\mathcal{G}}

\def\cY{\mathcal{Y}}

\def\l{\label}
\def\emph{}

\def\tb{{\bar{\tau}}}
\def\thisR{\thisR }

\def\theOm{\Omega}\def\theOm{\theOm}

\def\eee{\end{document}}
\def\textsl{}

\def\ttt{{t \in [0,\Ts]}}
\def\ttd{{t \in [0,\tb]}}

\def\Rt{\mathbf{R}}\def\Rt{\mathsf{R}}\def\Rt{R}\def\Rt{\theOm}

\def\Ts{\mathcal{T}}\def\Ts{\mathsf{T}}\def\Ts{\bar{T}}\def\Ts{T}

\def\proof{\noindent {\it {\textbf{Proof}}}.$\;\,$}
\def\finproof {\hfill $\Box$ \vskip 5 pt }\def\finproof {$\Box$}\def\finproof{\rule{4pt}{6pt}}

\def \be{\begin{eqnarray}}
\def \ee{\end{eqnarray}}
\def \b*{\begin{eqnarray*}}
\def \e*{\end{eqnarray*}}

\def \[{[\,\!\![}
\def \]{]\,\!\!]}

\def \1{{\bf 1}}

\def\ttt{t \in [0,\Ts]}

\def\R{{\mathbb R}}

\def\I{\mathds{1}}

\def\bal{\begin{aligned}}
\def\eal{\end{aligned}}

\def\hat{\widehat}

\def\myt{t}

\def\mym{m}

\def\xitau2{\xi_{(\thetau)}}\def\xitau2{\xi}
\def\xiitau2{\xi_i_{(\thetau)}}\def\xiitau2{\xi_i}
\def\xintau2{\tilde{\xi}_{(\thetau)}}\def\xintau2{\tilde{\xi}}

\def\Ltau2{\tilde{\xi}(\tau_{\mym},\tau_1)}
\def\chiitau2{P^i_{\tau}}

\def\mym{-}

\def\thelambda{\lambda^+}\def\thelambda{c}\def\thelambda{\lambda}
\def\thelambdam{\lambda}\def\thelambdam{\lambda^-}\def\thelambdam{\bar{c}}\def\thelambdam{\bar{\lambda}}
\def\theg{g}

\def\mygb{\bar{g}}\def\mygb{\tilde{g}}\def\mygb{f}\def\mygb{\hat{f}}

\def\tilde{\widetilde}

\def\emph{}

\def\G{{\cal G}}\def\G{{\cal F}}

\def\gg{{\cal G}}\def\gg{\mathbb{G}}
\def\G{{\cal G}}

\def\E{\mathbb{E}}

\def\Rf{\theOm}\def\Rf{\Rf}\def\Rf{R_{f}}\def\Rf{\bar{R}_{b}}\def\Rf{\bar{R}}

\def\thisR{R}\def\thisR{\thisR }\def\thisR{R_b}

\def\thetau{\vartheta}
\def\bthet a{\theta}
\def\tb{{\bar{\tau}}}\def\tb{\thetau}

\def\bthet a{\vartheta}
\def\tb{\thetau}\def\tb{{\bar{\tau}}}
\def\Ts{\mathcal{T}}\def\Ts{\mathsf{T}}\def\Ts{T}\def\Ts{\bar{T}}
\def\paragraph{\noindent\textbf}

\def\qqq{\quad\quad\quad}

\def\eee{\bibliographystyle{chicago}\bibliography{B}\end{document}} \def\eee{\end{document}}
\def\passhortciteN{\citeN}

\def\bthet{\boldsymbol\theta}\def\bthet{\mathbf{k}}

\def\theeta{\hat\tau}\def\theeta{\eta}\def\theeta{\tau}\def\theeta{\eta}

\def\gammat{\gamma_{\bullet}}\def\gammat{\gamma}

\def\td{\tau^\thisdelta}
\def\db{\tb^\thisdelta }

\def\qr#1{\eqref{#1}}
\def\fr#1{Fig.~\ref{#1}}

\def\sr#1{Sect.~\ref{#1}}

\newcommand{\indi}[1]{\I_{\{{#1}\}}}
\newcommand{\iend}{\end{itemize}}

\newcommand{\desb}{\begin{description}}
\newcommand{\dese}{\end{description}}

\newcommand{\dcb}{\begin{array}{lll}}
\newcommand{\dce}{\end{array}}
\newcommand{\ebe}{\begin{enumerate}\setlength{\baselineskip}{13pt}\setlength{\parskip}{5pt}}
\newcommand{\dbe}{\end{enumerate}}

\newcommand{\ibegin}{\begin{itemize}\setlength{\baselineskip}{19pt}\setlength{\parskip}{7pt}}

\makeatletter
\newenvironment{systeme*}{\arraycolsep=1.4pt\left\{\begin{array}{l}}{\end{array}\right.}

\makeatother

\def\Pb{\bar{P}}\def\Pb{\hat{P}}\def\Pb{P^{\delta} }\def\Pb{Q}

\def\vm{M}\def\vm{V}\def\vm{V\!M}
\def\ic{N}\def\ic{I}\def\ic{I\!M^c}
\def\ib{\mathfrak{\ic}}\def\ib{I\!M^b}
\def\cc{C}\def\cb{\mathfrak{\cc}}
\def\bc{b}\def\bc{c}

\def\Xd{X}\def\Xd{X^\delta}\def\Xd{X}\def\Xd{\bar{X}}\def\Xd{X}

\def\thisdelta{\delta}

\def\Rs{\Rt}

\def\bank{member\xspace}
\def\cure{liquidation\xspace}
\def\contribution{margin\xspace}\def\contribution{contribution\xspace}
\def\contributions{margins\xspace}\def\contributions{contributions\xspace}

\def\Zt{\tilde{Z}}\def\Zt{Z}

\def\xib{\bar{\xi}}

\def\xif{F}\def\xif{\xi_\star}
\def\Intens{\MMs }\def\Intens{X}\def\Intens{\gamma}
\def\INTENS{\boldsymbol{\MMs }}\def\INTENS{\mathbf{X}}
\def\intens{\boldsymbol{\gamma}}\def\intens{\mathbf{x}}
\def\thatmu{\lambda}\def\thatmu{\mu}
\def\og{\varpi}
\def\thea{a}
\def\Ms{\mathcal{C}}\def\Ms{\mathcal{C}^\star}
\def\MMs{\mathcal{C}}
\def\Cb{\MMs }\def\Cb{\mathcal{C}}

\newcommand{\bb}[1]{\mathbb{#1}}
\newcommand{\Exp}[1][]{\bb{E}^{\bb{#1}}}
\newcommand{\VaR}{\bb{V}a\bb{R}}

\def\theapim{\thea'_{im}}\def\theapim{\thea}
\def\theaim{\thea_{im}}\def\theaim{\thea}

\def\theOm{\theOm}\def\theOm{\chi}
\def\Cb{\Gamma}\def\Cb{\mathcal{C}}\def\Cb{C}
\def\cc{C}\def\cc{B}\def\cc{C^c}
\def\cb{\mathfrak{\cc}}\def\cb{C}\def\cb{C^b}
\def\ic{I}\def\ic{I^c}
\def\ib{\mathfrak{\ic}}\def\ib{I^b}

\def\LP{L\&P}\def\LP{L}
\def\price{fair value\xspace}\def\price{value\xspace}
\def\gain{e}

\def\b{}\def\b{\textcolor{blue}}
\renewcommand{\footnote}[1]{}
\begin{document}

\author{Yannick Armenti and St\'{e}phane Cr\'{e}pey\thanks{The research of St\'{e}phane Cr\'{e}pey benefited from the support of the EIF grant ``Collateral management in centrally cleared trading'', of the  ``Chair Markets in Transition'',  
F\'ed\'eration Bancaire Fran\c caise, and of the ANR 11-LABX-0019.
%The PhD grant of Yannick Armenti is funded by LCH.Clearnet Paris and the Chair Markets in Transition.
}
%and Shiqi Song\thanks{This research benefited from the support of the ``Chair Markets in Transition'' under the aegis of Louis Bachelier laboratory, a joint initiative of \'Ecole polytechnique, Universit\'e d'\'Evry Val d'Essonne and F\'ed\'eration Bancaire Fran\c caise.}
\\\\
stephane.crepey@univ-evry.fr
%, shiqi.song@univ-evry.fr
\\\\
Universit\'e d'\'Evry Val d'Essonne \\
Laboratoire de Math\'ematiques et Mod\'elisation d'\'Evry \\
91037 \'Evry Cedex, France
} 

\title{Central Clearing Valuation Adjustment
%and the Netting Benefit of CCPs
%Central Clearing, Regulatory Capital and the Compression Factor
%CCP Modeling: CCVA
}

\thispagestyle{empty}
\maketitle
\thispagestyle{empty}
 \begin{center}
%{\bf{DRAFT PAPER: DO NOT CIRCULATE}}
\end{center}
\begin{abstract}
We develop an XVA (costs) analysis of the clearance framework for a member of a clearing house. The systemic consequences of the default of the clearing house itself are outside the scope of such an XVA analysis. Hence the clearing house is assumed default-free.
We introduce a dynamic framework that incorporates the sequence of cash flows involved in the waterfall of resources of a clearing house. The overall XVA cost for a member, dubbed CCVA for central clearing valuation adjustment, is decomposed into CVA, MVA and KVA components. The CVA is the cost for a member of its losses on the default fund due to the defaults of other members.
The MVA is the cost of funding initial margin. The KVA mainly consists in the cost  
of the capital at risk that the member provides to the CCP through its default fund contribution. 
In the end the structure of the XVA equations for bilateral and cleared portfolios is similar, but the
data of the equations are of course not the same, reflecting
the different financial network structures.
The numerical experiments
emphasize the multilateral netting benefit of central clearing.  
However, it is known that this multilateral netting comes at the expense of a loss of netting across asset classes. If we compensate the first order multilateral netting effect by a suitable scaling factor accounting for the loss of netting across asset classes, then the bilateral and centrally cleared XVA numbers become comparable. 
The second more explanatory factor of the numerical results is the credit risk of the members and the ensuing MVA, especially in the bilateral setup, where even more initial margin is required. 
\end{abstract}

\begin{keywords}
Counterparty risk, central counterparty (CCP), margins, default fund, cost of funding, cost of capital,
netting.
%, backward stochastic differential equation (BSDE),
%dynamic copula.
\end{keywords}

\vspace{2mm}
\noindent
\textbf{Mathematics Subject Classification:} 
60G44, %Martingales with continuous parameter
91B25, %Asset pricing models
91B26, %Market models (auctions, bargaining, bidding, selling, etc.)
91B30, %Risk theory, insurance
91B70,  %Stochastic models
91B74,  %Models of real-world systems
91G20, %Derivative securities
91G40, %Credit risk
91G60, %Numerical methods (including Monte Carlo methods)
91G80, %Financial applications of other theories (stochastic control, calculus of variations, PDE, SPDE, dynamical systems)
%60H10, %Stochastic ordinary differential equations
%60G07.% Probability theory and stochastic processes
%60G44, %Martingales with continuous parameter
%91Gxx Mathematical finance
%91G10 Portfolio theory
%91G20 Derivative securities
%91G30 Interest rates (stochastic models)
%91G40 Credit risk
%91G50 Corporate finance
%91G60 Numerical methods (including Monte Carlo methods)
%91G70 Statistical methods, econometrics
%91G80 Financial applications of other theories (stochastic control, calculus of
%variations, PDE, SPDE, dynamical systems)
%91G99 None of the above, but in this section
\vspace{6pt}

\noindent
\textbf{JEL Classification:} {C63,%Computational Techniques • Simulation Modeling
C99	%Design of Experiments / Other
D52, %Incomplete Markets
D53,%Financial Markets
G12, %Asset Pricing • Trading Volume • Bond Interest Rates
G14, %	Information and Market Efficiency
G21, %	Banks • Depository Institutions • Micro Finance Institutions • Mortgages
G24, %	Investment Banking • Venture Capital • Brokerage • Ratings and Ratings Agencies
G28, %	Government Policy and Regulation
G33. %	Bankruptcy • Liquidation
}
%\vspace{2mm}
%\noindent
%\textbf{Mathematics Subject Classification:}
%91G40, %Credit risk.
%60H10, %Stochastic ordinary differential equations
%60G07.% Probability theory and stochastic processes
%%60G44, %Martingales with continuous parameter

%\tableofcontents
\section{Introduction}\label{s:intr}

To cope with counterparty risk,
the current trend in regulation 
is to push dealers to clear their trades via CCPs, i.e.~central counterparties (also known as clearing houses). Progressively, central clearing
is even becoming mandatory for vanilla products. Centrally cleared trading mitigates counterparty risk through an extensive netting of all transactions. Moreover, 
on top of the variation and initial margins that are used in the context of bilateral
transactions,
a CCP imposes its members to mutualize losses through an additional layer of protection, called the default or guarantee fund, which is pooled between the clearing members.
%on top of the variation and initial margins that are used in the context of bilateral
%transactions, a CCP deals
%with extreme and systemic risk on a mutualization basis, through
%an additional layer of protection, called the default or guarantee fund, which is pooled between the clearing members.

In this paper we develop the vision of a clearing house effectively eliminating counterparty risk (we do not incorporate the default of the clearing house in our setup), but at a certain cost for the members that we analyze. For this purpose, we develop an XVA (costs) analysis of centrally cleared trading, parallel to the one that has been developed in the last years for bilateral transactions.

\subsection{Review of the CCP Literature}\label{ss:lit}

\citeN{Duffie10} and
\passhortciteN{ContSantosMoussa2013} dwell upon the danger
of creating ``too big to fail'' financial institutions, including, potentially, clearing houses.
%Note that both papers ignore the counterparty risk of the clearing houses.
%{JPL pap ram syst\'emique aux concl oppos\'ees \`a \citeN{DuffieZhu} au titre que l'h\'et\'erog\'en\'eit\'e du r\'eseau rendrait le clearing finalement profitable: n\'eglige toutefois le risque de contrepartie de la CCP: ``fl\`eches'' d'exposures purement sortantes de la CCP alors qu'il en faudrait aussi dans l'autre sens, en r\'ealit\'e pur transfert sans aucune diminution du risque, chaises musicales entre quelques acteurs actionnaires et une CCP particuli\`erement peu capitalis\'ee);

Collateralization, whether in the context of centrally cleared trading or of
bilateral trading under ``standard CSA'' (credit support annex), which is the emerging bilateral trading alternative to centrally cleared trading,  
requires a huge amount of cash or liquid assets.
This
puts a high pressure on liquidity, an issue addressed in \citeN{SinghAitken2009}, \citeN{Singh2010},
 \citeN{LevelsCapel12} and \citeN{DuffieScheicherVuillemey2014}.  Relying on metrics \`a la \citeN{EisenbergNoe01}, \citeN{AminiFilipovicMinca15} assess the systemic risk and incentivization properties of a CCP design where, in order to spare the clearing members from liquidation costs, in situations of financial distress, the clearing members could temporarily withdraw from their default fund contributions to post variation margin.
%  liquidity squeezes characterize the CCP's equity, fee and guarantee fund policies that reduce systemic risk and are incentive compatible for banks.

\citeN{ContAvellaneda13} consider the optimal liquidation of the portfolio
of a defaulted member by the clearing house.
%%\\ JPL pap ramavel : soul\`eve 1 pbm important ms (probl\`eme peut-\^etre d'ailleurs insoluble) cds trait\'es \`a la liquidation of order book suppose une liquidit\'e qui n'existe pas sur les CDS, concentr\'es sur 3 ou 4 majors dt les difficult\'es de l'un impactent imm\'ediatement les autre(RAPH also note that the CDS margining scheme of rama and marco, although ``bought'' by ICE, is found almost impossible to implement by traders); n\'eglige enfin l'effet important d'impact n\'egatif du collat\'eral (I\!M dans le cas d'esp\`ece) sur les recovery (cf.~Kenyon), via la tentation des survivants de tirer le plus possible du collat\'eral en cas de coussin (marge) trop important}\\

%Margin schemes are studied under various respects in
%\citeN{HurlinPerignon12},
%\citeN{Cruz13} and \citeN{Menkveld2014}. The first paper proposes a validation framework. The other two papers (see also \passhortciteN{ArmentiCrepeyDrapeauPapapantoleon14}) propose methodologies to assess central clearing margin requirements.

Clearing is typically organized by asset classes, 
so that service closure of the CCP on one asset class does not harm its activity on other markets---and also because otherwise, in case of the default of a member, holders of less liquid assets (e.g.~CDS contracts) are penalized with respect to
holders of more liquid assets (e.g.~interest rate swaps). As a consequence, the multilateral netting benefit of CCPs comes at the expense of a loss of bilateral netting across asset classes
(see \citeN{DuffieZhu}).
\citeN{ContKokholm2011}
claim that the former effect typically dominates the latter.
% unless unrealistic homogeneity assumptions are made on the financial network.
But \citeN{GhamamiGlasserman16} show that, 
accounting for bilateral cross-asset netting, the higher regulatory capital
and margin requirements adopted for bilateral contracts do not necessarily create the intended cost incentive in favor of central
clearing.

\passhortciteN{ContMondescuYu11} and \citeN{PallaviciniBrigo13}
analyze the pricing implications of the differences between the margining procedures involved in bilateral and centrally cleared transactions.

Until recently, the cost analysis of CCPs, our focus in this paper, was only considered in an old business finance literature reviewed in \citeN{KnottMills2002}, notably \citeN{FennKupiec1993}. In the last years, new papers have appeared in this direction.
Under stylized assumptions,
\citeN{Arnsdorf12} derives an explicit approximation to a CCVA (using the terminology of the present paper), including effects such as wrong way risk (meant as procyclicality of the margins), credit dependence between members and left tailed distributions of their P\&Ls.
\citeN{Ghamami14} proposes a static one-period model where a CCVA can be priced by Monte-Carlo. 
%Consistent with the conclusions of the present study, 
\citeN{PallaviciniBrigo13bprel} 
%and \citeN{CrepeySong15} 
extend the bilateral counterparty risk dynamic setup of their previous papers to centrally cleared trading.
% viewed from the
%perspective of a client (as opposed to a member) of a clearing house.
However, they ignore the default fund
and the credit risk dependence issues that are inherent to the position of a clearing member.

\subsection{Contributions and Outline}

This paper develops an XVA (costs) analysis
of centrally cleared trading, parallel to the one that has been developed in the last years for bilateral transactions (see e.g.~\passhortciteN[Parts II and III]{BieleckiBrigoCrepeyHerbertsson13} or \citeN{BrigoMoriniPallavicini12}). A dynamic framework incorporates the sequence of cash flows involved in the waterfall of resources of the clearing house.
As compared with \citeN{Arnsdorf12} and \citeN{Ghamami14},
our CCVA accounts not only for the central clearing analog of the CVA, which is the cost for a member of its losses on the default fund in case of other members' defaults, but also for the cost of funding its margins (MVA) and for the cost of the capital (KVA) that is implicitly required from members through their default fund contributions (and for completeness and reference we also compute a DVA term).

%In order to ease the comparison between the centrally cleared and bilateral cases,
%we provide references to the corresponding bilateral developments
% available in a consistent notation
%in the monograph by \shortciteN{BieleckiBrigoCrepeyHerbertsson13}, along with reference to the original
% papers that can be downloaded from http://screpey.free.fr.
The framework of this paper can be used by a clearing house to find the right balance between initial margins and default fund in order to minimize the CCVA (subject to the regulatory constraints), hence optimize its costs to the members for a given level of resilience. A clearing house can also use it to analyze the benefit for a dealer to trade centrally as a member rather than on a bilateral basis, or to help its
members manage their CCVA (regarding the question for instance of how much of these costs they could consider passing to their clients).\\
%For readibility, we provide self-contained proofs for most statements, including
%the ones that are
%rather immediate adaptation of their bilateral counterparts.
%To demonstrate the practicality of our approach,
%% From an empirical point of view, the main contribution of this paper
%%is
%a numerical simulation assesses the netting benefit of central clearing and the impact of the credit risk of the members.

The paper is organized as follows. \sr{s:cc} presents our clearing house setup.
The waterfall of resources of the CCP is described in \sr{ss:md}.
The CCVA analysis is conducted in \sr{ph}.
%Under stronger assumptions, \sr{sss:Cureinterm} derives an alternative ``reduced'' CCVA equation \qr{red}.
\sr{s:DMO} introduces the common shock model that is used for the default times of the members of the clearing house.
\sr{num} provides an executive summary of the centrally cleared XVA analysis of this paper and recalls for comparison purposes the bilateral CSA XVA methodology of \citeN{CrepeySong15}.
\sr{s:num} designs an experimental framework used in the numerics of \sr{s:resu}.
\sr{ccva} concludes. Regulatory formulas are recalled in \sr{s:regul}. {\bf Proofs of all lemmas are deferred to \sr{s:proof}}.
%An index of symbols is provided after the bibliography.

\subsection{Basic Notation and Terminology}

\index{I@$\int_a^b$}$\int_a^b=\int_{(a,b]};$
\index{x@$x^+$}$x^{\pm} =\max(\pm x,0) $;
\index{d@$\boldsymbol\delta$}$\boldsymbol\delta_a$ represents a Dirac measure at a point $a$; $\boldsymbol\lambda$ denotes the Lebesgue measure on $\R_+$. 
%A time dependence is denoted in functional form by $\cdot(t)$, when deterministic, and as a subscript, by
%$\cdot_t$, for a stochastic process. 
Unless otherwise stated, 
% ``deterministic function'' of real arguments is measurable with respect to the corresponding Borel
%$\sigma$ field (and a function involving discrete arguments is always considered continuous with respect to these, in reference to the discrete topology);
a filtration satisfies the usual conditions;
a price
process
is a \emph{special semimartingale} in a
c\`adl\`ag version; all inequalities between random quantities
are meant
almost surely or
almost everywhere, as suitable; all the cash flows
are assumed to be
integrable whenever required; by ``martingale'' we mean local martingale unless otherwise stated, but true martingale is assumed whenever necessary. This means that we
only derive local martingale properties. Usually in applications
one needs true martingales, but this is not a real issue in our case, where even square integrability follows from additional assumptions postulated when dealing with BSDEs, which are our main pricing tool in this paper.

\section{Clearinghouse Setup}\label{s:cc}

We model a service of a clearing house dedicated to trading between its members,
labeled by
 $i\in N=\{ 0 ,\ldots, n\}.$ 

%We model a service of a clearing house dedicated to proprietary trading (typically on a given market) between its members,
%labeled by
% $i\in N=\{ 0 ,\ldots, n\}.$
%By comparison with proprietary trades, trades for external clients are treated on a trade-by-trade basis without any offsetting benefit. Hence they are heavily over-margined. As a consequence, proprietary trading between members is the most important risk management issue for the clearing house. 

\subsection{From Bilateral to Centrally Cleared Trading}%\label{s:liqproc}

\begin{figure}[H]
\begin{center}
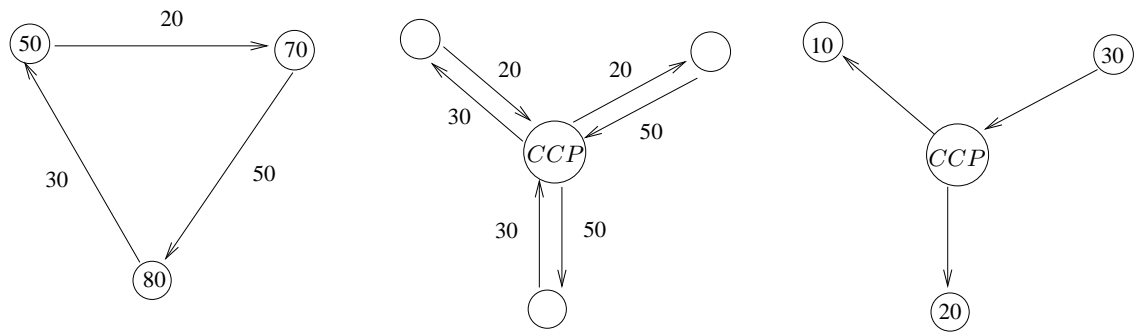
%\begin{picture}(0,0)%
%\includegraphics{biltoccp43.pdf}%
%\end{picture}%
%%
%%  Created by WinFIG version 5.01
%%  METADATA <version>1.0</version>
%%
%\setlength{\unitlength}{1697sp}%
%%
%\begingroup\makeatletter\ifx\SetFigFont\undefined%
%\gdef\SetFigFont#1#2#3#4#5{%
%  \reset@font\fontsize{#1}{#2pt}%
%  \fontfamily{#3}\fontseries{#4}\fontshape{#5}%
%  \selectfont}%
%\fi\endgroup%
%\begin{picture}(16423,4783)(513,-4464)
%%  METADATA <id>173</id>
%\put(8071,-1981){\makebox(0,0)[lb]{\smash{{\SetFigFont{9}{10.8}{\rmdefault}{\mddefault}{\updefault}{\color[rgb]{0,0,0}$CCP$}%
%}}}}
%%  METADATA <id>178</id>
%\put(13936,-2041){\makebox(0,0)[lb]{\smash{{\SetFigFont{9}{10.8}{\rmdefault}{\mddefault}{\updefault}{\color[rgb]{0,0,0}$CCP$}%
%}}}}
%\end{picture}%
 \end{center}
\caption{From bilateral to centrally cleared trading.}
 \label{f:biltoccp}
\end{figure}

In a centrally cleared setup, the clearing house interposes itself in all transactions, becoming
``the buyer to every seller and the seller to every buyer''. All the transactions between the clearing house and a given member are netted together. See Figure \ref{f:biltoccp} for an example, where the circled numbers in the left (respectively right) diagram show the gross positions of $n=3$ counterparties in a CSA setup
(respectively their net positions with the CCP after the introduction of the latter in the middle).

In addition to interfacing all trades, the clearing house asks for several layers of guarantee to be posted by the members against counterparty risk, including a default fund that is pooled between the clearing members.

The benefits of centrally cleared trading are multilateral netting benefit and mutualization of risk. The drawbacks are an increase of systemic risk, where ``too big to fail'' CCPs might be created, liquidity risk, due to the margin requirements, and a loss of bilateral netting across asset classes (cf. \citeN{Duffie10} and
\shortciteN{ContSantosMoussa2013}).

\subsection{Liquidation Procedure}\label{s:liqproc}

The mandate of a CCP is to liquidate  over a few days the portfolio of a defaulted member. During the liquidation period,
the CCP bears the risk of the portfolio. 
The trades with a defaulted member are typically reallocated by means of auctions among the surviving members and/or by a gradual liquidation of its assets in the market.

For ease of analysis in this paper,
we assume the existence of
a risk-free ``buffer'' 
 that is used by the clearing house for replacing defaulted members in their transactions with others at the end of a liquidation period of length $\delta$ (the defaulted transactions already involving the buffer as one counterparty are simply terminated). 
%The buffer can be viewed as an additional, risk-free member, \b{which therefore doesn't post any margins}. 
%But, in practice, the buffer need not correspond to an actual member. It can be implemented by the clearing house through back-to-back hedges of the defaulted transactions with a risk-free third-party. 
We assume that during the \cure period, the promised contractual cash flows and the hedge of a defaulted member are taken over by the CCP.

\subsection{Pricing Framework}\label{s:pfr}

Let $(\Omega,\gg,\mathbb{Q})$ 
represent a stochastic pricing basis,
with
$\gg=(\G_t)_{t\in\R_+},$
such that all our processes are $\gg$ adapted and all the random times of interest are $\gg$ stopping times. Expectation under $\mathbb{Q}$ and $(\mathcal{G}_t,\mathbb{Q})$ conditional expectation are denoted by $\E$ and $\E_{t}.$
We denote by $r$ a $\gg$ progressive
OIS rate process and by $\beta_t=e^{-\int_0^t r_s ds}$ the corresponding discount factor. An OIS (overnight index swap) rate is together the best market proxy for a risk-free rate and the reference rate for the remuneration of the collateral. 

For each member $i,$
we denote by
$D^i_t$ the process of the cumulative  contractual cash flows of its portfolio with the CCP (``promised dividend'' process ignoring counterparty  and funding risk), 
assumed of
finite variation. We denote by $P^i_t$ the mark-to-market of its portfolio ignoring counterparty and funding risk,
i.e.
\beqa\label{cleanf}{\beta} _t P^i_t =
{\E}_t \left( \int_t^{\Ts} {\beta} _s dD^i_s \right)\sp \ttt ,
\eeqa
where 
$\Ts$ is the final maturity of the CCP service portfolio, 
assumed held on a run-off basis
(as is standard in any pricing or risk model). 
All cash flows and values are considered from the point of view of the clearing house, e.g.
$P^i_t=1$ means that the member $i$ is short of a mark-to-market value equal to one (disregarding margins) toward the clearing house at time $t$.
Since all trades are between the members, we have 
$\sum_{i\in N} P^i = 0.$ 

\section{Margin Waterfall Analysis}\label{ss:md}

The mark-to-market pricing formula \qr{cleanf} ignores the counterparty risk of the
member $i$, with default time $\tau_i$ and survival indicator process $J^i=\ind_{[0,\tau_i)}$.
As a first counterparty risk mitigation tool, the members are required to exchange variation margins that
track the mark-to-market of their portfolios.
A clearing house can call for variation margins at every time of a margin grid of step $h$, e.g. twice a day.

However,
various frictions and delays,
notably the \cure period $\delta,$
imply gap risk, which is the risk of a gap between the variation margin and the debt of a defaulted
member at the time of liquidation of its portfolio. This is a special concern for certain classes of assets, such as credit derivatives, that may have quite unpredictable cash flows (see \citeN{CrepeySong15}).

This is why another layer of collateralization, called initial margins, is maintained in centrally cleared transactions as well as in bilateral transactions under standard CSA (the emerging bilateral trading alternative to centrally cleared trading).
Initial margins are also dynamically updated, based on some risk measure of the variation-margined P\&L of each member computed over the time horizon ${\delta'=}\delta+h$ of the so called margin period of risk (maximal time $h$ elapsed since the last margin call before the default plus \cure period $\delta$ between default and liquidation). 

Gap risk is magnified by wrong-way risk, which is~the risk of adverse dependence between the positions and the credit risks of the members. One may also face credit contagion effects between members (wrong-way and contagion risk are of special concern regarding credit derivatives).
Clearing houses deal with such extreme and systemic risk through a default fund mutualized between the clearing members. The default fund \contribution of each member is primarily intended to reimburse the losses triggered by its own default, but, if rendered necessary by exhaustion of the previous layers of the waterfall, it can also be used for reimbursing the losses due to the defaults of other members.

\subsection{Margins}\label{ss:marginsfirst}

Let $lh$, with $l\geq 0,$ represent the times of the variation and initial margin calls, and let
$lT$, with $T$ a multiple of $h$ (e.g.~$h=$ one day and $T=$ one month), represent the times of update of the default fund contributions.

Consistent with our sign convention that all cash flows and values are seen from the perspective of the clearing house, we count a margin positively when it is posted by a member and
we define the variation margin $V\!M^i$, initial margin $I\!M^i$ and default fund contribution $DFC^i$ of the member $i$ as the piecewise constant process reset at the respective grid times following, respectively (while the member $i$ is alive): \beql{mar}
&V\!M^i_{lh}=P^i_{lh-} \sp I\!M^i_{lh}=\rho^i_{lh}\sp
 DFC^i_{lT}= \varrho^i_{lT} ,
\eeql
where $\rho^i$ and $\varrho^i$ refer to suitable risk measures as explained below.
Note that \qr{mar} defines the level of reset of the respective cumulative amounts. Starting from $
V\!M^i_{0}=P^i_{0-}, I\!M^i_{0}=\rho^i_{0}$ and
$DFC^i_{0}= \varrho^i_{0} ,
$
the corresponding updates at grid times are $(P^i_{lh-}-P^i_{(lh-h)-}),$
%(with the convention $P^i_{0-}=0$),
$(\rho^i_{lh}-\rho^i_{lh-h})$ and $(\varrho^i_{lT}-\varrho^i_{lT-T}).$

\brem\label{rem:thresh}
In practice, the variation margin only tracks the mark-to-market of the portfolio up to some thresholds, or free credit lines of the members, and up to minimal transfer amounts devoted to avoiding useless updates. These features, which can be important in the case of bilateral transactions, are omitted here as negligible in the case of centrally cleared transactions.
\erem

Let
\beql{e:lp}&
 \LP ^i_{t,t+\delta' } ={P}^i_{t+\delta' }+ \int_{[ {t}, t+\delta' ]}e^{\int_s^{t+\delta' } r_{u}du} dD^i_s- P^i_{t-}.
\eeql
In particular, at margin call times $t=lh,$ 
we have, in view of the specification of the variation margin by the first identity in \qr{mar}:
\beql{e:lpbis}&
 \LP ^i_{lh,lh+\delta' } =
{P}^i_{lh+\delta' }+ \int_{[ {lh}, lh+\delta' ]}e^{\int_s^{lh+\delta' } r_{u}du} dD^i_s- V\!M^i_{lh},
\eeql
which is the variation-margined loss-and-profit of the member $i$ 
at the time horizon $\delta'=\delta +h $ of the margin period of risk (cumulative loss-and-profit also accounting for all the contractual cash flows capitalized at the risk-free rate during the margin period of risk $[ {t}, t+\delta' ]$). 
The risk measure used for fixing the initial margins
is a univariate risk measure computed
at the level of each member individually, which we write as
\beql{e:im}\rho^i_{lh}=\rho\left(\LP ^i_{lh,lh+\delta' }\right),\eeql
where $\rho$ can be value at risk,
%(until the crisis),
expected shortfall,
%(preferred since the crisis),
etc.. The dependence between the portfolios of the members is only reflected in the initial margins through the structural constraint that $\sum_{i\in N} P^i=0$.
\brem Historically, for computing initial margins, CCPs have been mostly using the SPAN methodology, introduced by the Chicago Mercantile Exchange in the 80s. 
This methodology is based, for each member, on the consideration of the most unfavorable among sixteen reference scenarios
(see \citeN{KupiecWhite96}). Nowadays, value at risk methodologies tend to become the standard.
% regarding initial margins.
\erem

Unless defaults happen, margins do not imply any transfer of ownership and can be seen in this sense as a loan by the posting member. 
By contrast, default fund contributions can be consumed
in case of other members' defaults, hence they should really be viewed as capital put at the disposal of the CCP by the clearing members. 
The ``cover two'' EMIR rule prescribes to size the default fund as, at least, the maximum of its largest exposure and of the sum of its second and third largest exposures to the clearing members (see \sr{ss:regulccva}). This is only a regulatory minimum and sometimes more conservative rules are used, such as a default fund set as the sum of the two largest exposures. It is then allocated between the clearing members by some rule, e.g. proportionally to their initial margins.
At a more theoretical level, the mutualization rationale of the default fund calls for the use of multivariate risk measures,
%for the determination of the default fund \contributions, 
which we write in an abstract fashion as
\beql{e:dfm} \varrho^i_{lT}= \varrho_i
\left(\Big(\LP ^j_{lT,lT+\delta' } 
-I\!M^j_{lT}\Big)_{j;J^j_{lT}=1}\right) \eeql 
%Indeed, the formula \qr{e:dfm} suggests a top down definition of the $\varrho^i_{lT}$ (as opposed to a bottom up approach as of \qr{e:im} for the $\rho^i_{lh}$),
(or an analog formula involving not only the $\LP ^j_{lT,lT+\delta' },$ but also intermediary $\LP ^j_{\cdot,\cdot+\delta' }$ between $(l-1)T$ and $lT$ to refrain members from temporarily closing their positions right before $lT$ in order to avoid to contribute to the default fund).
%all the daily 
%of the
%thinking for instance of the
%Euler allocation
% of a multivariate risk measure at the clearing house service level.

Regarding the distributions that are used for members loss-and-profits in all these risk measure computations, since the crisis, the focus has shifted from the cores of the distributions, dominated by volatility effects, to their queues, dominated by scenarios of crisis and default events. For the determination of the initial margins, Gaussian VaR models are generally banned since the crisis and CCPs typically focus on either Pareto laws or on historical VaR.
% (sometimes bootstrapped to make it a bit richer). 
Stressed scenarios and parameters are used for the determination of the default fund.

Note that margin schemes as above, even, in the case of the default fund contributions, possibly based on multivariate risk measures (cf. \qr{e:dfm}),
only account for asset dependence between the portfolios of the members, ignoring credit risk and contagion effects between members. This is in line
with the mandate of a clearing house to mitigate
(i.e.~put a cap on) its exposure to the members by means of the margins, in case a default would happen, where a defaults is viewed as a totally unpredictable event.
On top of the margins, add-ons are sometimes required from members with particularly high credit or concentration risk.

We refer the reader to \citeN{Ghamami14}, \passhortciteN{Cruz13}, \citeN{Menkveld2014} or
 \citeN{ArmentiCrepeyDrapeauPapapantoleon14}
for alternative margin schemes and default fund specifications.
Good margining schemes should guarantee
the required level of resilience to the clearing house at a bearable cost for the members. Two points of concern are procyclicality, in particular with haircuts that increase with the distress
of a member, and liquidity, given the generalization of central clearing and collateralization.
\citeN{CapponiCheng16} construct a model which endogenizes collateral, making it part of an optimization problem where the CCP maximizes profit by controlling collateral and fee levels.

%, an additional desirable feature is that margins increase sufficiently fast with the volatility of the market, without decreasing too quickly when the market gets more quiet, to limit procyclicality (especially via haircuts, which increase with the distress
%of the posting party). So, a certain asymetry of the margining process matters.
%Another concern is liquidity, given the generalization of central clearing and collateralization.

\subsection{Breaches}\label{s:breaches}

The default time of the member $i$ is
modeled as a stopping time $\tau_i$ with an intensity process $\gamma^i$. In particular,
%$\tau$ is a totally unpredictable stopping time and
any event $\{\tau_i =t\},$ for a fixed time $t,$ has zero probability and can be ignored in the analysis.
For every time $t\geq 0$, let
\beql{e:times}
{\bar{t}=t\wedge \Ts} \sp
t^\delta=t+\delta\sp
{\bar{t}^\delta= \ind_{t<\Ts} t^\delta + \ind_{t \geq\Ts} \Ts}
\eeql
and let $\hat{t}$ denote the greatest margin call time $lh\leq t$.
We denote by
\beql{c:fullcollat}
\MMs^i =V\!M^i +I\!M^i +DFC^i \eeql
the overall
 collateral process of the member $i$.
We assume that collateral posted is remunerated OIS and that the CCP substitutes itself to a defaulted member during its liquidation period, including regarding these collateral OIS remuneration cash flows. In our model collateral earns OIS but collateral OIS earnings are transferred as a remuneration to the posting member, they do not stay in the collateral accounts. Hence, the amount of available collateral for the liquidation of a defaulted member does not accrue at the OIS rate but stays constant during the liquidation period.
As a consequence, we have $\MMs ^i_t=\MMs ^i_{\hat{t}}$ for $t\leq \tau_i$ and the process $\MMs$ is stopped at time ${\hat{\tau}_i}.$
For each member $i,$ we write
\beql{c:eq:pasmark}
&\Delta^i_t=\int_{[\tau_i,t]}e^{\int_s^t r_{u}du} dD^i_s \sp \Pb^i_t={P}^i_t + \Delta^i_t\sp \varepsilon_i=(Q^i_{\td_{i}}
-\MMs^i_{\hat{\tau}_{i}})^+ ,\\
&\theOm_i=-\ind_{\varepsilon_i =0} Q^i_{\td_{i}}-\ind_{\varepsilon_i >0}(\MMs ^i_{\hat{\tau}_{i}}+R_i \varepsilon_i ) ,\\
&\xi_i = Q^i_{\td_{i}}+ \theOm_i=
 \ind_{\varepsilon_i>0}(Q^i_{\td_{i}}-\MMs ^i_{\hat{\tau}_{i}}-R_i \varepsilon_i)
 =(1-R_i)\varepsilon_i ,
\eeql
where $\Delta^i_t$ represents the cumulative contractual dividends  
capitalized at the risk-free rate that fail to be paid by member $i$ from time $\tau_i$ onwards. These dividends are promised but unpaid due to the default of the member $i$ at $\tau_i$. Hence, they also belong to the exposure of the CCP to the default of the member $i.$ More precisely, as will be understand  in more detail from the proof of Lemma \ref{l:brea} below,
$\chi_i$ corresponds to a terminal cash flow closing the position of the defaulted member $i$, paid by the CCP to the estate of the defaulted member at time $\tau^\delta_i$; $\varepsilon_i$ corresponds to
the raw exposure of the CCP to the default of the member $i$;
$\xi_i$ is the exposure accounting for
an assumed recovery rate $R_i$ of the member $i$. 
In fact, in the context of centrally cleared trading, by liquidation of a defaulted member, we simply mean the liquidation of its CCP portfolio, as opposed to the legal liquidation, by a mandatory liquidator, that can take several years (the New York branch of Lehman was legally liquidated in December 2013, more than five years after Lehman's default).
% and only matters in the case of bilateral OTC trading, with cleared trading,
In particular, 
%unless a defaulting member is also involved with the clearing house bilaterally, as it can happen
%through
%%rehyp?
%reinvestment of the margins by the clearing house, 
there is typically no recovery to expect on a defaulted member, i.e.~$R_i=0$. The reason why we introduce recovery coefficients is for the discussion regarding DVA and DVA2
%windfall benefits at own default (or DVA2)
 in \sr{ph} and for comparison with the bilateral trading setup of \sr{num}.

Note that we do not exclude joint defaults in our setup. In fact, joint defaults, which can be viewed as a form of ``instantaneous contagion'',
is the way we will introduce credit dependence between members in \sr{s:DMO}.
For $Z\subseteq N=\{ 0 ,\ldots, n\},$ we denote by $\tau_Z\in\R_+ \cup\{\infty\}$ the time of joint default of names in the subset $Z$ and only in $Z.$ At this stage we consider all the costs from the perspective of the CCP and the community of the surviving members altogether. The allocation of these costs between the CCP and the surviving members will be considered in \sr{ss:ef}. We call realized breach of a (possibly joint) default event the residual loss to the CCP after all the collateral of the defaulted member(s) has been consumed.
\bl\label{l:brea}
At each liquidation time $\td_{Z}=\tau_Z+\delta$ with $\tau_{Z}<\Ts$,
the realized breach of the CCP
is given by
% by $-\epsilon_{\td_{Z}}$, where
\beql{e:breach}
B_{\td_{Z}}=\sum_{i\in Z}\xi_i .
%(1-R_Z) (B^Z_{\td_{Z}})^+ ,
\eeql
\el

\subsection{Equity and Default Fund Replenishment Principle}

\l{ss:ef}
We proceed with the description of the next layers of the waterfall of resources of the clearing house, namely the equity and the default fund.

\begin{figure}[h!]
\begin{center}
\begin{picture}(0,0)%
\includegraphics{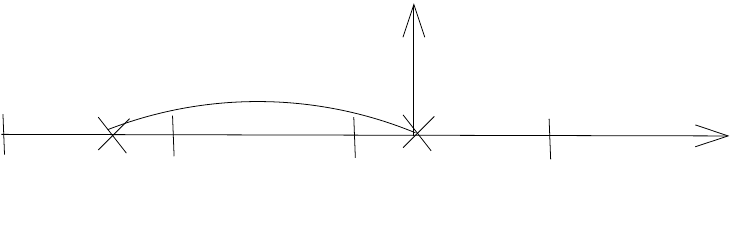}%
\end{picture}%
%
%  Created by WinFIG version 5.01
%  METADATA <version>1.0</version>
%
\setlength{\unitlength}{1973sp}%
\begingroup\makeatletter\ifx\SetFigFont\undefined%
\gdef\SetFigFont#1#2#3#4#5{%
  \reset@font\fontsize{#1}{#2pt}%
  \fontfamily{#3}\fontseries{#4}\fontshape{#5}%
  \selectfont}%
\fi\endgroup%
\begin{picture}(7014,2181)(1369,-1510)
%  METADATA <id>35</id>
\put(2176,-1111){\makebox(0,0)[lb]{\smash{{\SetFigFont{10}{12.0}{\rmdefault}{\mddefault}{\updefault}{\color[rgb]{0,0,0}$\tau_Z$}%
}}}}
%  METADATA <id>39</id>
\put(3804,-129){\makebox(0,0)[lb]{\smash{{\SetFigFont{10}{12.0}{\rmdefault}{\mddefault}{\updefault}{\color[rgb]{0,0,0}$\delta$}%
}}}}
%  METADATA <id>43</id>
\put(5611, -1){\makebox(0,0)[lb]{\smash{{\SetFigFont{10}{12.0}{\rmdefault}{\mddefault}{\updefault}{\color[rgb]{0,0,0}$\big(B_{t}- E_{t-}\big)^+=\sum_{i;J^i_t=1}\epsilon^i_{t}$}%
}}}}
%  METADATA <id>44</id>
\put(4936,-1404){\makebox(0,0)[lb]{\smash{{\SetFigFont{10}{12.0}{\rmdefault}{\mddefault}{\updefault}{\color[rgb]{0,0,0}$t=\tau^\delta_Z$}%
}}}}
\end{picture}%
\caption{Margin cash flows: resets at margin call grid times and refill of the default fund
%in case of breaches beyond equity
at liquidation times.}
\label{f:marks}
\end{center}
\end{figure}
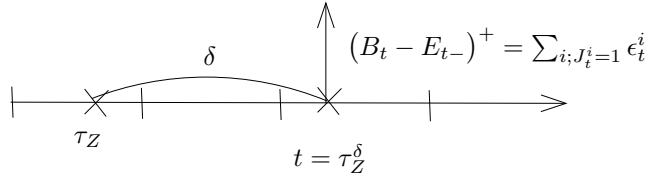

If the default of a member entails a positive breach, then the first payer (although to a typically quite limited extent) is the clearing house itself (before the surviving members),
via its equity $E$. 
\brem The regulation (e.g. EMIR) does not necessarily require that the CCP be the first payer in case of a realized breach. However, CCPs typically take the equity tranche of this risk, as a good management incentive.
\erem
Specifically,
at times $lY,$ $l \geq 0,$ where $Y$ is a multiple of $T$ (e.g. one year whereas $T$ is one month),
the equity process $E$ is reset by the clearing house at some target level $E_{lY}^\star,$ the ``skin in the game'' of the clearing house for the time period $[lY,(l+1)Y]$.
In the meantime, the equity is used as first resource for covering the realized breaches, i.e., at each $t=\td_{Z}$ with $\tau_Z<\Ts$, we have
\beql{e:eq}
\Delta E_{t}=-\big(B_{t}\wedge E_{t-}\big).
%\ind_{B_{\td_{Z}}>E_{\td_{Z}-}}\big(-E_{\td_{Z}-}\big)+\ind_{B_{\td_{Z}}\leq E_{\td_{Z}-}} ,
\eeql
The part of the realized breach left uncovered by the equity,
% (if any),
$(B_{t}- E_{t-})^+,$ is covered by the surviving members through the default fund, which they refill instantaneously by the following rule, at each $t=\td_{Z}$ with $\tau_Z<\Ts$ (see Figure \ref{f:marks}):
%breach.fig
%$\big(B_{t}- E_{t-}\big)^+=\sum_{i;J^i_t=1}\varepsilon^i_{t}$
\beql{e:dfb}
\epsilon^i_{t}=\big(B_{t}- E_{t-}\big)^+ \frac{{J^i_{t}}DFC^i_{t}}{\sum_{j\in N}J^j_{t}DFC^j_{t}},
\eeql
proportionally to their current default fund \contributions $DFC^i_{t}$
(or other keys of repartition such as their initial margins or the notionals
 of their positions).
%(the notation $\epsilon_{(\cdot)}$ emphasizes that $\epsilon^i$ does not need be be defined at all times as a process).
%In addition, the reset of its default margin at the level $DFC^i_{lT}$ prescribed by \qr{mar} implies an additional refill by the member $i$ (if alive) at ${lT}$ by $DFC^i_{lT}-DFC^i_{lT-T},$ i.e.
%\beql{e:dfq}
%\epsilon^i_{(lT)}={J^i_{lT}}(DFC^i_{lT}-DFC^i_{lT-T}).
%\eeql

In sum, the {margins} and the default fund contributions
$V\!M^i_{lh}, I\!M^i_{lh}$ and $DFC^i_{lT}$
are reset at their respective grid times by the {surviving} members according to \qr{mar};
the equity is reset at the times $lY$ by the clearing house and is used for covering the first levels of realized breaches at liquidation times according to \qr{e:eq};
%the {default margins}
%$DFC^i$ are reset at the $lT$ by the {surviving} members;
the losses in case of realized breaches above the residual equity are covered at liquidation times by the surviving members according to \qr{e:dfb} (see Figure \ref{f:marks}).
\brem\label{rem:fuf} 
The total size of the default fund is $\sum_{j\in N} J^j DFC^j,$
a quantity also referred to as the funded default fund. The unfunded default fund refers to the additional amounts members may have to pay via the above default fund replenishment principle in case of defaults of other members. More precisely,
\bel 
u^i_{lT} = \left(\sum_{lT-T<\td_{Z}<lT} \epsilon^i_{\td_{Z}}-DFC^i_{lT-T}\right)^+
\eel 
%\textbf{\r{To compare to the total unfunded default fund:
%\bel 
%u_{lT} = \left( \sum_{lT-T < \td_{Z} < lT} B_{\td_{Z}}- E_{\td_{Z}-} - DFC_{lT-T} \right)^+
%\eel
%}}
represents the unfunded default fund contribution of the member $i$ for the period $(lT-T,lT).$ The service closure, i.e.~the closure of the activity of the clearing house on a given market or service, is usually specified in terms of events such as the unfunded default fund 
%contribution of some (alive) member, $u^i_{lT},$
$\sum_{j\in N } J^j_{lT} u^j_{lT}$ 
reaching a cap given as, e.g., 
%$2 DFC^i_{lT-T}$ 
$2\sum_{j\in N }  J^j_{lT-T} DFC^j_{lT-T}$, i.e. twice the funded default fund.
Given the high levels of initial margins that are used in practice,
this is a very extreme tail event. 
Moreover, in case of service closure,
the risk of a member is bounded above by the sum
between its margins, three times its default fund contribution (assuming the above specification of service closure) and the cost of the liquidation of the service for this member. This cost is itself bounded by the notional of the member position, which would only be the actual cost in a scenario where all the assets of the CCP would jump to zero, also a very unlikely situation. In conclusion, 
the service closure event does not really matter regarding our present purpose of the XVA cost analysis of CCP membership.
The default of the CCP as a whole (i.e. the closure of all its services) is an even more unlikely event, especially because a central bank would hardly allow it to occur in view of its systemic consequences. 
Hence
we may and do ignore the service closure and the default of the clearing house in the context of this paper. See \citeN{ArmakolaLaurent15} about CCP resilience and see \citeN{Duffie2014} about
alternative
approaches
to
the
design
of
insolvency
and
failure
resolution
regimes
for
CCPs.
\erem

\section{Central Clearing Valuation Adjustment}\label{ph}
%Price and Hedge

We refer to the (generic) member 0 as ``the member'' henceforth, the other members being collectively referred to as ``the clearing house''.
For notational simplicity, we remove the index $0$ referring to the reference member. We call \price of the CCP portfolio of the
member its value inclusive of counterparty and funding risk (as opposed to the mark-to-market of the portfolio).

We assume that the member enters its portfolio at time 0, against an upfront payment of a certain amount $\Pi_0$, where the semimartingale $\Pi$ is a tentative \price process
of the CCP portfolio of the member.
We assume that profit-and-losses are marked to the model \price process $\Pi$ and realized in continuous time (the reader is referred to \shortciteN[Section 9.1]{AlbaneseCaenazzoCrepey15} for the discussion of other choices in this regard). 
%, so that the above risk-neutral pricing principle will result in a pricing BSDE for $\Pi$. 
%{The reader is referred to \shortciteN{AlbaneseCaenazzoCrepey15} regarding the more general FBSDE that appears for $\rho$ and $\Pi$ jointly if one postulates other losses realization schemes--unless replication holds and $\rho=0,$ in which case the FBSDE for $\rho$ and  $\Pi$ reduces to
%the BSDE \qr{martcond} for $\Pi$ again, independently of the realization schedule for the losses. }

In this section, 
%as in a bilateral trading counterparty risk setup, 
we derive a representation of the (no arbitrage) \price $\Pi$ of the CCP portfolio of a
member 
%{(value inclusive of counterparty risk and funding costs)} 
as the difference 
(cf.~the remark \ref{rem:sgn} below)
between the
mark-to-market of the portfolio and a correction
$\Theta.$
%corresponding to {the fair valuation of the cost of the clearance structure for that member}. 
We call $\Theta$ the 
%risk-neutral 
central clearing valuation adjustment (CCVA). The KVA-inclusive CCVA is obtained in a second step by adding to $\Theta$ a capital valuation adjustment (KVA) meant as the cost that it would require for remunerating the member at some hurdle rate for its CCP capital at risk (including its default fund contribution).

\subsection{DVA and DVA2 Issues}\label{ss:dvafva}

From the perspective of the \bank,
the effective time horizon of interest is $\db$ (cf.~\qr{e:times}).
The position of the \bank is closed at $\td$ (if $\tau<\Ts$), with a terminal cash flow from the member's perspective given, in view of \qr{c:eq:pasmark} and of the analysis developed in the proof of Lemma \ref{l:brea} (for $i=0$ here), by
\beql{e:T}
%\bar{\theOm}=
\theOm=-\ind_{\varepsilon =0} Q_{\td}-\ind_{\varepsilon >0}(\MMs _{\hat{\tau}}+R \varepsilon ) .
\eeql
In particular, if $\varepsilon >0,$ i.e.~$Q_{\td}>\MMs _{\hat{\tau}},$ then the member receives
\bel
%{e:Tsuite}
-\MMs _{\hat{\tau}}-R \varepsilon=-\MMs _{\hat{\tau}}-R (Q_{\td}-\MMs _{\hat{\tau}})
%=(1-R)(-\MMs _{\hat{\tau}})+R (-Q_{\td})
=(-Q_{\td})+(1-R)(Q_{\td}-\MMs _{\hat{\tau}}).\eel
%which is $(1-R)(Q_{\td}-\MMs _{\hat{\tau}})$ more than
However, for this amount to benefit to the member's shareholders, it needs to be hedged so that they can monetize it before $\tau$ (otherwise it is only a profit to the member's bondholders).  
But, in order to hedge this amount, the member would basically need to sell credit protection on itself, which is barely possible in practice. Consequently, from an entry (i.e. transaction)
price perspective, the member should 
ignore such a windfall benefit at own default and the ensuing debt valuation adjustment (DVA). This means
formally setting
$R=1$, which results in
$\theOm=- Q_{\td}$ in \qr{e:T} and $\xi=0$ 
later in \qr{nice}. Then ${R}$ becomes disconnected from what the clearing house would actually recover (if anything) from the member in case it defaults, but this is immaterial for analyzing the costs of this member itself, it only matters for the others. In sum, it is possible and convenient to analyze the no DVA case for the reference member just by formally setting $R=1$. 

If, however, 
some DVA is accounted for (i.e.~if $R<1$), then one may want to reckon likewise a funding benefit of the member at its own default, a windfall benefit called DVA2 in the terminology of \citeN{HullWhite13de},
corresponding to an additional
%terminal
cash flow to the member of the form
\beql{e:Tbis}
(1-\Rf ) ({\Pi_{\tau-}}+\Ms _{\hat{\tau}} )^{+}
\eeql
at time $\tau$ (if $<\Ts$). 
Here  
{$\Ms =V\!M  +I\!M$}  
and 
$\Rf$ is a recovery rate of the member to its funder, so that
the amount $({\Pi_{\tau-}}+\Ms _{\hat{\tau}} )^{+}$ in \qr{e:Tbis} represents
the funding debt of the member at its default (having assumed profit-and-losses marked-to-model and realized in real time, see the proof of Lemma \ref{l:selfie} below for more detail). The funder of the member corresponds to a third party, possibly composed in practice of several entities or devices and assumed default-free for simplicity, playing the role of lender/borrower of last resort after exhaustion
 of the {internal} sources of funding provided to the member through its collateral and its hedge.

More generally,
even if one considers that the ``true'' recovery rate of the member is simply zero, playing with formal recovery coefficients $R$ and $\bar{R}$ somewhere between 0 and 1 allows
reaching any desired level of interpolation between the entry price point of view $R=\bar{R}=1$ and the reference exit price
%reference-only fair valuation 
point
 of view $R=\bar{R}=0$.
On the DVA and DVA2 issues, see
{\citeN{HullWhite13de}, \citeN{BurgardKjaer13},
\citeN{AlbaneseAndersenRisk2015b}, \citeN{AAI2015}, 
\citeN{AndersenDuffieSong2016}
and \shortciteN{AlbaneseCaenazzoCrepey15}.

%Hedging and Funding PolicySelf-Financing Condition
\subsection{Gain Process}\label{ss:fhp}

The \bank can hedge
its collateralized portfolio and
needs to
fund its whole position (portfolio, margins and hedge).
%We postulate the hedging and funding specifications
%described in \shortciteN[Section 4.2.2 page 89]{BieleckiBrigoCrepeyHerbertsson13}.

Regarding hedging, we restrict ourselves to the situation
of a fully securely funded hedge, entirely implemented by means of swaps, short sales and repurchase agreements (all traded outside the clearing house, given our assumption of a constant CCP portfolio of the member), at no upfront payment.
%{hedge via the CCP or not matters or not?}.
As explained in \shortciteN[Section 4.2.1 page 87]{BieleckiBrigoCrepeyHerbertsson13}\footnote{Or \citeN[Part I, Section 2.1]{Crepey2012bc} in journal version.}, this assumption encompasses the vast majority of hedges that are used in practice. 
%In particular, it includes 
%CDS contracts that may be used \b{by the members} for hedging their counterparty-default-losses exposures. 
Consistent with arbitrage requirements and our terminology of a risk-neutral measure $\Q$,
we assume that the vector-valued gain process $\mathcal{M}$ of unit positions in the hedging assets is a $\Q$ martingale
 (see \shortciteN[Remark 4.4.2 pages 96-97]{BieleckiBrigoCrepeyHerbertsson13}\footnote{Or \citeN[Part I, Remark 4.1]{Crepey2012bc} in journal version.} or \citeN[Proposition 3.3]{BieleckiCrepeyRutkowski11}).
We assume that the member sets up a related hedge ($-\zeta$), i.e.~a predictable row-vector process with components yielding the
(negative of)
positions in the hedging assets. The ``short'' negative notation in front of $\zeta$ is used for consistency with the idea, just to fix the mindset, that the portfolio is ``bought'' by the \bank, which therefore ``sells'' the corresponding hedge.

Regarding funding,
we assume that  
variation margins $V\!M_t=P_{\hat{t}-}$ consist of cash re-hypothecable and remunerated at OIS rates,
{while initial margins  
%and default fund contributions 
consist of segregated liquid assets 
%(e.g.~government bonds) 
accruing at OIS rates. Initial margins and default fund contributions
 are supposed to be subject to CCP fees 
$\bc_t,$ e.g. 30 basis points.
We postulate that the member can invest excess-cash at a rate
$(r_t+\thelambda_t)$ and obtain unsecured funding at a rate $(r_t+\thelambdam_t)$.

 Let $\gain$ denote the gain process (or profit-and-loss, hedging error,..)
of the member's position, held by the member itself before
$\tb$ and then, if $\tau<\Ts$, by the clearing house (as liquidator of the member's position) on $[\tb,\db].$

%The equations \qr{jeq:selffinoconsbis}-\qr{eqfs} are the central clearing analogs of the equations (4.15) and (4.17) in \shortciteN[page 97]{BieleckiBrigoCrepeyHerbertsson13}\footnote{Or (4.2) and (4.4) in journal version \citeN[Part I]{Crepey2012bc}.}.
\bl\label{l:selfie} We have
$e_0=0$ 
%(assuming the portfolio entirely entirely formed from scratch at time 0, otherwise we have $\gain_0=y$, where $y$ represents the initial endowment of the portfolio) 
and, 
for $0<t\le\db$,
\beql{jeq:selffinoconsbis}
& d\gain_t = d\Pi_t-r_t \index{w@$\mathcal{W}_t$}\Pi_t dt -J_t \left(dD_t
+ \sum_{Z\subseteq N}\epsilon_{\td_{Z}}\boldsymbol\delta_{\td_{Z}}(dt) + g_t({\Pi}_t)dt \right)
\\&\qqq-\indi{\tau<\Ts} (1-\Rf ) ({\Pi}_{t-}+\Ms _{\hat{t}})^+ dJ_t - \zeta_t d\mathcal{M}_t, \eeql 
where, for any $\pi\in\R,$ 
\beql{eqfs}
\theg_t(\pi)&=
{  \bc_t (\MMs _t-P_{\hat{t}-} )}
% \MMs _t
 + {\thelambdam}_t \left(\pi + \Ms _t \right)^{+}
 -{\thelambda}_t \left(\pi+ \Ms _t \right)^{-}.
\eeql
\el

\brem \label{rem:g} The self-financing equation \qr{jeq:selffinoconsbis}  
holds for any funding coefficient
$\theg_t= \theg_t(\pi)$ there, not necessarily given by \qr{eqfs}, as soon as $( { r_t \Pi}_t +g_t({  \Pi}_t))dt$
represents the $dt$-funding cost of the member (whilst the member is alive, and net of the funding cost of
its hedge that is already comprised in the local martingale $\zeta_t d\mathcal{M}_t $). 
\erem

\subsection{Pricing BSDE}\label{ss:ch}

\bd
\rm
\label{d:sellp}
We call $\Pi$ a (no arbitrage) value process for the member's portfolio if $\Pi_{\db} = \indi{\tau<\Ts}\Rs$ and the ensuing gain process $\gain$ (cf. \qr{jeq:selffinoconsbis})
is a risk-neutral local martingale. 
\ed 
 
%The following result is the central clearing analog of
%\shortciteN[Lemma 4.4.7 page 99]{BieleckiBrigoCrepeyHerbertsson13}\footnote{Or Lemma 4.1 in journal version \citeN[Part I]{Crepey2012bc}.}.
\bp\label{l:justif}
A semimartingale $\Pi$ is a \price process for the member's portfolio 
%(\price from the member's perspective)  
if and only if it satisfies the following valuation
\index{BSDE!price-and-hedge}BSDE on $[0,\db]$:
\beql{martcond}
&\Pi_{\db} = \indi{\tau<\Ts} \Rs \mbox{ and, for }t\leq \db ,\\
& d\Pi_t =
 r_t \index{w@$\Pi_t$}\Pi_t dt +\indi{\tau<\Ts} (1-\Rf ) ( \Pi_{t-}+\Ms _{\hat{t}})^+ dJ_t\\&\qqq +J_t \left(dD_t
+ \sum_{Z\subseteq N}\epsilon_{\td_{Z}}\boldsymbol\delta_{\td_{Z}}(dt) + g_t(\Pi_t)dt \right)
+d\nu_t,
\eeql
for some local
%initially null
martingale $\nu$. 
%$\Pi$ is a \price process for the member's portfolio if and only if
%the corresponding hedging error $\gain$ is a risk-neutral local martingale.
%%the member, if $\tau>\Ts$ (hence $\db=\Ts$), or the clearing house (as its liquidator), if $\tau<\Ts$ (hence $\db=\td$), is left break-even.
\ep
\proof In view of \qr{jeq:selffinoconsbis}, \qr{martcond} is equivalent to $d\gain_t=d\nu_t- \zeta_t d\mathcal{M}_t.$ Since $\zeta_t d\mathcal{M}_t$ defines a local martingale, therefore $e$ and $\nu$ are jointly local martingales or not, which establishes the proposition.~\finproof \\

\noindent
Note that,  assuming $\nu$ a true martingale,
equivalently to the differential formulation \qr{martcond}, we can write (absorbing the $r_t \Pi_t dt$ term from \qr{martcond} into the risk-neutral discount factor $\beta$ in \qr{c:niceequivsplit})  :  
\beql{c:niceequivsplit}
&\beta_t {\Pi}_t =\E_{t} \Big[ \indi{\tau<\Ts} \left(\beta_{\td} \Rs +\beta_\tau (1-\Rf ) (\Pi_{\tau-}+\Ms _{\hat{\tau}})^{+} J_t \right)
\\&\qqq- \sum_{t< \td_{Z} <\tb} \beta_{\td_{Z}}\epsilon_{\td_{Z}} - \int_t^{\tb}
\beta_s J_s \Big(dD_s  + g_s(\Pi_s)ds  \Big)
 \Big]\sp  0\leq t\leq \db.
\eeql

\subsection{CCVA Representation}\label{ss:gencasecva}

In
this section we define the central counterparty valuation adjustment (CCVA) and derive the corresponding
BSDE.

\begin{defi} Given a
\price $\Pi$ for the member,
 the corresponding \index{CCVA}CCVA is the process defined on $[0,\db]$
as $\Theta  = -( Q + \Pi).$
%\beql{e:newdefth}{\Theta}_t = Q_t - \Pi_t ,\,\ttd.\eeql
\end{defi}
\brem\label{rem:sgn}
Recall from \qr{c:eq:pasmark} that $Q=P+\Delta ,$ with all values viewed from the perspective of the clearing house.
%Hence, the mark-to-market from the perspective of the member is $(- Q).$
Consistent with the usual definition of a valuation adjustment (see \shortciteN{BrigoMoriniPallavicini12} or \shortciteN{BieleckiBrigoCrepeyHerbertsson13}), we have $\Theta =(- Q )- \Pi,$ where $(- Q)$ corresponds to the perspective of the member.
\erem
Let
\beql{e:xib}
\xib _{t}= \E( {\beta}_{t}^{-1} {\beta}_{\tau+\delta} \xi \,|\, \G_{t}),
\eeql
where $\xi=(1-R ) (Q_{\td}-
\MMs _{\hat{\tau}})^+$ as before (cf.~\qr{c:eq:pasmark}). Let
$\hat{\xi}$ be a $\gg$ predictable process, which exists by Corollary 3.23 2) in \passhortciteN{He1992}, such that
\beql{thephi}
%\indi{\tau<\Ts}
\hat{\xi} _{\tau}=%\indi{\tau<\Ts}
\E(\beta_\tau^{-1} {\beta}_{\td} \xi \,|\, \G_{\tau-})= %\indi{\tau<\Ts}
\E( \bar{\xi} _{\tau} \,|\, \G_{\tau-}).
\eeql
In particular, in the no-DVA case with $R=1,$ then
$\xi=\bar{\xi}=\hat{\xi}=0.$ 
%which will imply that $\Theta=0$ on $[\tb,\db]$ (see the beginning of the proof of Proposition \ref{c:supgen}), as expected.

%The following result is the central clearing analog of \shortciteN[Proposition 4.5.2 page 104]{BieleckiBrigoCrepeyHerbertsson13}\footnote{Or Proposition 2.1 in journal version \citeN[Part II]{Crepey2012bc}.}
%in the case of bilateral transactions.

\bp \label{p:supgen} Let there be given semimartingales $\Pi$ and $\Theta$ such that $\Theta=-(Q+\Pi)$ on $[0,\db]$.
The process $\Pi$ is a
\price process for the member's portfolio if and only if
the process $\Theta$
satisfies the following BSDE:
\beql{nice}
\beta_t {\Theta}_t& =\E_{t}
\Big[
\sum_{t< \td_{Z} <\tb} \beta_{\td_{Z}}\epsilon_{\td_{Z}}
 - \indi{\tau<\Ts} \left(
\beta_{\td} \xi
+\beta_{\tau} (1-\Rf ) (P_{\tau-}-\MMs _{\hat{\tau}}+\Theta_{\tau-} )^{-}J_t\right)
\\&\qqq +
\int_t^{\tb}
\beta_s \big(  g_s(-P_s -\Theta_s)  \big) ds
 \Big]\sp t\in [0,\db].
\eeql

\ep
\proof Assuming $\Theta$ defined as $-(Q+\Pi)$ for some \price process $\Pi$ on $[0,\db]$,
then
the terminal condition ${\Theta}_{\db}=-\indi{\tau<\Ts} \xi$ that is implicit in \qr{nice} results from
 \qr{c:eq:pasmark} and the terminal condition for $\Pi$ in \qr{martcond}. Moreover, we have, for $t\in[0,\db]$,
\beql{e:fortheta}
&-\beta_t {\Theta}_t =\beta_t {Q}_t+\beta_t
{\Pi}_t=
\beta_t {P}_t+ \int_0^t\beta_s dD_s+(\beta_t
{\Pi}_t-
\int_0^{t}
%\int_0^\tau
 \beta_s J_s dD_s),
\eeql
hence
\bel
&-\beta_t {\Theta}_t
- \int_0^{t}
\beta_s J_s \Big( \sum_{Z\subseteq N}\epsilon_{\td_{Z}}\boldsymbol\delta_{\td_{Z}}(ds) + g_s(-P_s -\Theta_s)ds  \Big) \\
&\quad\quad-\indi{\tau<\Ts}\int_0^{t}
%\boldsymbol\delta_{\tau}(ds)
(1-\Rf ) ( -P_{s-} -\Theta_{s-} +\Ms _{\hat{s}})^+ dJ_s=\Big(\beta_t {P}_t+ \int_0^t\beta_s dD_s \Big)
+\int_0^t \beta_s d\index{e@$\varepsilon_t$}\nu_s,
\eel
by the pricing BSDE \qr{martcond} satisfied by $\Pi$. In view also of \eqref{cleanf} (used for $i=0$ here), this is a {(local)} martingale, hence it coincides with the conditional expectation
of its terminal condition (assuming it is a true martingale), which establishes \qr{nice}.
%Hence, \eqref{martcond} implies \qr{nice}.
The converse implication is proven similarly.~\finproof
\brem
As an alternative argument equivalent to the above, one can substitute
the right-hand side in \qr{c:niceequivsplit} for
$ \beta_t {\Pi}_t$ in \qr{e:fortheta}, which, after an application of the tower rule, yields \qr{nice}. One can proceed similarly
to show \qr{c:niceequivsplit} if \qr{nice} is assumed.
\erem
Let, for $\vartheta\in\R$,
\beql{e:prelfin}
\mygb_t( \vartheta )
&=
 g_t(-P_t -\vartheta)
 -\gamma_t \hat{\xi}_{t}
-(1-\Rf ) \gamma_t (P_t -\Ms _t+ \vartheta)^-
\\& = \underbrace{-\gamma_t \hat{\xi}_{t}}_{dva_t} +
\underbrace{\big( {\bc_t ( \MMs _t-P_{\hat{t}-})}
 + {\thelambdat}_t \left(P_t -\Ms _t + \vartheta\right)^{-}
 -{\thelambda}_t \left(P_t -\Ms _t + \vartheta \right)^{+} \big)}_{fva_t(\vartheta)} 
 , \eeql
by definition \qr{eqfs} of $g$,
where $\thelambdat ={\thelambdam} - (1-\Rf) \gamma $
(recall
$\gamma=\gamma^0$ is the assumed intensity of $\tau$).
%{By a (left-limited) process $Y$ stopped at ${\theomega-}$, we mean the process $Y^{\theomega-}=JY +(1-J)Y_{-} .$}
From the perspective of the \bank,
the two terms
in the decomposition \eqref{e:prelfin}
 of the coefficient $\mygb_t ( \vartheta)$
 can respectively be interpreted as a beneficial
 debt valuation adjustment coefficient ($dva_t$ that can be ignored by setting $R=1$)
and a funding valuation adjustment coefficient ($fva_t(\vartheta)$ in which the DVA2 component can be ignored by setting $\Rf=1$).

\bp \label{c:supgen} The ``full CCVA BSDE'' \qr{nice}
for a semimartingale $\Theta$ on $[0,\db]$ is equivalent to the following ``reduced CCVA BSDE'' for a semimartingale $\hat{\Theta}$
on $[0,\tb]:$
\beql{red}
\beta_t {\hat{\Theta}}_t &={\E}_{t}
\Big[
\sum_{t< {\tau}^{{\delta}}_{\Zt} < \tb}\beta_{{\tau}^{{\delta}}_{\Zt}} {\epsilon}_{{\tau}^{{\delta}}_{\Zt}}+
\int_t^{\tb}
\beta_s   \mygb_s( \hat{\Theta}_s) ds
 \Big]  \sp \ttd,
\eeql
equivalent in the sense that if $\Theta$ solves \qr{nice}, then
 $\hat{\Theta}=J{\Theta}$ solves \qr{red}, whilst
if $\hat{\Theta}$ solves \qr{red}, then
${\Theta}=J\hat{\Theta}-(1-J)\indi{\tau<\Ts}\xib$ solves \qr{nice}.
\ep

\proof The full CCVA BSDE
\eqref{nice} is obviously equivalent to
${\Theta}=-\indi{\tau<\Ts}\xib$ on $[\tb,\db]$ and 
\bel%{nicebi}
\beta_t {\Theta}_t &=\E_{t}
\Big[
\sum_{t< \td_{Z} <\tb} \beta_{\td_{Z}}\epsilon_{\td_{Z}}
 - \indi{\tau<\Ts}\beta_{\tau} \left(
\xib_{\tau}
+ (1-\Rf ) (P_{\tau-}-\Ms _{\hat{\tau}}+\Theta_{\tau-} )^{-} \right)
\\&\qqq +
\int_t^{\tb}
\beta_s g_s(-P_s -\Theta_s) ds 
 \Big]
\eel on $[0,\tb),$
which is in turn equivalent to
\beql{e:fullbis}
&{\Theta}=-\indi{\tau<\Ts}\xib\mbox{ on }[\tb,\db]\mbox{ and, on } [0,\tb), \\
&
\beta_t {\Theta}_t ={\E}_{t}
\Big[
\sum_{t< {\tau}^{{\delta}}_{\Zt} \leq \tb}\beta_{{\tau}^{{\delta}}_{\Zt}} {\epsilon}_{{\tau}^{{\delta}}_{\Zt}}+
\int_t^{\tb}
\beta_s   \mygb_s( {\Theta}_s) ds
 \Big]  ,
\eeql because on $[0,\tb):$
\bel
\E_{t}
\Big[& \indi{\tau<\Ts}\beta_{\tau} \left(
\xib_{\tau}
+ (1-\Rf ) (P_{\tau-}-\Ms _{\hat{\tau}}+\Theta_{\tau-} )^{-} \right)\\\qqq
&=\E_{t}
\Big[ \indi{t<\tau<\Ts}\beta_{\tau} \left(
\hat{\xi}_{\tau}
+ (1-\Rf ) (P_{\tau-}-\Ms _{\hat{\tau}}+\Theta_{\tau-} )^{-} \right)\Big]
\\\qqq
&=-
\E_{t}
\Big[ \int_t^{\Ts} \beta_{s} \left(
\hat{\xi}_{s}
+ (1-\Rf ) (P_{s-}-\Ms _{\hat{s}}+\Theta_{s-} )^{-} \right)dJs\Big]
\\\qqq
&=
 \E_{t}
\Big[ \int_t^{\Ts} \beta_{s} \gamma_s\left(
\hat{\xi}_{s}
+ (1-\Rf ) (P_{s-}-\Ms _{\hat{s}}+\Theta_{s-} )^{-} \right)ds\Big].
\eel
Here the last identity holds by consideration of the (local, assumed true) martingale
\bel&\beta_{t}(\hat{\xi}_{t} + (1-\Rf ) (P_{\tau-}-\Ms _{\tau-}+ \Theta_{\tau-} )^{-}) dJ_t + \beta_{t}\gammat _t( \hat{\xi}_t
+ (1-\Rf ) (P_{t}-\Ms _{t}+\Theta_{t} )^{-} )
 dt.\eel
One readily checks that if $\Theta$ solves \qr{e:fullbis}, then
 $\hat{\Theta}=J{\Theta}$ solves \qr{red}, whilst
if $\hat{\Theta}$ solves \qr{red}, then
${\Theta}=J\hat{\Theta}-(1-J)\indi{\tau<\Ts}\xib$ solves \qr{e:fullbis}.
\finproof\\
 
Note that, provided $r$ and $\tilde{\lambda}$ are bounded fro below, the reduced BSDE coefficient
$\mygb_t(\vartheta)$ in \eqref{e:prelfin}
 satisfies the monotonicity assumption
$$
\big(\hat{f}_t(\vartheta) - \hat{f}_t(\vartheta')\big)
(\vartheta - \vartheta')
 \leq C(\vartheta-\vartheta')^2 ,
$$
for some constant $C$. Then,
under mild integrability conditions,
the reduced CCVA BSDE \eqref{red} is well-posed in
the space of
square integrable solutions (see e.g.  \citeN[Sect.~5]{KrusePopier14}). By virtue of
Proposition \ref{c:supgen}, so is in turn
the full CCVA BSDE \eqref{nice}. 

\brem In the terminology of \citeN{CrepeyNguyen15},
\qr{red} is the ``partially reduced'' CCVA BSDE (cf. also Lemma 2.3 in \citeN{Crepey13a1}),
while the ``fully reduced'' BSDE
(simply called ``reduced''  in  \citeN{CrepeySong15})
is the BSDE on the time interval $[0,T]$ obtained from \qr{red} by projection on a smaller filtration (the market or reference filtration myopic to the defaults of the two parties). In this paper we only work with the partially reduced BSDE in order to avoid the enlargement of filtration technicalities.
%, which also corresponds to Theorem 4.3 in \citeN{Crepey13a1}.
\erem

\subsection{Cost of Capital}\label{ss:coc}
The capital at risk of the member is
composed of its default fund contribution $DFC_{t},$ which represents implicit
capital at risk, and of its regulatory CCP capital $K^{cm}_t$ 
as of \qr{e:kcm}.
Along the lines of \shortciteN{AlbaneseCaenazzoCrepey15},
we define the capital valuation adjustment (KVA) of the member as the cost of remunerating its capital at risk 
$K_t =DFC_{t}+K^{cm}_t$
at some hurdle rate $k$ throughout the whole life of the portfolio (or until the member defaults). Such a KVA
is given  by the following formula (cf.~\shortciteN{AlbaneseCaenazzoCrepey15}):
\beql{e:kexpl}
\mbox{KVA}_t=k \E_t \int_t^{\tb} e^{-\int_t^s   (r_u  +k)
 du} K_s ds ,\,t\in[0,\tb].
\eeql 
The KVA-inclusive CCVA is then defined as the sum between our previous 
CCVA $\Theta$ and this KVA. 
%The reader is referred to \shortciteN{AlbaneseCaenazzoCrepey15} for a more detailed KVA story that underlies the formula \qr{e:kexpl}.
  
\section{Common Shock Model of Default Times}\label{s:DMO}
%\section{CCVA Common-Shock Model}\label{s:DMO}

\def\e{Z}
\def\mys{0}

We use a dynamic Marshall-Olkin (DMO) copula model
of the default times $\tau_i$
(see \shortciteN[Chapt.~8--10]{BieleckiBrigoCrepeyHerbertsson13}\footnote{Or \shortciteANP{BieleckiCousinCrepeyHerbertsson11} (\citeyearNP{BieleckiCousinCrepeyHerbertsson11},\citeyearNP{Bielecki2012ab}) for the journal versions.} and \citeN{CrepeySong15}).
As demonstrated in \shortciteN[Sect.~8.4]{BieleckiBrigoCrepeyHerbertsson13}\footnote{Or \shortciteN[Part II]{Bielecki2012ab} in journal version.}, such a model can be efficiently calibrated to marginal and portfolio credit data, e.g.~CDS and CDO data (or proxies) on the members.
The joint defaults feature of the DMO model is also interesting
in regard of the EMIR ``cover two'' default fund sizing rule (cf. \sr{ss:regulccva}).

Let there be given a family $\cY$
of
``shocks'', i.e.~subsets $Y$ of members, typically the singletons
$\{0\},\{1\},\ldots,\{n\}$
and a
small number of ``common shocks'' representing simultaneous
defaults.
%including the \emph{shock intensities}, labelled $\gamma^Y$.
%Values
%of $\mathbf{X}$ (real vectors of the same dimension as $\mathbf{X}$) are denoted by $\mathbf{x}$.
%, including the components corresponding to the $\gamma^Y$ labelled as $x_Y$.
%%For instance,
%%the $\gamma^Y$ can be
%%given in the form of extended CIR processes
%%\begin{eqnarray} \label{varX}
%%dX^{Y}_t=a(b_Y (t)-\gamma^Y_t)dt+c \sqrt{\gamma^Y_t} dW^Y_t,
%%\end{eqnarray}
%%for nonnegative constants $a$ and $c$, functions $b_Y (t)$
%%and for independent Brownian motions $W^Y$ in their own completed filtration $\mathbb{W}=(\mathcal{W}_t)_{t\geq 0}$,
%%under the pricing measure $\mathbb{Q}$
%%(the case of deterministic intensities $X^{Y}_t= b_Y (t)$ can be embedded in this framework as the limiting case of an ``infinite mean-reversion speed''
%%$a$).
%%%; see \shortciteN{BieleckiCousinCrepeyHerbertsson12}).
For $Y\in {\cal Y},$ we define
\bel%{e:dtau}
\theeta_Y =\inf \{\myt>\mys;\,
\int_0^\myt \gamma_Y(s) ds > E _{Y}\}\sp
{J}^{Y} =\ind_{[0,\eta_Y)} ,
\eel
for a shock intensity function $\gamma_Y(t)$ and
an independent standard exponential random variable $ E _{Y}$.
We then set
   $$\tau_i=\min_{\{Y\in\cY; i\in Y\}}\theeta_Y\sp i\in N.$$
%\begin{frame}
%\bb{Example: $n=5$ and ${\cal Y}=\{\{0\},\{1\},\{2\},\{3\},\{4\},\{3,4\},\{1,2,3\}, \{0,1\}\}.$}
%\begin{center}
% \includegraphics[scale=0.5]{Model_example_5_names_cp_risk_uni}%
%\end{center}
%\eb
%\begin{itemize}
%\item[\fl] Default intensity of the member:
%$\emph{\gamma_{\bullet}}= \sum_{Y\in\cY_{\bullet}}\Intens_Y ,$
%where $\cY_{\bullet}=\{Y\in\cY; 0\in Y\}.$
%\end{itemize}
%\end{frame}
\bex\label{extraj} \fr{fig:Model_example_5_names} shows one possible default path
in a common shock model with $n=5$ and ${\cal Y}=\{\{0\},\{1\},\{2\},\{3\},\{4\},\{3,4\},\{1,2,3\}, \{0,1\}\}.$
The inner ovals show which shocks happen and cause the observed defaults at successive default times. First, the default of name $1$ occurs as the consequence of the shock $\{1\}$. Second, names $3$ and $4$ default simultaneously as a consequence of the shock $\{3,4\}$. Third, the shock $\{1,2,3\}$ triggers the default of name $2$ alone (as name $1$ and $3$ have already defaulted). Fourth, the default of name $0$ alone occurs as the consequence of shock $\{0,1\}$.
\begin{figure}[H]
\begin{center}
\includegraphics[scale=0.8]{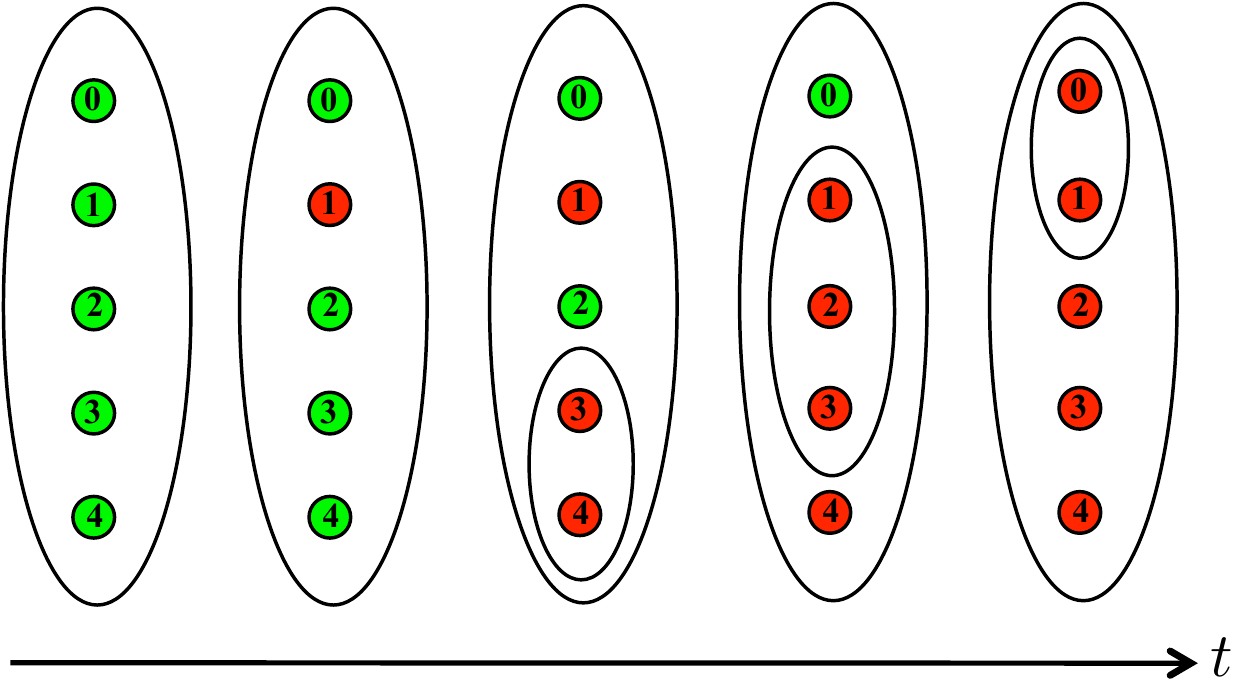}%
 \end{center}
\caption{One possible default path
in a model with $n=4$ and ${\cal Y}=\{\{0\},\{1\},\{2\},\{3\},\{4\},\{3,4\},\{1,2,3\}, \{0,1\}\}.$}
 \label{fig:Model_example_5_names}
\end{figure}
\eex
Again, in the case of the reference member (labeled 0), we omit the superscript 0 in the notation. In particular, we have
$J=\ind_{[0,\tau)}=\prod_{Y\in\cY_\bullet} J^Y,$ where $\cY_\bullet =\{Y \in \cY;\,0 \in Y\}$,
hence the intensity $\gamma$ of $\tau$ is given as
\beql{e:gamtau}\gamma=J_{-}\gamma_{\bullet}\mbox{, where }\gamma_{\bullet}= \sum_{Y\in\cY_{\bullet}}\Intens_Y .\eeql
We assume that all the market risk factors
are gathered in a vector process $\mathbf{X}$ without jump at $\tau$ and that the processes $\mathbf{X}$ and $\Xd=(\mathbf{X },\mathbf{J}),$
where
$\mathbf{J} =
(J^Y)_{Y\in\cY},$
are Markov in the full model filtration $\gg$ given as the
filtration
of $\mathbf{X }$
 progressively enlarged by the random times $\eta_Y, Y\in\cY$
(in
\sr{s:num}-\ref{s:resu}, $\mathbf{X }$ is simply a Black-Scholes stock $S,$ 
%possibly 
augmented by additional factors in order to cope with
the potential path dependence of dividends and collateral).
%,  and see \citeN[Section 7]{CrepeySong15} for possible wrong-way risk extensions of this setup).
Setting $\hat{\Delta}_t =\int_0^t e^{ \int_s^t r_u du} dD_s$ so that $\beta_t\Delta_t=\beta_t\hat{\Delta}_t-
\beta_\tau\hat{\Delta}_{\tau-}$ for $t\geq \tau$,
%and $\Xd=(\mathbf{X },\mathbf{\theH }),$
we assume, consistent with the interpretation of each respective quantity, that
\bel%{e:ac}
&{\epsilon}_{t}
={\epsilon}(t,\Xt_{t}) \mbox{ for } t={\tau}_{\Zt}^\delta\sp
Z\subseteq N \\
&P_t=P(t,\mathbf{X }_t)\sp \hat{\Delta}_t=\hat{\Delta}(t, \mathbf{X }_t)\sp \MMs_t=\MMs(t,X_t)\sp \ttd
\eel
(having augmented $\mathbf{X }$ by $\hat{\Delta}$ and/or $\MMs$ if need be), for continuous functions
${\epsilon}(t,x)$, $P(t,\mathbf{x}),$ $\hat{\Delta}(t, \mathbf{x})$ and $\MMs(t,x).$
In particular, it holds that
$$\Delta_\tau=\hat{\Delta}_\tau-
 \hat{\Delta}_{\tau-}=\hat{\Delta}(\tau,\mathbf{X }_{\tau})-
 \hat{\Delta}({\tau},\mathbf{X }_{\tau-}) =0,$$
by continuity of $\mathbf{X }$ at $\tau$ (as opposed to $\Delta_\tau\neq 0$ in
the gap risk model of \citeN{CrepeySong15}). 
\noindent
\bl\label{l:dva} We have
\bel
dva_t= dva (t,\Xt_t )= -J_t
 \xib \big(t, {X}_t\big) \gamma_{\bullet} \sp \Q\times\boldsymbol\lambda ~ a.e.,\eel
for a function ${\bar{\xi}} (t, x)$ such that
$\bar{\xi} _{\tau}= {\bar{\xi}}(\tau, {\Xd}_{\tau-}).$
\el

\section{XVA Engines}\label{num}
%\section{Numerical Implementation}\label{num}

%With CCVA, the scope of the system passes from a CSA portfolio of assets with one counterparty, in the case of bilateral OTC trading, to all the transactions of a service (or market) and all the members of a clearing house, in the case of centrally cleared trading.
%Hence, CCVA represents an unprecedented computational challenge for banks.
In this section, we summarize in algorithmic terms the central clearing XVA methodology of this paper, as well as a bilateral trading XVA methodology recalled for comparison purposes from \citeN{CrepeySong15}. In both cases we use the common shock model of \sr{s:DMO} for modeling the default times involved. 
%TVA is used as a unified acronym
%for CCVA (central clearing valuation adjustment) in the CCP setup and BVA (bilateral valuation adjustment) in the CSA setup.

\subsection{CCVA Engine}\label{ss:ccvaengine}
In spite of the nonlinearity inherent to the funding component of the CCVA,
standard Monte Carlo loops can be used for estimating a linearized first order CCVA obtained
replacing 
%$g_s(P_s -\Theta_s)$ by $g_s(P_s)$ and in turn
$fva_s( \Theta_s)$ by $fva_s(0)$ in \qr{e:prelfin},
i.e. $\mygb_s( \hat{\Theta}_s)$ by $\mygb_s( 0)$ in \qr{red}. A nonlinear correction can be estimated based on the Monte Carlo expansion of \citeANP{FujiiTakahashi12a} (\citeyearNP{FujiiTakahashi12a},\citeyearNP{FujiiTakahashi12b}) (further studied in \citeN{GobetPagliarani15})  in vanilla cases, with explicit formulas for $P_t$,
or by the branching particles scheme of \citeN{HenryLabordere12} in more exotic situations. In the bilateral trading setup of \citeN{CrepeySong15} (see also \citeN{CrepeyNguyen15}), the nonlinear correction is consistently found less than 5\% to 10\% of the linear part. Hence, in this paper, we just use the linear part. 
We obtain by first order linear approximation 
in the reduced CCVA BSDE \qr{red}: 

\beql{e:exec}
%{redo}
&{\Theta}_0 =\hat{\Theta}_0 \approx {\E}
\Big[
\sum_{0<{\tau}^{{\delta}}_{\Zt}< \tb}\beta_{{\tau}^{{\delta}}_{\Zt}} {\epsilon}_{{\tau}^{{\delta}}_{\Zt}}+
\int_0^{\tb}
\beta_s   \mygb_s( 0
) ds
 \Big] =\underbrace{{\E}
\sum_{0<{\tau}^{{\delta}}_{\Zt}< \tb}\beta_{{\tau}^{{\delta}}_{\Zt}} {\epsilon}_{{\tau}^{{\delta}}_{\Zt}}}_{\mbox{{CVA}}}+\underbrace{
\E\int_0^{\tb}
\beta_s dva_s ds}_{\mbox{{DVA}}}
\\&\qqq
+\underbrace{{\E}
\int_0^{\tb}
\beta_s \Big( \tilde{\lambda}_s \left(\Ms_s
-  P_s \right)^{+}
 -{\thelambda}_s \left(\Ms _s
-P_s \right)^{-}  \Big)  ds}_{{\mbox{{MVA}}}}+\underbrace{{\E}
\int_0^{\tb}
\beta_s  
{ \bc_s ( \MMs _s -P_{\hat{s}-})}  ds}_{{\mbox{{MLA}}}}
 ,
\eeql
where
$\beta_t=e^{-\int_0^t  r_s  ds} ,$
$\tilde{\lambda}=\bar{\lambda}-(1-\Rf)\gamma_{\bullet}$,
$\Ms=V\!M  +I\!M   $
and, for each $t={\tau}^\delta_{\Zt}< \tb$,
\bel
&{\epsilon}_{t}=
\big(B_{t}- E_{t-}\big)^+ \frac{DFC_{t}}{\sum_{j\in N}J^j_{t}DFC^j_{t},
}\mbox{ in which }B_{t}=
\sum_{i\in \Zt}(P^i_{t} +\Delta^i_{t}
%{ -P_0}
-
\MMs ^i_{t})^+  
%\sp \MMs ^i_{t}=\MMs ^i_{\hat{\tau}_{\Zt}}
\eel
with, for each member $i,$ $\MMs^i =V\!M^i +I\!M^i +DFC^i  $
(cf.~\qr{e:dfb} and \qr{c:fullcollat}-\qr{e:breach}).
In addition,
%\item[$\bullet$]
$dva =-\gamma
 \hat{\xi} ,$
where $\hat{\xi}$ is a predictable process such that
$
\hat{\xi} _{\tau}= \E( {\beta}_{\tau}^{-1} {\beta}_{\td} \xi \,|\, \G_{\tau-})$
(cf.~\qr{thephi}),
with
$
\xi
=
(1-R)(P_{\td}+\Delta_{\td} -\MMs_{\tau})^+ .
$
%The DVA can be ignored by setting $R=1$\index{debt valuation adjustment (DVA), hence
%$\xi=\hat{\xi}=0$}.

The ${\epsilon}$ terms
in \qr{e:exec} give rise to a CVA paid
by the member through
its contributions to the refill of realized breaches.
%contribution to the covering of other members realized breaches whilst it is alive.
{The terms dubbed MVA and MLA in \qr{e:exec}, where
\bel%{e:mvl}
\Ms _s-P_{s}= P_{\hat{s}-}+ I\!M_s  -P_{s}\approx
I\!M_s  \mbox{ and } \MMs _s -P_{\hat{s}-}=
I\!M_s +DFC_s ,\eel
are interpreted as a margin valuation adjustment (cost to the member of funding its initial margins, essentially)
%, as 
%$P_{\hat{s}-}$ is essentially compensated by $P_{s}$ in the frictionless variation margins setup of this paper commented upon in the remark \ref{rem:thresh}, so that there is  virtually no cost of funding the variation margins)
and a margin liquidity adjustment (cost to the member of the CCP margin fees).}
The positive (respectively negative) terms in
\qr{e:exec} can be considered as deal adverse (respectively deal friendly) as they increase (respectively decrease) the CCVA $\Theta$. The DVA and the DVA2 can be ignored in $\Theta$ by setting $R=1$ and $\Rf=1$, respectively.

For numerical purposes, we use the following randomized version of \qr{e:exec}:
\beql{e:execnum}
& {\E}
\Big[
\sum_{0<{\tau}^{{\delta}}_{\Zt}< \tb}\beta_{{\tau}^{{\delta}}_{\Zt}} {\epsilon}_{{\tau}^{{\delta}}_{\Zt}}+
 \indi{\zeta<\tb}\frac{e^{\mu\zeta}}{\mu}
\beta_{\zeta}   \mygb_\zeta( 0
)
 \Big] ,
\eeql
where $\zeta$
denotes an independent exponential time of parameter $\mu$.
Moreover, to deal with the $dva_\zeta$ term in
$\mygb_\zeta( 0
) $,
we use the following result.

\bl\label{l:nested} For any predictable process $h$ and independent atomless random variable $\zeta,$
%with density $p$,
we have:
\beql{e:curepe}
 &\mathbb{E}[\ind_{\{\zeta< \tb\}}h_\zeta \beta_\zeta dva (\zeta,\Xt_\zeta)]=
-\mathbb{E}\Big[\ind_{ \{\zeta< \tb\}} h_\zeta
\beta_{\zeta+\delta} (1-R)
\Intens_{\bullet}(\zeta)
\big(Q_{\zeta^\delta}-\Ms_{\zeta}\big)^+\Big].
\eeql 
\el
 
Plugging $h_\zeta=\frac{e^{\thatmu \zeta }}{\thatmu }$
in \qr{e:curepe} to deal with
the $dva_\zeta$ term in
$\mygb_\zeta( 0
) $,
\qr{e:execnum} is rewritten as
\beql{e:finalhat}
& \hat{\Theta}_0\approx {\E}\Big\{
\sum_{0<{\tau}^{{\delta}}_{\Zt}< \tb}\beta_{{\tau}^{{\delta}}_{\Zt}} {\epsilon}_{{\tau}^{{\delta}}_{\Zt}}+\indi{\zeta<\tb}\frac{e^{\mu\zeta}}{\mu}\times
 \\&
\Big[
- \beta_{\zeta^\delta }
\Intens_{\bullet}(\zeta)(1-R)
\big(Q_{\zeta^\delta}-\MMs_{\zeta}\big)^+
 +\beta_{\zeta} \Big(  \tilde{\thelambda}_\zeta (\Ms _\zeta 
- P_\zeta)^{+}
 -{\thelambda}_\zeta  (\Ms _\zeta 
-P_\zeta)^{-}  \Big)  \Big]
\Big\}.
\eeql
The KVA-inclusive CCVA is then defined as the sum between
\qr{e:finalhat} and a $\mbox{KVA}$ as of \qr{e:kexpl}, valued at time $t=0$ by simulation and randomization of the time integral there.

\subsection{BVA Engine}\label{ss:bil} Here we provide an executive summary of a bilateral CSA trading setup recalled for comparison purposes from
\citeN{CrepeySong15} (cf. also \citeN{PallaviciniBrigo13bprel}  or \citeN{BichuchCapponiSturm16}
for related bilateral counterparty risk analyses with asymmetric funding costs).

\brem In \citeN{CrepeySong15}, 
the cash flows are viewed from the perspective of the bank, which will be taken as the reference member here, whereas we view them in this paper from the perspective of the clearing house, i.e. opposite to the one of the member. Hence, the sign conventions are opposite, i.e.~$P, \Delta, Q,$ etc... in this paper correspond to their opposites in \citeN{CrepeySong15}, which is why we see
$\cdot^{\mp}$ here whenever we have $\cdot^{\pm}$ there.
\erem

In the context of bilateral trading between a bank, taken as the reference member labeled by 0 in the previous CCP setup, and a counterparty taken as another member $i\neq 0,$
let \index{v@$\vm$}$\vm$ denote the variation margin,
 where $\vm\geq 0$ (resp.~$\leq 0$) means collateral posted by the bank and received by the counterparty (resp.~posted by the counterparty and received by the bank). Let \index{i@$\ib$}$\ib\geq 0$
and $\ic\leq 0$ represent the initial margin
posted by the bank and~the negative of the initial margin posted by the counterparty. Hence,
\beql{e:margins}\index{c@$\cb$}\cb=\vm+ \index{i@$\ib$}\ib
\mbox{~~and~~}
\index{c@$\cc$}\cc=\vm+ \index{i@$\ic$}\ic
 \eeql
are the total collateral guarantee for the counterparty and the negative of the total collateral guarantee for the bank. {Assuming the variation margins re-hypothecable and the initial margins segregated (as typically so in practice),
the collateral funded by the bank is \index{c@$\Cb$}
$\Cb= \vm+ \ib $}. For consistency with our CCP setup, $\vm_t$ will be taken as $P_{\hat{t}-}$. So, in the spirit of a standard CSA,
 we are considering full
collateralization, and even overcollateralization through the initial margins. {We assume that $\vm$ and $\ib$ are remunerated at the OIS rate $r$.
Following \citeN{CrepeySong15}, at time 0,
the difference ${\Theta}_0$ between the mark-to-market of the portfolio and its 
value inclusive of counterparty and funding risk 
(both from the perspective of the bank, cf.~the remark \ref{rem:sgn}), 
difference
dubbed BVA for bilateral valuation adjustment, can be linearized as follows:
\beql{e:bilat}
%{redo}
&{\Theta}_0 =\bar{\Theta}_0 \approx {\E}
\Big[
\int_0^{\tb}
\beta_s   \bar{f}_s( 0
) ds
 \Big]  = \underbrace{{\E}
\int_0^{\tb} \beta_s cdva_s ds
}_{\mbox{CDVA}} +
\\&\qqq
+\underbrace{{\E}
\int_0^{\tb}
\beta_s \Big(   \tilde{\lambda}_s \left(\Cb _s
%-\vartheta
-  P_s\right)^{+}
 -{\thelambda}_s \left(\Cb _s
%-\vartheta
- P_s\right)^{-}  \Big)  ds}_{{\mbox{{MVA}}}}.
\eeql
Here:
\begin{itemize}
\item[$\bullet$] $P$ means the mark-to-market of the position of the member with the counterparty (viewed from the perspective of the
latter),
\item[$\bullet$] the meaning of $\beta,$ $\tilde{\lambda}$ and $\lambda$ is as in the CCP setup, 
\item[$\bullet$]
$\tau=\tau_b\wedge\tau_c$
is the first-to-default time of the bank and the counterparty,
% (as opposed to the default time of the member, i.e.~the bank, in the CCP setup), 
\item[$\bullet$] $cdva =\gamma
 \hat{\xi} ,$
where $\hat{\xi}$ is a predictable process such that
$
\hat{\xi} _{\tau}= \E( {\beta}_{\tau}^{-1} {\beta}_{\td} \xi \,|\, \G_{\tau-}), $
with
$$
\xi
=
\ind_{\{\tau_c\leq \tau^\delta_b\}}(1-R_c)(P_{\td}+\Delta_{\td} -\cc_{\tau})^- - \ind_{\{\tau_b\leq \tau^\delta_c\}}(1-R_b)(P_{\td}+\Delta_{\td} -\cb_{\tau})^+ ,
$$
in which the recovery rates
$R_c$ of the counterparty to the bank and $R_b$ of the bank to the counterparty
are usually taken in a bilateral trading setup as $40\%$.
% (and the DVA2 is typically ignored in practice).
\end{itemize}
For numerical purposes, we use the following randomized version of \qr{e:bilat}:
\beql{e:execnumbil}
& {\E}
\Big[
 \indi{\zeta<\tb}\frac{e^{\mu\zeta}}{\mu}
\beta_{\zeta}   \bar{f}_\zeta( 0
)
 \Big] ,
\eeql
where $\zeta$
denotes an independent exponential time of parameter $\mu$.
The $cdva_\zeta$ term in $\bar{f}_\zeta( 0
)$ is treated by the following bilateral analog of Lemma \ref{l:nested}. 
We write $\cY_b =\{Y \in \cY;\, 0  \in Y\},$
$\cY_c =\{Y \in \cY;\, i\in Y\}$ 
and we recall that $X=(\mathbf{X},\mathbf{J})$ denotes 
the market risk and common shock factor process introduced in
\sr{s:DMO}, assumed without jump at $\tau$. Similar to Lemma \ref{l:dva}, it holds that 
$ cdva_t= cdva (t,\Xt_t ).$
In addition (see Lemma 8.2 and its proof in \citeN[hal version 2]{CrepeySong15}, in a slightly more general setup where $\mathbf{X}$ may jump at $\tau$):

\bl\label{l:nestedbil} For any predictable process $h$ and independent atomless random variable $\zeta,$ 
we have:
\beql{e:toterbil}
&\mathbb{E}\big[\ind_{\{\zeta< \tb\}}h_\zeta \beta_\zeta cdva (\zeta,\Xt_\zeta)\big]=
\mathbb{E}
\Big[\ind_{ \{\zeta< \tb\}} h_\zeta \beta_{\zeta^\delta}
 \times\\&
\Big(\big(\sum_{Y\in\cY_{c}}\Intens_{Y}(\zeta) +\indi{\tau_c \leq \zeta^{\delta}}\sum_{Y\in\cY_{b}\setminus \cY_{c}}
\Intens_{Y}(\zeta) \big)
(1-R_c)\big(Q_{\zeta^\delta}-C_{\zeta}\big)^-
 \\&
 \quad\quad\quad\quad
- \big(\sum_{Y\in\cY_{b}}\Intens_{Y}(\zeta) +\indi{\tau_b \leq \zeta^{\delta}}\sum_{Y\in\cY_{c}\setminus \cY_{b}}
\Intens_{Y}(\zeta) \big)
 (1-R_b)\big(Q_{\zeta^\delta}-\cb_{\zeta}\big)^+\Big)
\Big].~\finproof
\eeql 
\el
Plugging $h_\zeta=\frac{e^{\thatmu \zeta }}{\thatmu }$
in \qr{e:toterbil} to deal with
the $cdva_\zeta$ term in
$\bar{f}_\zeta( 0
)$,
\qr{e:execnumbil}
 is rewritten as (compare \qr{e:finalhat}):
\beql{e:finalbar}
& \bar{\Theta}_0\approx {\E}\Big\{
\indi{\zeta<\tb}\frac{e^{\mu\zeta}}{\mu}
\Big[
\beta_{\zeta^\delta}\Big(\big(\sum_{Y\in\cY_{c}}\Intens_{Y}(\zeta) +\indi{\tau_c \leq \zeta^{\delta}}\sum_{Y\in\cY_{b}\setminus \cY_{c}}
\Intens_{Y}(\zeta) \big) 
(1-R_c)\big(Q_{\zeta^\delta}-C_{\zeta}\big)^-
 \\&
 \quad\quad\quad\quad
- \big(\sum_{Y\in\cY_{b}}\Intens_{Y}(\zeta) +\indi{\tau_b \leq \zeta^{\delta}}\sum_{Y\in\cY_{c}\setminus \cY_{b}}
\Intens_{Y}(\zeta) \big) 
 (1-R_b)\big(Q_{\zeta^\delta}-\cb_{\zeta}\big)^+\Big)
\\& \quad\quad +\beta_{\zeta} \Big(  \tilde{\thelambda}_\zeta (P_\zeta
- \MMs _\zeta )^{-}
 -{\thelambda}_\zeta  (P_\zeta
-\MMs _\zeta )^{+}   \Big)
\Big]\Big\}.
\eeql
{Such adjustments are then computed counterparty by counterparty and added over $i=1,\ldots,n$ to obtain the BVA of the bank.
\brem In practice, netting sets typically merge into a unique funding set, meaning that
one should solve for a single MVA at the level of the whole portfolio of the bank. However, in the present frictionless variation-margining case 
(cf.~the remark \ref{rem:thresh}), 
$$\MMs _\zeta-P_{\zeta}= P_{\hat{\zeta}-}+ I\!M_\zeta  -P_{\zeta}\approx
I\!M_\zeta \geq 0$$ holds
counterparty by counterparty, so that a unique funding set or funding by netting sets makes a negligible difference in practice.
\erem}

Similar as in the CCP setup, the KVA-inclusive BVA
is obtained by adding to
\qr{e:finalbar} a 
$\mbox{KVA}$ in the sense of the formula
\qr{e:kexpl} (valued at $t=0$),
except that $K$ is now the bilateral regulatory capital given by the formulas of \sr{ss:regulbva}.

\section{Experimental Framework}
\label{s:num}

In this section we design an experimental framework that is used for the XVA comparative numerical analysis of \sr{s:resu}.
% for studying numerically
%the netting benefit of central clearing. 
%This study is illustrative of one among many possible uses
%of our ``XVA engines''. It is simplistic in many respects. In particular,
%we only consider one clearing house trading one single asset, whereas 
%\citeN{DuffieZhu} explain how the netting benefit of central clearing can be compromised
%\r{by 
%%fragmentation, namely 
%dispersion of trades 
%among different CCPs}. 

%the multilateral netting benefit of centrally cleared trading  (see Figure \ref{f:biltoccp}) is compensated, at least partially, by the loss of bilateral netting across asset classes (see \citeN{DuffieZhu} and \passhortciteN{ContSantosMoussa2013}). To proxy this compensation, we scale our bilateral XVA metrics by the compression factor of the reference name.
% also implies a loss of netting between vanillas and exotics, which in
%some cases might balance  .
%And we are not even talking about
%wrong-way and systemic risks that would also have a strong impact if duly incorporated into the model.

\subsection{Driving Asset}

{Given an interest rate process $S$, we consider a stylized swap of strike $\bar{S}$ with cash flows
$ h_l (\bar{S} -S_{T_{l-1}})$ at increasing times
$T_l,$ $l=1,\ldots, d ,
$
where $h_l= T_{l} - T_{l-1}.$ 
We suppose a 
%constant OIS short term interest rate $r$ and 
stylized Black-Scholes dynamics
with risk-neutral drift $\kappa$
and volatility $\sigma$ for the interest rate process $S$.
Denoting by $T_{l_t}$ the smallest $T_l >t ,$
the mark-to-market of the swap for the party receiving the above cash flows
is given, for $T_0=0\leq t \leq T_d=\Ts$,
by
${P}_t = \beta_{t}^{-1} \beta_{T_{l_t}} h_l (\bar{S}-S_{T_{l_t-1}} ) +P^{\star}_t,$
where
\beql{e:as}
&P^{\star}_t
%=  \beta_{t}^{-1} \bar{S} \sum_{l=l_t+1}^d \beta_{T_l}h_l -
%S_t \sum_{l=l_t +1}^d
%{e^{(\kappa -r)(T_{l-1}-t)}}
%\beta_{T_{l-1}}^{-1} {\beta_{T_l}} h_l
%\\&\qqq 
= \beta_{t}^{-1} \bar{S} \sum_{l=l_t+1}^d \beta_{T_l}h_l - \beta_{t}^{-1} S_t \sum_{l=l_t +1}^d  {\beta_{T_l}} h_l e^{\kappa (T_{l-1}-t)}
=P_{\star}(t, S_t).
\eeql
We 
choose the notional $Nom$ of the swap and its strike 
$\bar{S}$ in such a way that each leg of the swap
has a mark-to-market equal to one at time 0.
%strike the swap to $$\bar{S}= S_0 \frac{\sum_{l=1}^d {\beta_{T_l}} h_l e^{\kappa T_{l-1} }}{\sum_{l=1}^d \beta_{T_l}h_l}$$ 
%so that $P_0 = 0$.  
%Moreover, we assume a notional $Nom$ such that 
%each leg of the swap has a mark-to-market equal to one at time 0.
%, i.e.~
%$\mbox{Nom} = (\sum_{l=1}^{d}  \beta_{T_l} h_l \bar{S} )^{-1}$.
} 
\begin{figure}[H]
\begin{center}
\includegraphics[scale=0.3]{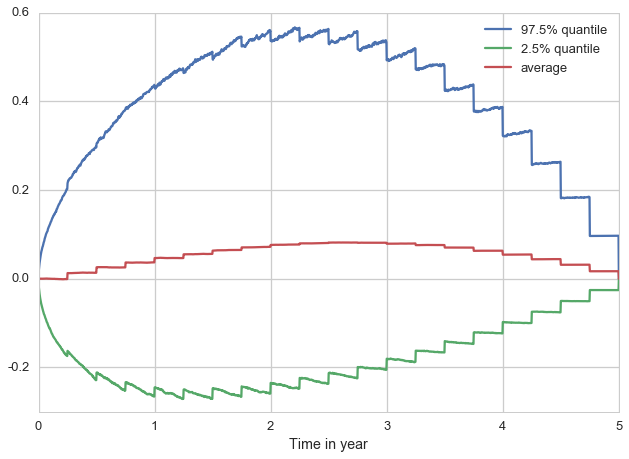}%
\end{center}
\caption{Mark-to-market process
of the swap viewed from the point of view of a party receiving floating
and paying fix in the swap (party
with a long unit position in the swap).
The mean and quantiles as a function of time are computed by Monte Carlo simulation of the process
($-P_t$)
%=-\beta_{t}^{-1} \beta_{T_{l_t}} h_l (\bar{S}-S_{T_{l_t-1}} ) -P_{\star}(t,S_t),$ 
based on the formula
\qr{e:as} for $P_{\star}$, used along $m=10^4$ simulated trajectories of $S.$}
\label{swap-head}
\end{figure}

The following numerical values are used:
\bel%{dataprel}
&r=2\%\sp  S_0=100\sp \kappa=12\% \sp \sigma=20\%\sp h_l = 3\mbox{ months} \sp \Ts = 5\mbox{ years} ,
\eel
resulting in the mark-to-market process displayed in Figure \ref{swap-head} from the point of view of a party receiving floating
and paying fix, which we call a long unit position in the swap. 
Figure \ref{swap-head} exhibits the typical profile of an interest rate swap in an increasing term structure of interest rates, where expectations of increasing rates make the swap in the money on average (i.e.~the average curve is in the positive in Figure \ref{swap-head}). This yields to the product the XVA flavor that would be absent in a flat interest rates environment where the mark-to-market process of the swap would be zero and not give rise to any adjustments. The present Black--Scholes setup and values of the parameters for the process $S$ allow us to obtain this stylized pattern without having to introduce a full flesh interest rate model, 
which would add useless complexity with respect to our goal in this paper. 

\subsection{Structure of the Clearing house}

We consider a clearing house with $(n+1)$ members chosen
among the
%61 riskiest of the
125 names of the CDX index
as of 17 December 2007, a particular day toward the beginning of the
global financial crisis.
The default times of the 125 names are modeled by a common shock model with
piecewise constant intensities $\gamma_Y$ constant on the time intervals $[0,3]$ and $[3,5]$ years,
calibrated
to the corresponding 3 and 5 year CDS and 5 year CDO
data.
With five nested common shocks $Y$ on top of an idiosyncratic shock $Y=\{i\}$ for each of the 125 names,
a nearly perfect calibration can be achieved,
as developed in \shortciteN[Sect.~8.4.3]{BieleckiBrigoCrepeyHerbertsson13}\footnote{Or \citeN[Part II]{Bielecki2012ab} in journal version.}

We consider a subset of nine representative members of the index, with
increasing
CDS spreads shown in the first row of Table \ref{t:w}.
%{do also for 2 and 4 members},
\begin{table}[H]
 \begin{center}
\begin{tabular}{|c|c|c|c|c|c|c|c|c|c|}
\hline
$\Sigma_i$ & 45 & 52 & 56 & 61 & 73 & 108 & 176 & 367 & 1053\\
\hline
$\alpha_i$ & (0.46) & 0.09 & 0.23 & (0.05) & 0.34 & (0.04) & 0.69 & (0.44) & (0.36)\\
\hline
\end{tabular}
%\begin{tabular}{|c|c|c|c|c|c|c|c|c|c|c|c|}
%%\hline
%%$i$&0&1&2&3&4&5&6&7&8
%%%&9
%%\\
%%\hline
%%areski's index &27&30&18&65&51&
%%14&5&6&10\\
%\hline
%$\Sigma$ &1053 &  367 &  176  & 108 &  73 &
%% 38 &   39 &
%   61 &56 &   52 &   45
% \\
%\hline
%$\alpha$ &
% (0.36)& (0.44)& 0.69&  (0.04)& 0.34& (0.05)& 0.23& 0.09& (0.46)
% \\
%\hline
%\end{tabular}
 \caption{({\em Top}) Average 3 and 5 year CDS spreads $\Sigma_i$, in basis points (bp), for a representative subset of nine members of the CDX index
as of 17 December 2007.
 ({\em Bottom}) Coefficients $\alpha_i$ summing up to 0 used for determining the swap positions of the nine members.}
\label{t:w}
\end{center}
\end{table}
\noindent
The coefficients $\alpha_i$ in the second row, where parentheses mean negative numbers, will be used in a way explained below
for determining the positions in the swap of the nine members in the simulations.
These coefficients were obtained as the difference between a vector of nine uniform
numbers and its cyclic shift, so that $\sum_{i\in N} \alpha_i=0.$ 

%As we shall see below,
%the so called compression factor $\nu_0 =\frac{\sum_{i\in N}|\alpha_i|}{|\alpha_0|} -1$
%will correspond to the gross position of the reference member when trading bilaterally, i.e.~the size of its position before netting through the CCP, whereas its net, centrally cleared position will be equal to one.

\subsection{Member Portfolios}\label{ss:data}
We represent in an antisymmetric matrix form
%with upper triangular part
 \bel%{e:matrix}
&
 \varpi
%(\varpi_{i,j})_{i,j\in N}
=
\bordermatrix{\cr &0&1&2&3& \cdots & n \cr
 0& 0 &\og_{0,1}&\og_{0,2}&\og_{0,3}& \cdots & \og_{0,n}\cr
1&\cdot&0 &\og_{1,2} &\og_{1,3}& \cdots & \og_{1,n}\cr
2&\cdot&\cdot&0 &\og_{2,3} & \cdots& \og_{2,n} \cr
3&\cdot&\cdot&\cdot&0 & \cdots & \og_{3,n}\cr
\vdots &\vdots &\vdots &\vdots &\vdots &\ddots & \vdots \cr n &\cdot &\cdot &\cdot &\cdot &\cdots& 0\cr},
\eel
where each ``$\cdot$'' represents the negative of the symmetric entry in the matrix,
the positions of each member $i$ with respect to each member $j$ (or short positions of $j$ with respect to $i$) in the swap.
% with mark-to-market process depicted in Figure \ref{swap-head}. 
Note that the data of the CCVA BSDE related to the member 0, or of the linearized time-0 CCVA formula \qr{e:finalhat},
 only depend on the matrix
$\varpi$ through the sums of each of its rows, corresponding to the vector of the short positions of the different clearing members against the CCP.
By contrast, the data of the BVA BSDE related to the member 0, or of the linearized time-0 BVA formula \qr{e:finalbar}, only depend on the matrix
$\varpi$ through its first row (vector of the short positions of the
different counterparties $i=1,\ldots,n$ against the reference member 0). Hence, we can
forget about the detail of the above matrix, focusing on the $\omega^{csa}_i:=\varpi_{0,i}$
and
{$ \omega^{ccp}_i := \sum_{l\neq i}\varpi_{l,i},$}
$i\neq 0,$
for comparing two trading setups:
\begin{itemize}
\item A CSA setup as of \sr{ss:bil}, where each member $i\neq 0$ trades a short $\omega^{csa}_i \in\R$ position in the swap with the member 0,
whichever other trades members $i\neq 0$ may have between each others. \begin{itemize}
%\item I.e. $\omega_i=\omega_{0,i}, i\neq 0$
\item For instance, but non necessarily, each member $i\neq 0$ has a short $\omega^{csa}_i \in\R$ position with the member $0$ and there are no other trades between members (at least after netting at the level of each pair of members), which corresponds to the situation where only the first row and column are nonzero in the matrix $\varpi.$
\item In any case, the netted long position of the member 0 is $\sum_{i\neq 0}\omega^{csa}_i.$ However, netting does not apply across different counterparties in the CSA setup. We call compression factor
$\nu_0$ the gross position of the reference member 0, i.e.~the number $\nu_0=\sum_{i\neq 0}|\omega^{csa}_i|$
of trades the member 0 is engaged into in the CSA setup.
\end{itemize}
\item A CCP setup as of \sr{ss:ccvaengine}, where each member $i\neq 0$ trades a short $\omega^{ccp}_i \in\R$ position in the swap through the CCP ($\omega^{ccp}_i\leq 0$ effectively means a long position of member $i$),
whichever way this position may be distributed among other members. \begin{itemize} 
\item For instance, but non necessarily, each member $i\neq 0$ has a short $\omega^{ccp}_i \in\R$ position with the member 0 and there are no other trades between members, which again corresponds to the situation where only the first row and column are nonzero in $\varpi.$
\item In any case, since members trade between themselves, the member has a
$\sum_{i\neq 0}\omega^{ccp}_i$ position in the driving asset after netting through the CCP, instead of a non netted position of size $\nu_0$ before clearing through the CCP.
\end{itemize}
\end{itemize}
Moreover, in order to obtain diverse while comparable setups,
we will alternately consider
as reference member 0
each of the nine members in Table \ref{t:w}, for positions
in the driving asset determined by the coefficients $\alpha_i$ (summing up to zero) in the second row of Table \ref{t:w} through the following rule:
$\omega_i=-\frac{\alpha_i}{\alpha_0},$ $i\neq 1$ (where $\omega=\omega^{csa}$ or $\omega^{ccp}$, as suitable).
Since the coefficients $\alpha_i$ add up to 0, this specification ensures $\sum_{i\neq 0} \omega_i=1,$
i.e. a netted position of the member 0 (whoever it is), always equal to 1 in the CCP setup.
We also define $\omega_0=-\frac{\alpha_0}{\alpha_0}=-1,$ consistent with the member 0 being long a $+1$, i.e.~short a $-1$, net position in the swap in the CCP setup (in the CSA setup this value of $\omega_0$ is purely conventional).

Note that
\bel%{e:nualp}
\nu_0=\sum_{i\neq 0}|\omega_i|=\sum_{i\neq 0}\frac{|\alpha_i|}{|\alpha_0|} =\frac{\sum_{i\in N}|\alpha_i|}{|\alpha_0|} -1,\eel so the smaller $|\alpha_0|$, the larger the compression factor $\nu_0$
(gross position of the reference member when trading bilaterally, whereas its net, centrally cleared position is equal to one).
 
%where we emphasize once more that netting across different counterparties is only effective in the CCP setup.

\bex\label{e:w}
Table \ref{t:member} shows the resulting values of the $\omega_i$ of the different members $i\neq 0$ when the name with CDS spread
$61$ bp (name with the second smallest $|\alpha_i|$ in Table \ref{t:w}, with corresponding entries emphasized in bold in Table \ref{t:member})
%, referred to as ``Name I'' henceforth,
is taken as reference member 0 (prototype of a name with a large gross position). 
Hence, the $\omega_i$ in Table \ref{t:member} are proportional to the $\alpha_i$ in Table \ref{t:w}, modulo a
scaling factor so that the $\omega_i$ of this particular name (then labeled as 0) is $-1$. In this case $\nu_0=\sum_{i\neq 0} |\omega_i|=53.00.$
\begin{table}[H]
\begin{center}
\begin{tabular}{|c|c|c|c|c|c|c|c|c|c|}
\hline
$\Sigma$ & 45 & 52 & 56 &{\bf 61 } & 73 & 108 & 176 & 367 & 1053\\
%1053 & 367 & 176 & 108 & 73 & {\bf 61 } & 56 & 52 & 45\\
\hline
$\omega$ & (9.20) & 1.80& 4.60& {\bf (1.00) }& 6.80 & (0.80) & 13.80& (8.80) &
(7.20)  \\
\hline
\end{tabular}
\caption{Positions $\omega_i$ in the swap of the nine members with CDS spreads $\Sigma_i$,
in the respective $\omega_i=\omega_i^{csa}$ or $\omega_i^{ccp}$ meaning,
when the reference member 0 is
the name with CDS spread 61 bp and the second smallest $|\alpha_i|$ in Table \ref{t:w}.}
\label{t:member}
\end{center}
\end{table}
\eex

\bex\label{e:wbis} Table \ref{t:memberbis} is the analog of Table \ref{t:member} when the member with spread $367$ bp
(name with the second largest credit spread in Table \ref{t:w}, with corresponding entries emphasized in bold in Table \ref{t:member})
%, referred to as ``Name II'' henceforth,
is taken as reference member 0 (prototype of a risky name).
In this case $\nu_0=\sum_{i\neq 0} |\omega_i|=5.14.$
\begin{table}[H]
\begin{center}
\begin{tabular}{|c|c|c|c|c|c|c|c|c|c|c|}
\hline
 $\Sigma_i$ &45 & 52 & 56 & 61 & 73 & 108 & 176 & {\bf 367 }& 1053\\
\hline
$\omega_i$& (1.05) & 0.20 & 0.52 & (0.11) & 0.77 & (0.09) & 1.57 & {\bf (1.00)} & (0.82)\\
\hline
\end{tabular}
 \caption{Analog of Table \ref{t:member}
when the reference member 0 is 
%Name II 
the name with CDS spread 367 bp (name with the second largest credit spread $\Sigma_i$) in Table \ref{t:w}.}
\label{t:memberbis}
\end{center}
\end{table}
\eex

\subsection{Margins}\label{ss:margins}

\paragraph{CCP setup} The initial margin $I\!M^i$ posted by each member $i\in N$ is set through \qr{e:im}, using
as risk measure
$\rho$ 
the {risk-neutral} value at risk of some level $\theaim$ ``close to 1'' .
Since the pricing function $P_\star$ in \qr{e:as} is decreasing in $S,$ therefore $I\!M^i$ can be
 proxied, at each simulated time $\zeta$ in
\qr{e:finalhat}
or
\qr{e:finalbar}, by
\beql{l:improx}
I\!M^i_{\zeta}= Nom \times |\omega_i |\times\left\{
\begin{array}{ll}P_\star(\zeta, S_{\zeta})- P_\star (\zeta,S_{\zeta}e^{\sigma\sqrt{\delta'}\Phi^{-1}(\theaim)+ (\kappa-\frac{\sigma^2}{2}) \delta'  })  , & \omega_i \geq 0\\ P_\star (\zeta, S_{\zeta}e^{\sigma\sqrt{\delta'}\Phi^{-1}(1-\theaim)+ (\kappa-\frac{\sigma^2}{2}) \delta'  })-P_\star(\zeta, S_{\zeta})   , & \omega_i \leq 0,\end{array}
\right. \eeql
 where $\Phi$ is the standard normal cdf and where we recall that $ \delta' =\delta+h $ is the margin period of risk.
\begin{itemize}
\item[] For instance, the reference member 0, with $\omega^{ccp}_0=-1,$ is long one unit in the swap with mark-to-market profile shown
in Figure \ref{swap-head}, hence the exposure of the CCP to member 0 is the opposite profile. Accordingly (recalling that Figure \ref{swap-head} shows $(-P_t)$),
the CCP asks initial margins to the member 0 based on $P_\star (\zeta, S_{\zeta}e^{\sigma\sqrt{\delta'}\Phi^{-1}(1-\theaim)+ (\mu-\frac{\sigma^2}{2}) \delta'  })- P_\star(\zeta, S_{\zeta}) $, consistent with the second line in \qr{l:improx} in case
$\omega_{i=0}\leq 0$.
\end{itemize}

Consistently with a ``cover two'' EMIR rule (see \sr{ss:regulccva}), the default fund contributions are set as
the sum of the two largest exposures of the clearing members (exposures in the sense of their EADs as explained in \sr{ss:ead}), allocated between them proportionally to their initial margins.\\

\paragraph{CSA setup} The initial margin $-\ic \geq 0$ required by the member 0 from the member $i\neq 0$ (cf.~\qr{e:margins}) is given by the right-hand side formula in \qr{l:improx} valued at some quantile level $\theapim$ (possibly different from the one used in the CCP setup). 
\begin{itemize}
\item[] For instance, if $\omega^{csa}_i=+2$, meaning that the member 0 has a ``double Figure \ref{swap-head} exposure'' with regard to counterparty $i$,
then
the member 0 asks the counterparty $i$ to post initial margins
based on $P_\star(\zeta, S_{\zeta})- P_\star (\zeta,S_{\zeta}e^{\sigma\sqrt{\delta'}\Phi^{-1}(\theaim)+ (\kappa-\frac{\sigma^2}{2}) \delta'  })$ (recall again that Figure \ref{swap-head} shows $(-P_t)$),
consistent with the use of the first branch in \qr{l:improx} in the case where
$\omega^{csa}_i\geq 0$ (for $i\neq0$).
\end{itemize}
Symmetrically, the formula for the initial margin $\ib\geq 0$ required by the member $i$ from
the member 0 reads
\bel
\ib_{\zeta}=-\omega_i\times Nom \times\left\{
\begin{array}{ll} P_\star(\zeta, S_{\zeta})- P_\star (\zeta,S_{\zeta}e^{\sigma\sqrt{\delta'}\Phi^{-1}(\theapim)+ (\kappa-\frac{\sigma^2}{2}) \delta'  }) , & \omega_i \leq 0\\ P_\star(\zeta, S_{\zeta}) - P_\star (\zeta, S_{\zeta}e^{\sigma\sqrt{\delta'}\Phi^{-1}(1-\theapim)
+ (\kappa-\frac{\sigma^2}{2}) \delta'  }), & \omega_i \geq 0.\end{array}
\right. \eel

\subsection{Exposure-at-defaults}\label{ss:ead}
 
The prime motivation for the Black--Scholes model used for $S$ and for our risk-neutral value-at-risk for the IMs is that these give rise to an explicit formula for the exposure-at-defaults (EAD), which are the basic primitive of all the 
regulatory capital formulas. This
avoids the computational burden of nested Monte Carlo simulations (see the introductory paragraph to \sr{s:regul}).
We also use EADs as a proxy of the exposures of the members in the context of our EMIR ``cover two'' default fund computations (cf. \sr{ss:regulccva}).

%In the case of our Black--Scholes model with 
%value-at-risk of level $a$ for the initial margins, we have an explicit
%formula for the conditional expectation in the right-hand side
%of \qr{e:ead}. 
In fact, 
for any grid time 
\def\thes{v}$\thes=t+ \epsilon p$ involved in EAD computations (cf.~\qr{e:ead}, \qr{e:lp} and \qr{e:as}, with $\epsilon$ taken as one month in the numerics), we have in our model:
%$\Exp_{t}\big[ (\LP ^i_{\thes,\thes+\delta' }-I\!M^i_{\thes}  ) ^+ $
\begin{align*} &\Exp_{t}   \Big[\big({P}_{\thes + \delta'}+ \int_{[ {\thes},\thes+ \delta']}e^{\int_s^{\thes+ \delta'} r_{u}du} dD_s-P_{\thes- }-I\!M_{\thes }  \big) ^+\Big]
\\& =
 \Exp_{t}  \Big[  \Big({P}^\star({\thes+ \delta'},S_{\thes+ \delta'}) - {P}^\star({\thes },S_{\thes })-  \VaR_{t} \big( {P}^\star({\thes+ \delta'},S_{\thes+ \delta'}) - {P}^\star({\thes },S_{\thes }\big) \Big) ^+\Big]
\\& =
 \Exp_{t}  \Exp_{v}\Big[  \Big({P}^\star({\thes+ \delta'},S_{\thes+ \delta'}) - {P}^\star({\thes },S_{\thes })-  \VaR_{t} \big( {P}^\star({\thes+ \delta'},S_{\thes+ \delta'}) - {P}^\star({\thes },S_{\thes }\big) \Big) ^+\Big],
 \end{align*}
where $\VaR$ represents the risk-neutral value-at-risk of level $a$. Denoting by $\bb{ES}$ the corresponding expected shortfall, the conditional version of the identity $\mathbb{E} [ X  \ind_{X \geq \VaR (X)}] = {(1-a)} \bb{ES}(X)$ yields
\begin{align*} &  \Exp_{v}\Big[  \Big({P}^\star({\thes+ \delta'},S_{\thes+ \delta'}) - {P}^\star({\thes },S_{\thes })-  \VaR_{t} \big( {P}^\star({\thes+ \delta'},S_{\thes+ \delta'}) - {P}^\star({\thes },S_{\thes }\big) \Big) ^+\Big]
 \\& =(1-\thea)   \Big( 
\bb{ES}_{v} \left(  {P}^\star({\thes+ \delta'},S_{\thes+ \delta'}) - {P}^\star({\thes },S_{\thes })\right) 
- \VaR_{v} \left(  {P}^\star({\thes+ \delta'},S_{\thes+ \delta'}) - {P}^\star({\thes },S_{\thes }) \right) \Big)
%\\& { = -(1-\thea)  \beta_{\thes + \delta'}^{-1} e^{-\kappa (\thes + \delta')} S_{\thes} e^{(\kappa - \frac{\sigma^2}{2})\delta'}
%\sum_{l = l_{\thes+\delta'}}^{d} \beta_{T_l} h_l e^{\kappa T_{l-1}} \left[ e^{\sigma \sqrt{\delta'} \frac{\phi( \Phi^{-1}(\thea) )}{\thea}} - e^{\sigma \sqrt{\delta'} \Phi^{-1}\left(\thea\right)} \right]}
\\& { = (1-\thea) \left(e^{\sigma \sqrt{\delta'} \Phi^{-1}\left(\thea\right)} - e^{\sigma \sqrt{\delta'} \frac{\phi( \Phi^{-1}(\thea) )}{\thea}} \right) \beta_{\thes + \delta'}^{-1} e^{-\kappa \thes -\frac{\sigma^2}{2}\delta'} S_{\thes}
 \sum_{l = l_{\thes+\delta'}}^{d} \beta_{T_l} h_l e^{\kappa T_{l-1}}},
 \end{align*}
where $\Phi$ and $\phi$ are the standard normal cdf and density.
Hence,
\beql{e:eads} &\Exp_{t} \left[\left({P}_{\thes + \delta'}+ \int_{[ {\thes},\thes+ \delta']}e^{\int_s^{\thes+ \delta'} r_{u}du} dD_s-P_{\thes- }-I\!M_{\thes }  \right)^+ \right]=f_v^{a,\delta'}\times (1-a) e^{-\kappa t}S_t,
 \eeql
where
\beql{e:eadss} f_v^{a,\delta'}= { \left(e^{\sigma \sqrt{\delta'} \Phi^{-1}\left(a\right)} - e^{\sigma \sqrt{\delta'} \frac{\phi( \Phi^{-1}(a) )}{\b{1-a}}} \right) \beta_{\thes + \delta'}^{-1} e^{ -\frac{\sigma^2}{2}\delta'} 
 \sum_{l = l_{\thes+\delta'}}^{d} \beta_{T_l} h_l e^{\kappa T_{l-1}}}.
\eeql
Likewise, we have
\beql{e:eadsbis} &\Exp_{t} \left[\left({P}_{\thes + \delta'}+ \int_{[ {\thes},\thes+ \delta']}e^{\int_s^{\thes+ \delta'} r_{u}du} dD_s-P_{\thes- }-I\!M_{\thes }  \right)^- \right]= g_v^{a,\delta'}\times (1-a) e^{-\kappa t}S_t,
 \eeql
where
\beql{e:eadssbis} g_v^{a,\delta'}= { -\left(e^{\sigma \sqrt{\delta'} \Phi^{-1}\left(1-a\right)} - e^{-\sigma \sqrt{\delta'} \frac{\phi( \Phi^{-1}(a) )}{\b{1-a}}} \right) \beta_{\thes + \delta'}^{-1} e^{ -\frac{\sigma^2}{2}\delta'} 
 \sum_{l = l_{\thes+\delta'}}^{d} \beta_{T_l} h_l e^{\kappa T_{l-1}}}.
\eeql
Based on \qr{e:eads} through \qr{e:eadss}, explicit formulas for the EADs follows.
Figure \ref{f:eads} shows the time-0 EADs of the nine CCP members  
for their positions in the swap corresponding to the choice of
the name of examples \ref{e:w} or \ref{e:wbis} as reference member.
\begin{figure}[htbp]
\begin{center}  
\includegraphics[width=0.49\textwidth,height=0.27\textheight]{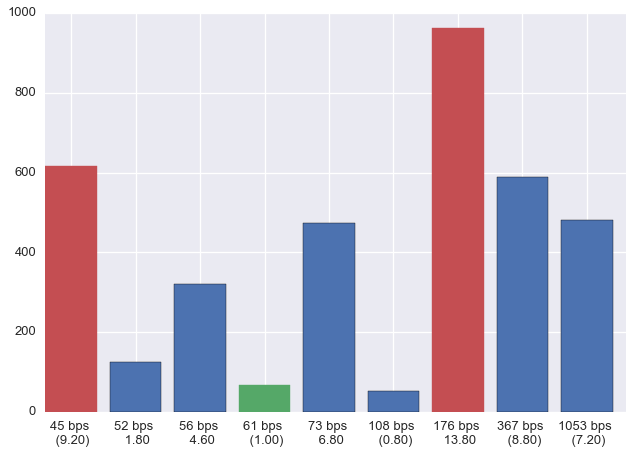} \includegraphics[width=0.49\textwidth,height=0.27\textheight]{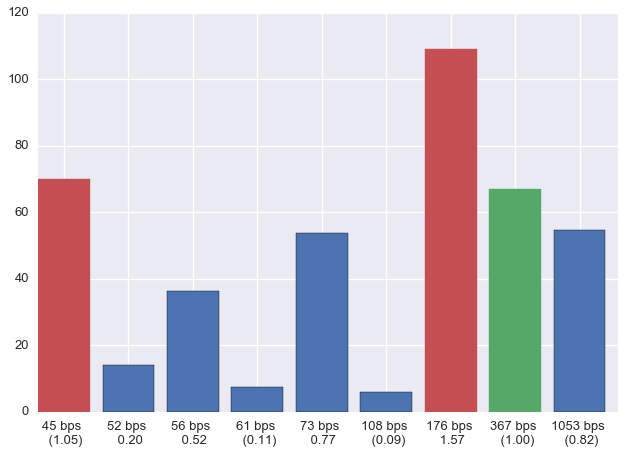}  
\end{center}
\caption{Time-0 EADs  in basis points (IM quantile level $a=70\%,$ liquidation period $\delta=5$ days).
The two largest EADs, in red, size the default fund. The reference member EAD is in green. The corresponding positions $\omega_i$ of the member are displayed at the bottom.
{\it Left}: Reference member with $\Sigma_0=$61 bps and $\nu_0 =53.00$. {\it Right}: Reference member with $\Sigma_0=$367 bps and $\nu_0 =5.14$.}\label{f:eads}
\end{figure}

\subsection{XVA Data}\label{ss:datanum}

The following numerical values are used in the sequel:
\beql{data}
& \bar{R}=1 \sp \bar{\lambda}=\frac{1}{2}\Sigma_0
\sp \lambda =0 \sp k=10\%
\sp h=1\mbox{ day}\sp \mu = \frac{2}{\Ts} \sp m= 10^4 
%\sp \thea_{ead} = 85\% 
,
\eeql
where
$m$ is the number of simulations used for estimating the expectations in
\qr{e:finalhat}
and
\qr{e:finalbar}. The level of $10\%$ used for $k$ is consistent with reference orders of magnitude for a hurdle rate.  

Moreover, in a CCP setup, unless otherwise stated, we set
\beql{dataccva}
&R=0\sp
\delta=5 \mbox{ days } \sp
\theaim = 70 \%
 \sp T=1 \mbox{ month}   %\thea_{m} = 80\% 
\sp Y=1 \mbox{ year},\\& E^\star=25\% K^{ccp}
\sp {c= 30}\mbox{~bp},
\eeql
where $K^{ccp}$ is defined in \qr{e:kccp}.
The low quantile level used to set the initial margins is meant to compensate the excessive simplicity of the Black--Scholes setup without wrong-way risk used for $S$ (it also leads to moderate standard errors with a relatively small number $m=10^4$ of simulations).
%Wrong-way risk add-ons to the present stylized setup could be made following the lines of \citeN{CrepeySong15}.
Margin fees of $c=30$ bp are consistent with current CCP practices. These margin fees are distinct from the commission fees, not included in our setup, that a CCP is also charging to its members. 
%This means that the difference between the resulting BVA and CCVA could be interpreted as the break-even value of this commission ensuring equal costs to centrally cleared and bilateral trading.
In practice, commission fees are of the order of a few basis points of the size of the positions, i.e.~a few basis points in the case of a unit position in our swap with each leg equal to one at time 0.

In a CSA setup,
alternatively to \qr{dataccva}, unless otherwise stated,
we set
\bel%{databva}
&R_b=R_c=40\%
\sp\delta=15 \mbox{ days}
\sp \theapim = 80\% \sp c=0.
\eel
The value $\theapim = 80\%$ used in the bilateral case is higher than the value $\theaim = 70\%$ used in the CCP setup, where the protection offered by the default fund allows requiring less initial margins.

\section{Numerical Results}\label{s:resu}

All our XVA numbers are stated in basis points
(recall that both legs of the swap are worth one at time 0).
For comparability purposes, common random inputs are used in all our Monte Carlo estimates, i.e.~we use the same sampled
trajectories of $S$ and sampled sets of default times $\tau_i$
%{$\zeta, W_\zeta$, $\tau_i,$ etc.}
in all cases, it is only the way these $m=10^4$ random input sets are used which changes. The computation times are proportional to the number of members $n$ and model trajectories $m$, e.g.~about 5 minutes on a standard laptop to compute
a full set of XVAs in Table \ref{t:0} (four or five XVA components and their sum), where $n=8$ and $m=10^4,$
using pre-simulated values for all the random inputs. Negative (e.g.~DVA) numbers are displayed in parentheses.
{Regarding the aggregated XVA numbers in the tables, i.e.~BVA in the CSA setup, CCVA in the CCP setup and TVA sometimes used as a common acronym for covering both cases, they are all KVA-inclusive,	
but they do not include the corresponding DVA numbers, which are only showed for reference. In other words, all the displayed TVA numbers correspond to entry price TVAs. The CCP MLA number are consistently found one order of magnitude smaller than the other XVA numbers, which is a sanity check that the CCP margin fees do not drive the comparison between the CCP and the CSA setup.}

Note that, for simplicity, we are comparing a situation where all the trading is centrally cleared with a situation where all the trading is
bilateral. In practice, vanilla products (hedges) tend to be cleared
and exotics tend to be bilaterally traded. Therefore,
in a more realistic
setup, the multilateral netting benefit
that CCPs provide is balanced
by a loss of bilateral netting across asset classes (see \citeN{DuffieZhu} and \passhortciteN{ContSantosMoussa2013}).
 To correct this bias, we will also show
bilateral XVA figures scaled by the compression factor $\nu_0$ of the reference name.
%\b{Given that we use higher initial margins in 
%the bilateral setup to compensate for the absence of a default fund there,} 
%we
%expect to obtain not so different CVA(/DVA) in both setups (at least after the above-mentioned scaling), {but to see a shift \b{from MVA in the CSA case,
%where larger initial margins need be funded, to KVA in the CCP case, where default fund contributions are implicit capital at risk that deserves remuneration at a hurdle rate}. 

\subsection{Multilateral Netting Benefit}\label{ss:basecase}

Table \ref{t:0} shows the XVA numbers obtained by considering alternately
each of the nine members in Table \ref{t:w}
as reference member, using the $\alpha_i$ coefficients for setting the positions of the members in each case as explained in \sr{ss:data}
(cf.~the examples \ref{e:w}
and \ref{e:wbis}). The different cases in Table \ref{t:0} are ordered by increasing values of the compression factor $\nu_0$, i.e.~by decreasing $|\alpha_0|$. We can see from Table \ref{t:0} that the {MVA and the KVA are the main contributors
in the respective CSA and CCP setup}. 
Moreover, the CSA XVA numbers vary roughly proportionally to the compression factor $\nu_0,$ whereas the CCP XVA numbers are essentially not impacted by
$\nu_0$. This illustrates the 
multilateral
netting benefit
provided by the CCP, especially for members with a large compression factor $\nu_0$.

\begin{table}[H]
\begin{center} 
\small 
\begin{tabular}{|c|c|c|c|c|c|c|c|c|c|}
\hline
$\nu_0$ & 2.91 & 4.87 & 5.14 & 6.50 & 6.94 & 10.74 & 29 & 53 & 66.50\\
$\alpha_0$ & 0.69 & (0.46) & (0.44) & (0.36) & 0.34 & 0.23 & 0.09 & (0.05) & (0.04)\\
$\Sigma_0$ & 176 & 45 & 367 & 1053 & 73 & 56 & 52 & 61 & 108\\
\hline
CVA  & 11.07 & 25.06 & 19.34 & 14.06 & 28.37 & 42.69 & 111.38 & 238.22 & 299.37\\
DVA  &  (8.76) &  (4.49) &  (30.85) &  (90.10) &  (8.08) &  (13.59) &  (28.77) &  (52.70) &  (111.33)\\
MVA  & 30.38 & 13.63 & 110.50 & 339.69 & 31.41 & 39.34 & 98.46 & 204.72 & 449.68\\
KVA  & 11.17 & 21.16 & 19.40 & 21.14 & 29.26 & 46.28 & 122.20 & 221.63 & 275.87\\
BVA & 52.62 & 59.85 & 149.24 & 374.90 & 89.04 & 128.31 & 332.05 & 664.57 & 1024.92\\
% BVA' & 43.85 & 55.36 & 118.39 & 284.79 & 80.96 & 114.72 & 303.28 & 611.87 & 913.58\\
\hline
CVA  & 7.88 & 11.33 & 6.54 & 3.57 & 10.85 & 11.73 & 11.91 & 11.60 & 9.23\\
DVA  &  (2.57) &  (0.69) &  (5.43) &  (13.03) &  (1.07) &  (0.89) &  (0.81) &  (0.90) &  (1.57)\\
MVA  & 5.19 & 1.39 & 10.33 & 24.24 & 2.22 & 1.76 & 1.61 & 1.86 & 3.23\\
MLA  & 1.17 & 1.22 & 1.09 & 0.89 & 1.22 & 1.22 & 1.22 & 1.22 & 1.20\\
KVA  & 10.79 & 11.59 & 10.00 & 7.97 & 11.44 & 11.52 & 11.54 & 11.58 & 11.21\\
CCVA & 25.03 & 25.54 & 27.95 & 36.67 & 25.73 & 26.23 & 26.27 & 26.26 & 24.87\\
%CCVA' & 22.46 & 24.85 & 22.52 & 23.64 & 24.66 & 25.34 & 25.47 & 25.36 & 23.30\\
\hline
\end{tabular}
\end{center}

\caption{XVA numbers obtained by considering alternately
each of the nine members in Table \ref{t:w}
as reference member 0, using the $\alpha_i$ for setting the positions of the members in each case as explained in \sr{ss:data}. (\textit{Up}) Credit spread $\Sigma_0$, coefficient $\alpha_0$ and compression factor $\nu_0$ of the reference member in each case (ordered by increasing $\nu_0$, i.e.~decreasing $|\alpha_0|$). (\textit{Middle}) CSA XVA numbers. (\textit{Bottom}) CCP XVA numbers.
}
 \label{t:0}
\end{table}

Table \ref{t:0suite} shows the percentage standard errors corresponding to the Monte Carlo estimates of Table \ref{t:0}.
As we can see from the table,
the standard errors are typically no more than a few percents in relative terms.
Standard errors of Monte Carlo estimates are no longer shown in the sequel.

\begin{table}[H]
\begin{center}																		
\small																		
\begin{tabular}{|c|c|c|c|c|c|c|c|c|c|}																		
\hline																		
$\nu_0$	&	2.91	&	4.87	&	5.14	&	6.50	&	6.94	&	10.74	&	29	&	53	&	66.50\\
$\alpha_0$	&	0.69	&	(0.46)	&	(0.44)	&	(0.36)	&	0.34	&	0.23	&	0.09	&	(0.05)	&	(0.04)\\
$\Sigma_0$	&	176	&	45	&	367	&	1053	&	73	&	56	&	52	&	61	&	108\\
\hline																		
CVA	&	3.40	&	2.87	&	3.40	&	4.97	&	3.22	&	3.22	&	3.22	&	2.90	&	2.89\\
DVA	&	5.66	&	10.38	&	4.08	&	2.58	&	8.92	&	9.21	&	9.49	&	9.28	&	7.05\\
MVA	&	0.79	&	0.78	&	0.75	&	0.96	&	0.77	&	0.64	&	0.63	&	0.84	&	0.80\\
KVA	&	0.58	&	0.54	&	0.64	&	0.81	&	0.54	&	0.54	&	0.54	&	0.54	&	0.55\\
\hline																		
CVA	&	2.55	&	2.93	&	3.13	&	4.49	&	2.69	&	2.71	&	2.70	&	2.91	&	2.66\\
DVA	&	3.11	&	3.02	&	3.05	&	3.42	&	3.15	&	2.92	&	2.94	&	3.27	&	3.21\\
MVA	&	0.86	&	0.78	&	0.77	&	0.96	&	0.91	&	0.67	&	0.69	&	0.95	&	0.93\\
MLA	&	0.65	&	0.60	&	0.71	&	0.88	&	0.61	&	0.61	&	0.60	&	0.60	&	0.62\\
KVA	&	0.58	&	0.58	&	0.65	&	0.84	&	0.57	&	0.59	&	0.59	&	0.59	&	0.58\\
\hline
\end{tabular}
\end{center}

\caption{Percentage standard errors corresponding to the Monte Carlo estimates of Table \ref{t:0}.}
 \label{t:0suite}
\end{table}

\subsection{Impact of the Credit Spread of the Reference Member}\label{ss:creditspread}

The CCP multilateral netting benefit dominates the comparison between our CSA and CCP XVA numbers. However,  in our stylized setup, we cannot see the netting benefit across assets of bilateral trading.
In order to compensate for this bias and obtain comparative results net of the first order CCP multilateral netting benefit,
Table \ref{t:0bis} shows the same results as Table \ref{t:0}, but with all the CSA XVA numbers scaled by the corresponding
compression factor $\nu_0$ {(we will present in this way all the CSA XVA results in the sequel)} and ordered by increasing credit spread $\Sigma_0$ of the reference name, instead of increasing $\nu_0$ in Table \ref{t:0}. 
%This scaling by $\nu_0$ is meant to proxy a situation where there would be netting of the positions of the reference member across its different counterparties
%also in the CSA setup.

From Table \ref{t:0bis} we can see that, if we get rid of the CCP multilateral netting benefit through this scaling, then the CSA and CCP XVA numbers become of a similar order of magnitude. The aggregated TVA numbers even become
 in favor of the CSA setup, except
for the reference name with the largest (actually huge) credit spread of 1053 bp. These results can be put in perspective with the ones in 
\citeN{GhamamiGlasserman16} (see \sr{ss:lit}).

Regarding the comparison between the nine different cases within the CCP setup, as also within the CSA setup after scaling by the compression factor, 
Table \ref{t:0bis} shows that the main explanatory factor of the results is the credit spread of the reference member, risky members being heavily penalized in terms of MVA, especially in the CSA setup.
In both cases, the dominant patterns
are a logarithmic decrease of the CVA numbers and a linear increase of the $|\mbox{DVA}|$ and MVA numbers with respect to the credit spread of the reference name.

\begin{table}[H]
\begin{center}
\small
\begin{tabular}{|c|c|c|c|c|c|c|c|c|c|}
\hline
$\nu_0$ & 4.87 & 29 & 10.74 & 53 & 6.94 & 66.5 & 2.91 & 5.14 & 6.5\\
$\alpha_0$ & (0.46) & 0.09 & 0.23 & (0.05) & 0.34 & (0.04) & 0.69 & (0.44) & (0.36)\\
$\Sigma_0$ & 45 & 52 & 56 & 61 & 73 & 108 & 176 & 367 & 1053\\
\hline
CVA $/ \nu_0$ & 5.15 & 3.84 & 3.97 & 4.49 & 4.09 & 4.50 & 3.80 & 3.76 & 2.16\\
DVA $/ \nu_0$ &  (0.92) &  (0.99) &  (1.27) &  (0.99) &  (1.16) &  (1.67) &  (3.01) &  (6.00) &  (13.86)\\
MVA $/ \nu_0$ & 2.80 & 3.40 & 3.66 & 3.86 & 4.53 & 6.76 & 10.44 & 21.50 & 52.26\\
KVA $/ \nu_0$ & 4.34 & 4.21 & 4.31 & 4.18 & 4.22 & 4.15 & 3.84 & 3.77 & 3.25\\
BVA $/ \nu_0$ & 12.29 & 11.45 & 11.95 & 12.54 & 12.83 & 15.41 & 18.08 & 29.03 & 57.68\\
% BVA' $/ \nu_0$ & 11.37 & 10.46 & 10.68 & 11.54 & 11.67 & 13.74 & 15.07 & 23.03 & 43.81\\
\hline
CVA  & 11.33 & 11.91 & 11.73 & 11.60 & 10.85 & 9.23 & 7.88 & 6.54 & 3.57\\
DVA  &  (0.69) &  (0.81) &  (0.89) &  (0.90) &  (1.07) &  (1.57) &  (2.57) &  (5.43) &  (13.03)\\
MVA & 1.39 & 1.61 & 1.76 & 1.86 & 2.22 & 3.23 & 5.19 & 10.33 & 24.24\\
MLA  & 1.22 & 1.22 & 1.22 & 1.22 & 1.22 & 1.20 & 1.17 & 1.09 & 0.89\\
KVA  & 11.59 & 11.54 & 11.52 & 11.58 & 11.44 & 11.21 & 10.79 & 10.00 & 7.97\\
CCVA & 25.54 & 26.27 & 26.23 & 26.26 & 25.73 & 24.87 & 25.03 & 27.95 & 36.67\\
%CCVA' & 24.85 & 25.47 & 25.34 & 25.36 & 24.66 & 23.30 & 22.46 & 22.52 & 23.64\\
\hline
\end{tabular}
\caption{XVA numbers obtained by considering alternately
each of the nine members in Table \ref{t:w}
as reference member 0, using the $\alpha_i$ for setting the positions of the members in each case as explained in \sr{ss:data}. (\textit{Up}) Credit spread $\Sigma_0$, coefficient $\alpha_0$ and compression factor $\nu_0$ of the reference member in each case (ordered by increasing $\Sigma_0$). (\textit{Middle}) CSA XVA numbers scaled by the compression factors $\nu_0$. (\textit{Bottom}) CCP XVA numbers.
}
 \label{t:0bis}\end{center}
\end{table}
 
\subsection{Impact of the Liquidation Period}

Focusing on the reference members 
of the examples 
\ref{e:w} and \ref{e:wbis}, respectively dubbed ``safe member'' and ``risky member'' henceforth (with respective credit spread of $\Sigma_0=61$
and $367$ bp),
Table \ref{t:days}
shows the impact of changing the length $\delta$ of the \cure period
from 5 days to 15 days in the CSA setup and
vice versa in the CCP setup.  
The CSA 15 day and CCP 5 day numbers in Table \ref{t:days} are simply retrieved from Table \ref{t:0bis}, for comparison purposes with
the additional CSA 5 day and CCP 15 day numbers.
The results are consistent with a $\sqrt{\delta}$ pattern in line with the distributional properties of the Black--Scholes model used for $S.$

\begin{table}[H]
\begin{center}
\begin{tabular}{|c|c|c|c|c|}
\hline
Member & \multicolumn{2}{c|}{61 bps, $\nu_0 =53.00$} & \multicolumn{2}{c|}{367 bps, $\nu_0 =5.14$}\\
\hline
$\delta$ & 5 d & 15 d & 5 d & 15 d\\
\hline
CVA $/ \nu_0$ & 2.17 & 4.49 & 1.82 & 3.76\\
DVA $/ \nu_0$ & (0.50) & (0.99) & (2.90) & (6.00)\\
MVA $/ \nu_0$ & 2.34 & 3.86 & 13.14 & 21.50\\
KVA $/ \nu_0$ & 2.43 & 4.18 & 2.18 & 3.77\\
BVA $/ \nu_0$ & 6.94 & 12.54 & 17.15 & 29.03\\
%BVA' $/ \nu_0$ & 6.43 & 11.54 & 14.25 & 23.03\\
\hline
CVA  & 11.60 & 19.54 & 6.54 & 10.78\\
DVA  & (0.90) & (1.41) & (5.43) & (8.91)\\
MVA  & 1.86 & 3.40 & 10.33 & 18.96\\
MLA  & 1.22 & 2.25 & 1.09 & 2.00\\
KVA  & 11.58 & 21.60 & 10.00 & 18.59\\
CCVA & 26.26 & 46.79 & 27.95 & 50.34\\
% CCVA' & 25.36 & 45.38 & 22.52 & 41.44\\
\hline
\end{tabular}

\end{center}
\caption{Impact of the \cure period.
(\textit{Left}) Safe reference member.
%Reference member Name I with $\Sigma_0=61 bp.$
%%credit spread $\Sigma_0=$ 61 bp and compression factor $\nu_0=53.00$.
(\textit{Right}) Risky reference member. %Reference member Name II with $\Sigma_0=367 bp.$ 
(\textit{Top}) CSA XVA numbers scaled by $\nu_0$.
%(\textit{Middle}) Genuine XVA numbers in the CSA setup.
(\textit{Bottom}) CCP XVA numbers.}
\label{t:days}
\end{table}

\subsection{Margin Optimization}

Table \ref{tablequantiles} shows the impact of using higher quantile levels $a$ for the initial margins, which were only 80\% and 70\% in the respective CSA and CCP setup so far (with the motivation exposed in \sr{ss:datanum}). The left column in each of the two main panels, retrieved from Table \ref{t:0bis}, corresponds to our base case where $\theaim=70\%$ and $\theapim=80\%.$ 
%and $\thea_{ead}=85\%.$
When higher values are used for the quantile levels,
i.e.~going from left to right in each panel,
we observe the same qualitative patterns as before in terms of the comparison between the CSA and the CCP setup. Considering now the impact of higher quantile levels inside each CSA or CCP setup, we can see a shift from CVA(/DVA) and KVA into MVA.

Ultimately, for very high quantiles, the CVA(/DVA) and KVA would reach zero whereas the MVA would keep increasing, since excessive margins
become useless and a pure cost to the system, in the CSA as in the CCP setup. 
Figure \ref{fig:quantiles} 
illustrates this further by showing the aggregated TVA numbers and the relative weight of their CVA, FVA and KVA contributions when the quantile level $a$ used for setting the IM goes from 55\% to 100\%, where FVA means MVA in the CSA setup (left graphs) and MVA$+$MLA in the CCP setup (right graphs).
In each of the four cases considered in the upper panels (left CSA vs. right CCP curve and blue safe vs. green risky reference member curve), the numerical values of the TVA exhibit a convex
dependence with respect to $\thea$ (although, mathematically speaking, this depends on the values of the numerical parameters that are used, see for instance the CVA curve in the left graph of Figure \ref{f:3names}, which shows a more detailed XVA decomposition of the safe reference member CCVA curve in the upper right graph of Figure \ref{fig:quantiles}).
In the case of the risky reference member in the CSA setup, the level of initial margins is too high already with a 55\% quantile level: The risky reference member (green) BVA curve in the upper left graph of Figure \ref{fig:quantiles} keeps increasing when $a$ increases from 55\% to 100\%.
In each of the other three cases, the TVA  has a minimum at some value $a<1.$
For both reference names, the optimal quantile level is larger in the CCP than in the CSA setup. This is because, in the CCP setup, the member is happy to post more initial margins, which ``cost''  her $\bar{\lambda}=\frac{1}{2}\Sigma_0$,  in order to reduce her default fund contribution, which ``costs'' her a greater $k=10\%$ 
(cf.~\qr{data}).
In each of the four considered cases, the FVA becomes preponderant and even hegemonic (as it tends to infinity) when $a$ goes to 100\%.
 
%These numerical results are also consistent with the conclusions of
\citeN{CapponiCheng16} construct a model which endogenizes collateral, making it part of an optimization problem where the CCP maximizes profit by controlling collateral and fee levels. They conclude that the 
%optimal 
collateral level should decrease with funding costs, on top of increasing with market volatility. The above numerical results are quite in line with such statements.

\begin{table}[H]
\begin{center}
\small
\begin{tabular}{|c||c|c|c||c|c|c|} 
\hline
Member 
& \multicolumn{3}{c||}{$\Sigma_0=$ 61 bp, $\nu_0=$ 53.00} 
& \multicolumn{3}{c|}{$\Sigma_0=$ 367 bp, $\nu_0=$ 5.14} \\
\hline
\theapim& $ 80\%$ & $ 90\%$ & $ 99 \%$ 
& $ 80\%$ & $ 90\%$ & $ 99 \%$  \\
\hline
CVA $/ \nu_0$ & 4.49 & 2.64 & 0.74 & 3.76 & 2.23 & 0.62\\
DVA $/ \nu_0$ &  (0.99) &  (0.56) &  (0.15) &  (6.00) &  (3.51) &  (1.02)\\
MVA $/ \nu_0$ & 3.86 & 5.87 & 10.66 & 21.50 & 32.99 & 60.18\\
KVA $/ \nu_0$ & 4.18 & 1.78 & 0.13 & 3.77 & 1.61 & 0.12\\
BVA $/ \nu_0$ & 12.54 & 10.29 & 11.53 & 29.03 & 36.83 & 60.92\\
% BVA' $/ \nu_0$ & 11.54 & 9.73 & 11.38 & 23.03 & 33.32 & 59.90\\
\hline
\theaim & $  70\%$ & $  80\%$ & $  95\%$ 
& $  70\%$ & $  80\%$ & $  95\%$\\
\hline
CVA  & 11.60 & 9.15 & 4.64 & 6.54 & 5.17 & 2.62\\
DVA  &  (0.90) &  (0.66) &  (0.22) &  (5.43) &  (4.02) &  (1.43)\\
MVA  & 1.86 & 2.83 & 5.32 & 10.33 & 15.71 & 29.53\\
MLA  & 1.22 & 1.54 & 2.56 & 1.09 & 1.38 & 2.31\\
KVA  & 11.58 & 6.55 & 1.19 & 10.00 & 5.66 & 1.03\\
CCVA & 26.26 & 20.07 & 13.72 & 27.95 & 27.91 & 35.49\\
% CCVA' & 25.36 & 19.40 & 13.49 & 22.52 & 23.88 & 34.06\\
\hline
\end{tabular}

\end{center}

\caption{Impact of the level of the quantile level $\thea$ that is used for setting the initial margins.
(\textit{Left}) Safe reference member.
%Reference member Name I with $\Sigma_0=61$ bp
(\textit{Right}) Risky reference member.
%Reference member Name II with $\Sigma_0=367$  bp. 
(\textit{Top}) CSA setup with all XVA numbers scaled by $\nu_0$.
(\textit{Bottom}) CCP setup.}
\label{tablequantiles}
\end{table}

\begin{figure}[htbp]
\begin{center} 
\includegraphics[width=0.49\textwidth,height=0.24\textheight]{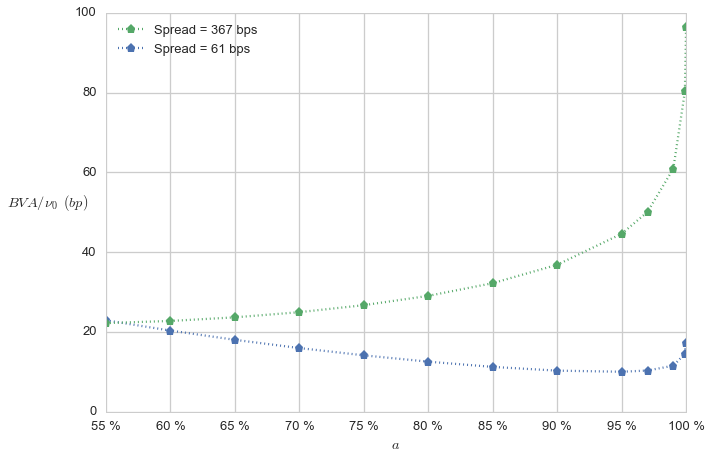} 
\includegraphics[width=0.49\textwidth,height=0.24\textheight]{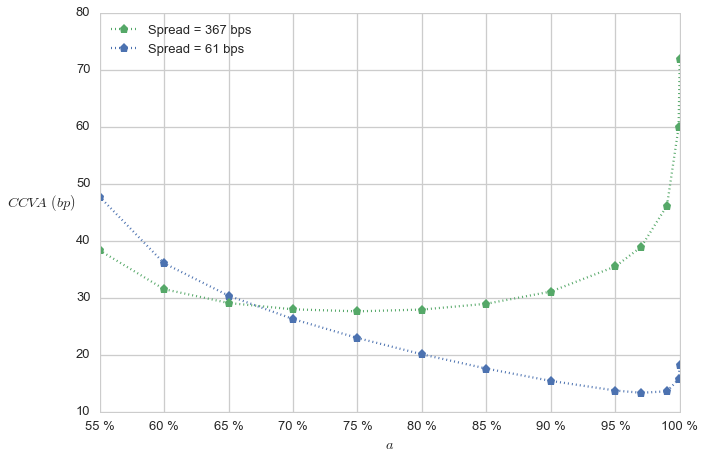}\\
\includegraphics[width=0.49\textwidth,height=0.24\textheight]{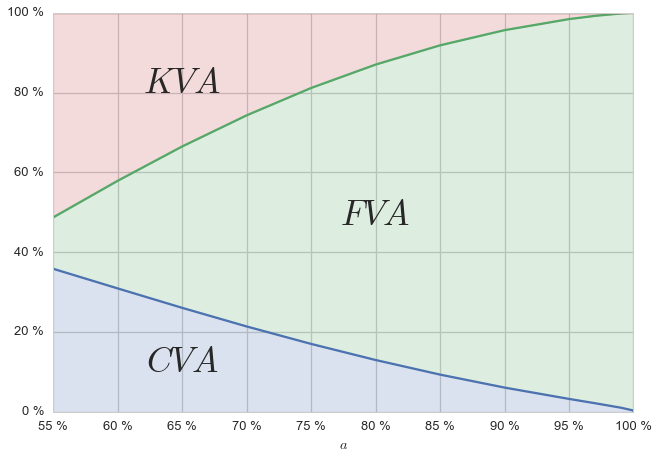}\includegraphics[width=0.49\textwidth,height=0.24\textheight]{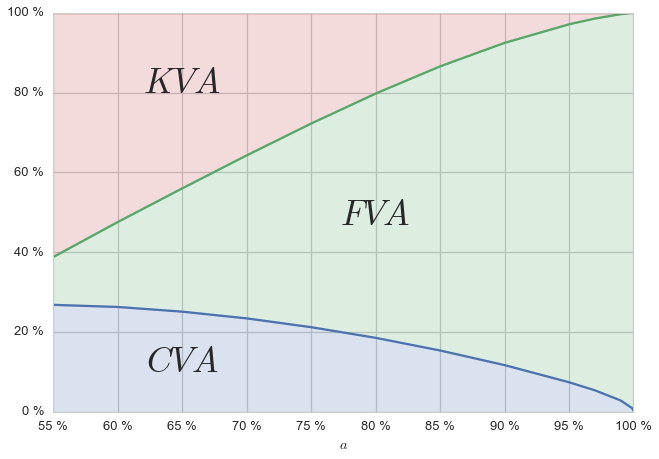} \\
\includegraphics[width=0.49\textwidth,height=0.24\textheight]{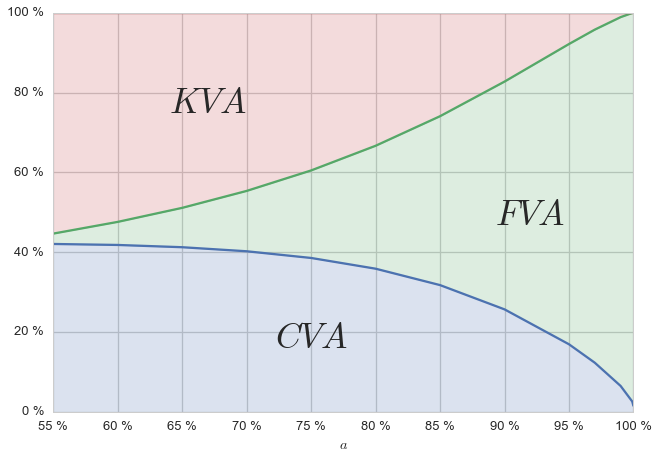}\includegraphics[width=0.49\textwidth,height=0.24\textheight]{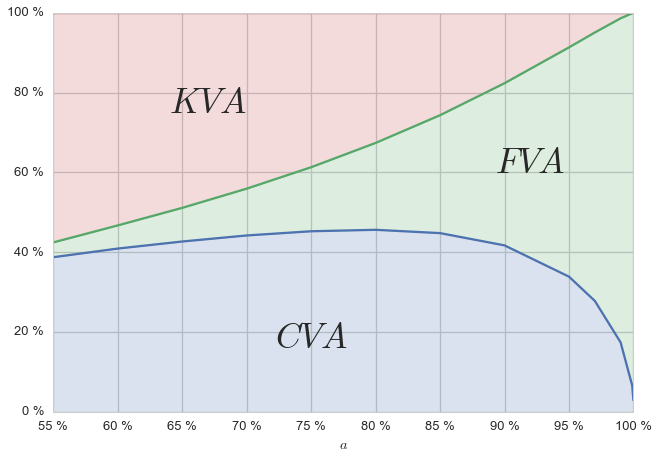}
\end{center}
\caption{Varying the initial margins quantile level $a$. {\it Left}: CSA setup. {\it Right}: CCP setup. {\it Top}: BVA$/\nu_0$ vs. CCVA.  
{\it Bottom}: XVA relative contributions
in the case of the safe reference member. {\it Middle}: XVA relative contributions in the case of the risky reference member.}
\label{fig:quantiles}
\end{figure}

\subsection{Impact of the Number of Members}

Another interesting question is what happens when we vary the number of members of the CCP. Obviously, more members means more mutualization of risk. However, the main effects in a CCP are already visible with nine members as above: with more members, things would mainly happen as in the projection of the system onto the ten (or so) greatest members anyway. Figure \ref{f:3names}
 illustrates that, if there are now not enough members, a regulatory ``cover two'' default fund specification sized to the two largest exposures of the clearing members may result in very heavy default fund contributions and KVA for the small members in the common
situation of heterogeneous members' exposure. 
\begin{figure}[htbp]
\begin{center}
\includegraphics[width=0.49\textwidth,height=0.25\textheight]{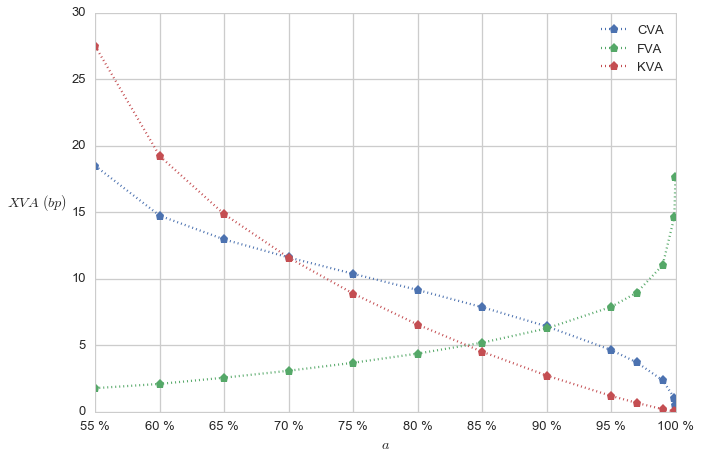} \includegraphics[width=0.49\textwidth,height=0.25\textheight]{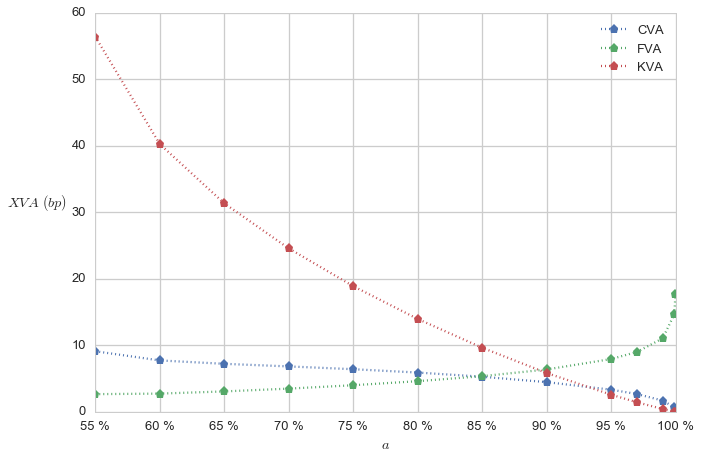}  
\end{center}
\caption{CCP XVA results for the reference member with $\Sigma_0=$ 61 bp and $\nu_0=$ 53.00. {\it Left}: Results in our previous CCP with nine members. {\it Right}: Results in a CCP restricted to three members: the reference member and two other members. The reference member, with $\omega_i=-1$ by definition,
corresponds to the member with time-0 EAD displayed in green in the left panel of Figure \ref{f:eads}. The two other members are the members of the original CCP with the greatest time-0 EADs, i.e. the members with the time-0 EADs displayed in red in the left panel of Figure \ref{f:eads}. Moreover, we modified the positions of these two members as $\omega_i=-9$ and 10, instead of $-9.2$ and 13.8 in the left panel of Figure \ref{f:eads},  for being in line with the clearing condition
$\sum_{i\in N}P^i=0$.}\label{f:3names}
\end{figure}

\section{Conclusions}\label{ccva}

\appendix
\section{Regulatory Capital and Default Fund Formulas}\label{s:regul}
%\subsection{exposure-at-default}\label{ss:regul-ead}
%\sr
A primitive of all the regulatory capital formulas 
are the so-called exposure-at-defaults given,
for $i\in N=0,1\ldots,n$ and $t\in [0,\Ts],$ as
\begin{align}\label{e:eadalph}
EAD^i_{t}
&=& 1.4  \, \epsilon   \sum_{  \epsilon p<1 \wedge (T-t)} EEE^i_t (t_p) ,
\end{align}
where (see \citeN[formulas (1)-(2)-(3) pages 26-27]{Basel2-2005}):
\begin{itemize}
\item the factor 1.4 is a wrong-way risk multiplier, 
\item $\epsilon$ is a time-integration step (e.g.~one month),
\item $t_p = t+ \epsilon p,$ 
\item the effective expected exposures $EEE^i_t (t_p)$ are defined through the following iteration: $EEE^i_t(t_{-1})=0$ and, for $p\ge 0,$
\begin{align} 
\label{e:ead} 
&EEE_t^i(t_p) = 
\max \Big( EEE^i_t(t_{p-1}), \Exp_{t}\big[ (\LP ^i_{t_p,t_p+\delta' }-I\!M^i_{t_p}  ) ^+ \big]\Big)  
\end{align} 
where $\LP ^i_{t_p,t_p+\delta' }$ has been defined in~\qr{e:lp}.
\end{itemize}

In our case, we also use EADs as a proxy of the exposure of the CCP to the members in the context of EMIR ``cover two'' default fund computations (see \sr{ss:regulccva}).
For our default fund and KVA computations, such EADs must then be computed at any randomization time $t=\zeta$ used in \qr{e:finalhat} or for simulating the time integral in \qr{e:kexpl}. Unless an explicit formula is available for the conditional expectations in the right-hand side of \qr{e:ead},
such EAD exposures can only be done by means of nested Monte Carlo simulations.

Note that in both our centrally cleared and bilateral trading setups, 
%As explained in \shortciteN{GreenKenyonDennis14},
we neglect capital for market risk in the paper,
as if the 
reference member (or bank) was perfectly hedged in terms of market risk. Otherwise one more capital term is required for market risk.
 
\subsection{CCP Setup}\label{ss:regulccva}

Under centrally cleared trading, the ``cover two'' EMIR rule prescribes to size the default fund  as, at least,  the maximum of the greatest and of the sum of the second and third greatest exposures ``under extreme but plausible market conditions'' (see \citeN[article 42, paragraph 3, page 37]{EUl2}).
This total amount is then allocated between the clearing members according to some repartition key, e.g. proportional to their initial margins.

As explained in the paper, default fund contributions are ``implicit capital'' that the clearing members put at the disposal of the CCP.
In addition, to cover their residual risk beyond the guarantee provided by the different margin layers of the CCP, 
the regulatory capital $K=K^{cm}$ of a generic reference member is defined, following 
%\citeN[2014, page 11]{BCBSall}, as:
\citeN[page 11]{BCBS14}, as:
\beql{e:kcm}
& K^{cm} = \max \left( K^{ccp} \times \frac{DFC }{{E} + \sum_{i\in N} J^i DFC^{i}} , 8\% \times 2\% \times DFC \right),
\eeql
where DFC is the default fund contribution of the reference member and where
\beql{e:kccp}K^{ccp} = RW \times Cap_{Ratio} \times \sum_{i\in N}J^i  EAD^i\eeql
with
%$EAD^i=(E\!B\!R\!M^i - I\!M^i - DFC^i)^+$ and
$RW = 20\%$ and $Cap_{Ratio}=8\% .$
%, hence we finally decided to simply ignore $K^{cm}$ in the body of this article.}

\brem\label{e:ccprc}
%Since the default fund is meant to be depleted by realized breaches, but no more in principle, the unfunded default fund contributions of a member can be interpreted as its unexpected losses. Accordingly, 
\citeN{Ghamami14} argues that the CCP regulatory capital $K^{cm}$
of a member 
should rather be based on its expected future unfunded default fund contributions (see the remark \ref{rem:fuf}), which represent the losses of the member beyond the level already funded via its default fund contribution.
%instead of its funded default fund contribution $DFC$ as in \qr{e:kcm}. 
%under the current regulation. 
%%\b{Related to this or not, fact is that
%%the explicit capital $K^{cm}$ as of \qr{e:kcm} was always found negligible in our
%%numerical experiments as compared with the implicit capital $DFC$}. 
\erem

%Conversion Factor used in the calculation of the add-on for financial derivatives
%Residual Maturity
%Add-on factor
%Interest rates
%Equal to 1 year or less
%0.00\%
%Over 1 year and not exceeding 5 years
%0.50\%
%Over 5 years
%1.50\%
%Currency rates and gold
%Equal to 1 year or less
%1.00\%
%Over 1 year and not exceeding 5 years
%5.00\%
%Over 5 years
%7.50\%
%Equities
%Equal to 1 year or less
%6.00\%
%Over 1 year and not exceeding 5 years
%8.00\%
%Over 5 years
%10.00\%
%Precious metals
%Equal to 1 year or less
%7.00\%
%Over 1 year and not exceeding 5 years
%7.00\%
%Over 5 years
%8.00\%
%Investment grade
%credit default swaps
%Equal to 1 year or less
%5.00\%
%Over 1 year and not exceeding 5 years
%5.00\%
%Over 5 years
%5.00\%
%Non-investment grade credit default swaps
%Equal to 1 year or less
%10.00\%
%Over 1 year and not exceeding 5 years
%10.00\%
%Over 5 years
%10.00\%
%Other Commodities
%(including energy products)
%Equal to 1 year or less
%10.00\%
%Over 1 year and not exceeding 5 years
%12.00\%
%Over 5 years
%15.00\%

\subsection{CSA Setup}\label{ss:regulbva}

In the bilateral setup, the capital at risk $K$ of the bank reduces to its regulatory capital (there is no bilateral trading analog of the default fund),
which comprises a first
contribution for
counterparty default losses and a second one for the volatility of the CVA (the market risk of the bank being supposed to be hedged out).
Since we focus on the reference member 0 with $n$ counterparties $i\in N^\star=\{1,2,\ldots,n\},$
the capital formulas below all need to be summed over $i\in  N^\star.$

\subsubsection{$K^{ccr}$}

The Basel II regulatory capital specified for \emph{counterparty risk} is defined as $$K^{ccr} = Cap_{Ratio}\sum_{i\in N^\star} R W\!A^i,$$
where
\begin{equation*}
R W\!A^i = 12.5 \times w_i \times 	1.4\times EAD^i.
\end{equation*}
Here $Cap_{Ratio} \geq 8\%$ (which is the value that we use in the numerics) is a chosen capital ratio that the bank must hold.
The capital weight $w_i$ is given by the internal ratings-based formula 
\begin{equation*}
w_i = (1-R_i) \left(
\Phi \left( \frac{\Phi^{-1} \left( {DP}_i \right)}{\sqrt{1-corr_i}} + \sqrt{\frac{ {corr}_i}{1-corr_i}} \Phi^{-1}(0.999) \right) - {DP}_i \right)     \frac{1 + ( \hat{T}^i - 2.5) b({DP_i)}}{1 - 1.5  b({DP_i)}} 
\end{equation*}
(see \citeN[page 7]{BISbasel05}), where:
\begin{itemize}
\item $R_i$ is the recovery rate of the counterparty $i$,

\item $\Phi$ is the standard normal cdf,

\item ${DP}_i$ is the one year default probability of the counterparty $i$, historical in principle, proxied in our numerics by the
risk-neutral default probability extracted from the corresponding CDS spread,
\item ${corr_i}$ is the
asset--counterparty $i$ correlation in the sense of
$${corr_i} = 0.12 \frac{1-e^{-50 DP_i}}{1-e^{-50}} + 0.24 \frac{1 - (1 - e^{-50 DP_i})}{1-e^{-50}}$$

\item $\hat{T}^i$ is the effective time to maturity of the netting set $i$,
i.e.~the time to maturity of the swap in our numerical case study where a single derivative is considered,
%$$\hat{T} = \min(5.0, {\max(1.0, \frac{\sum_{l=1}^d m_l \, \mbox{Nom}_l}{\sum_{l=1}^d \mbox{Nom}_l}}),$$

\item $b(p) = \big( 0.11852 - 0.05478  \ln (p) \big)^2.$
%$b$ is the smoothed (regression) maturity adjustment (smoothed over $DP$), i.e.
%$b(p) = \big( 0.11852 - 0.05478  \ln (p) \big)^2.$

\end{itemize}

\subsubsection{$K^{cva}$}

The standardized CVA risk capital charge in \citeN[$\S 104$]{BISbasel3} reads as
\begin{equation*}
{K}^{cva} = 2.33 \sqrt{Y} \left[ \left( 0.5 \sum_{i\in N^\star} w_i \hat{T}^i \tilde{EAD}^i \right)^2 + 0.75 \sum_{i\in N^\star} \left( w_i \hat{T}^i \tilde{EAD}^i \right)^2 \right]^{0.5} ,
\end{equation*}
which we approximate as in
\citeN{GreenKenyonDennis14} by
\begin{equation*}
 \frac{2.33}{2}
\sqrt{Y}\sum_{i\in N^\star} w_i \hat{T}^i \tilde{EAD}^i,
\end{equation*}
where:
\begin{itemize}
\item $Y$ is the one year risk horizon, i.e.~$Y=1,$
\item $\hat{T}^i$ is above,
\item $\tilde{EAD}^i=\frac{1 - \exp(0.05\hat{T}^i)}{0.05\hat{T}^i} EAD^i,$
\item $w_i$ is a weight based on the external rating extracted from
the one year default probability $DP_i$
as of the following table,
where the left part comes from {Moody's}
and the right part is taken from
\citeN[$\S 104$]{BISbasel3}:
%\begin{table}[H]
\begin{center}
\begin{tabular}{|c|c|c|}
\hline
 {Default Prob}& {Rating} & {Weight} \\
\hline
0.00\% & {AAA} & 0.7\% \\
0.02\% & {AA} & 0.7\% \\
0.06\% &{A} & 0.8\% \\
0.17\%&{BBB} & 1.0\% \\
 1.06\%&{BB} & 2.0\% \\
3.71\%&{B} & 3.0\% \\
12.81\%&{CCC} & 10.0\% \\
\hline
\end{tabular}
 \label{t:1bis}
 \end{center}
%\end{table}
\end{itemize} 
\section{Proofs of Auxiliary Results}\label{s:proof}.
%Technical Proofs / Proofs of Auxiliary Lemmas
\subsection{Proof of Lemma \ref{l:brea}}
Under our stylized model of the liquidation procedure,
during the \cure period $[\tau_{Z},\td_{Z}]$, where $\tau_{Z}=\tau_{i}$ 
if and only if $i\in Z,$
the clearing house substitutes itself to the defaulting members, taking care of all their dividend cash flows, which represent a cumulative cost
of $\sum_{i\in Z}\Delta^i_{\td_i}$ (including a funding cost at the risk-free rate comprised in the $\Delta^i_{\td_i}$). 
%\b{The CCP also passes to the estate of the defaulted member the OIS remuneration of its collateral from other members during its liquidation period. Hence, the amount of available collateral for the liquidation itself does not accrue at the OIS rate, but stays constant during the liquidation period.}
At the liquidation time $\td_{Z}$, the clearing house substitutes the buffer to itself as counterparties in all the concerned contracts (or simply puts an end to the contracts that were already with the buffer), which represents a supplementary cost
$\sum_{i\in Z}P^i_{\td_{i}}.$ In addition, for any
$i\in Z:$
\begin{itemize}
\item If $\varepsilon_i=0,$ meaning that $Q^i_{\td_{i}}\leq\MMs ^i_{\hat{\tau}_{i}},$ then either
$Q^i_{\td_{i}}\leq 0$ and an amount $(-Q^i_{\td_{i}})$ is paid by the clearing house to the member $i$ (who keeps ownership of all its collateral), or $Q^i_{\td_{i}}\geq 0$
and the ownership of an amount $Q^i_{\td_{i}}$ of collateral
is transferred to the clearing house. In both cases, the clearing house gets $Q^i_{\td_{i}};$
\item Else, i.e.~if $\varepsilon_i>0,$ meaning that the overall collateral $\MMs ^i$ of a member $i\in Z$ does not cover the totality of its debt to the clearing house, then, at time $\td_{i}$, the ownership of $\MMs ^i$ is transferred in totality to the clearing house. If $R_i >0$ then the  clearing house also gets a recovery
$R_i \varepsilon_i.$
\end{itemize}
In conclusion, the realized breach of the CCP is the sum over $i\in Z$ of the
\bel
&P^i_{\td_{i}}+\Delta^i_{\td_i}-\ind_{\varepsilon_i>0}(\MMs ^i_{\hat{\tau}_{i}}+R_i \varepsilon_i)-\ind_{\varepsilon_i=0} Q^i_{\td_{i}}= Q^i_{\td_{i}}
-\ind_{\varepsilon_i=0} Q^i_{\td_{i}}-\ind_{\varepsilon_i>0}(\MMs ^i_{\hat{\tau}_{i}}+R_i \varepsilon_i)\\&\qqq =
\ind_{\varepsilon_i>0}(Q^i_{\td_{i}}-\MMs ^i_{\hat{\tau}_{i}}-R_i \varepsilon_i) =(1-R_i)\varepsilon_i=\xi_i .~\finproof
\eel

\subsection{Proof of Lemma \ref{l:selfie}}
To formulate in mathematical terms the above-described margining, hedging and funding policy of the member, 
we introduce
three funding assets $B^0$, $B^{f}$ and $\bar{B}^{f}$ evolving on $[0,\db]$ as
\beqa \label{eq:bf}
dB^{0}_t = r_t  B^{0}_t dt\sp
dB^{f}_t =(r_t + {\lambda}_t) B^{f}_t dt\sp d\bar{B}^{f}_t =
(r_t + \bar{\lambda}_t) \bar{B}^{f}_t dt + (1-\Rf)\bar{B}^{f}_{t-} dJ_t . \eeqa
These represent the
risk-free OIS deposit asset and
the assets used by the bank for its respective investing and unsecured funding purposes. Under our
continuous-time mark-to-model and realization assumption on profit-and-losses, the amount on the funding accounts of the bank is
%Consider the decomposition, for $t\in[0,\db],$
\begin{eqnarray*} 
%\label{jeq:soof} 
-\Pi_t  = -(  \Pi_t + \Ms_t )+ \Ms_t,\end{eqnarray*}
where $\Ms_t=V\!M  +I\!M$ is the amount of margins that need to be funded by the member (its default fund contribution is assumed to be taken on its uninvested equity, hence does not need to be funded), so that 
the terms in the parenthesis represent the
amount invested or 
%funded 
%via  
borrowed unsecured
(depending on its sign) by the bank,
and where we recall that collateral is remunerated OIS by the receiving party.
 Defining
\beql{jeq:selffiproofsuite}
\eta^{f}_t=\frac{(   \Pi_t + \Ms_t )^{-}}{B^{f}_t}
\sp
\bar{\eta}^{f}_t=-\frac{(  \Pi_t + \Ms_t )^{+}}{\bar{B}^{f}_t}
\sp\eta^{0}_t=\frac{\Ms_t}{B^0_t}
\sp
\bar{\eta}^0_t=-\frac{(  \Pi_t + \Ms_t )}{B^0_t},\eeql
we can write
\begin{eqnarray}
 \label{jeq:selffiproof}\bal
 -\Pi_t  &= J_{t } {\eta}^{f}_t B^{f}_t
+ J_{t } \bar{\eta}^{f}_t \bar{B}^{f}_t+ \eta^0_t B^0_t 
+ (1-J_{t }) \bar{\eta}^0_t B^{0}_t,
\eal
\end{eqnarray}
where, by self-financing condition,
\beql{jeq:selffiproofcont}
&d\left(J_{t } {\eta}^{f}_t B^{f}_t
+ J_{t } \bar{\eta}^{f}_t \bar{B}^{f}_t+
\eta^0_t B^0_t 
+ (1-J_{t }) \bar{\eta}^0_t B^{0}_t\right)\\&\qqq\qqq=
 J_{t}{\eta}^{f}_t dB^{f}_t
+ J_{t-}\bar{\eta}^{f}_{t-} d\bar{B}^{f}_t+ \eta^0_t dB^0_t 
  +  (1-J_{t})\bar{\eta}^0_t dB^{0}_t  .
\eeql
A left-limit in time is required
in $J_{t-}\bar{\eta}^{f}_{t-}$ 
because $\bar{B}^{f}_t $ in \qr{eq:bf} jumps at time $\tau,$
so that the process $\bar{\eta}^{f},$ which is defined through \qr{jeq:selffiproofsuite}, is not predictable.
 
In view of  
\eqref{jeq:selffiproof}-\qr{jeq:selffiproofcont} and of the additional cash flows that affect the member (contractual cash flows, margin fees, realized breaches refills and hedging cash flows),
the gain process $e$ associated with the member's valuation-and-hedge policy $({\Pi} ,\zeta)$ satisfies the following forward SDE:
$\gain_0=0$ and, for $0<t\le\db$,
\bel
 d \gain_t &=\underbrace{d\Pi_t}_{\mbox{gain on the derivative portfolio}} -\underbrace{J_t dD_t}_{\mbox{contractual dividends}}
-\underbrace{J_t  {  \bc_t ( \MMs _t -P_{\hat{t}-}) }  dt}_{\mbox{margin fees}}  \\&\qqq
-\underbrace{J_t \sum_{Z\subseteq N}\epsilon_{\td_{Z}}\boldsymbol\delta_{\td_{Z}}(dt)}_{\mbox{refill of realized breaches}} 
- \underbrace{ \zeta_t d\mathcal{M}_t  }_{\mbox{loss on the hedge}}
\\
&\qqq 
 + J_{t}{\eta}^{f}_t dB^{f}_t
+ J_{t-}\bar{\eta}^{f}_{t-} d\bar{B}^{f}_t +\eta^0_t dB^0_t  + (1-J_{t})\bar{\eta}^0_t dB^{0}_t  .  
\eel
Substituting \eqref{eq:bf} into the above 
yields  
\bel 
& d\gain_t = {d\Pi_t -r_t \Pi}_t dt - \zeta_t d\mathcal{M}_t
-\indi{\tau<\Ts}(1-\Rf ) ({\Pi}+\Ms _{\hat{t}})^+ dJ_t\\& -J_t \left(dD_t
+ \sum_{Z\subseteq N}\epsilon_{\td_{Z}}\boldsymbol\delta_{\td_{Z}}(dt) + \left({  \bc_t  ( \MMs _t -P_{\hat{t}-}) }
 + {\thelambdam}_t \left({\Pi}_t + \Ms _t \right)^{+}
 -{\thelambda}_t \left({\Pi}_t+ \Ms _t \right)^{-} \right)dt  
\right),
\eel
which is \qr{jeq:selffinoconsbis}, by definition \qr{eqfs} of $g$. 
\subsection{Proof of Lemma \ref{l:dva}}
 
Since
$\xi=(1-R ) (Q_{\td}-
\MMs_{\hat{\tau}})^+$ (cf.~\qr{c:eq:pasmark}),
where
\bel 
& \MMs_{\hat{\tau}}=\MMs_{\tau-}=\MMs(\tau,X_{\tau-})\mbox{ and }\\
&Q_{\tau^\delta}=P_{\tau^\delta}+\Delta_{\tau^\delta}=
P(\tau^\delta,\mathbf{X}_{{\tau^\delta}})+ \hat{\Delta} (\tau^\delta,\INTENS_{{\tau^\delta}}) - e^{\int_\tau^{\td} r(u,\mathbf{X}_u)du} \hat{\Delta}(\tau ,\INTENS_{{\tau}}),\eel
we have by definition \qr{e:xib} of $\xib:$
\beql{e:bt}
&\xib _{\tau}= (1-R ) \E \left[ e^{-\int_\tau^{\td} r(u,\mathbf{X}_u)du} \times\right.\\
&\left.\Big(P(\tau^\delta,\mathbf{X}_{{\tau^\delta}})+ \hat{\Delta} (\tau^\delta,\INTENS_{{\tau^\delta}}) - e^{\int_\tau^{\td} r(u,\mathbf{X}_u)du} \hat{\Delta}(\tau ,\INTENS_{{\tau}})-
\MMs(\tau,X_{\tau-})\Big)^+ \,\Big|\, \G_{\tau}\right].
\eeql 
Therefore, the Markov property of $\mathbf{X }$
and the continuity of $\mathbf{X }$ at time $\tau$
imply that
$
\xib _{\tau}$ can be represented in functional form as
$\xib (\tau, {X }_{\tau-}).$
Hence (cf.~\citeN[Lemma 5.1]{CrepeySong15}), it holds that
$$
\gamma_t \hat{\xi}_t =
\gamma_t{ \xib \big(t, {X}_t\big)} \sp \Q\times\boldsymbol\lambda-a.e.,$$
where
\qr{e:gamtau} yields
$\gamma =J_{-}\gamma_{\bullet}$.
This gives the result since $dva=
- \gamma \hat{\xi}.$ 
\subsection{Proof of Lemma \ref{l:nested}}
We denote by $\mathcal{T}_{\delta}$ the transition function of the homogeneous
%s{strong}
Markov process $(t,\mathbf{X}_t, \beta_t)$ over the time horizon $\delta$,
i.e.~
\bel (\varphi, (t, \intens , b) )\to\mathcal{T}_{\delta}
[\varphi](t, \intens, b)&=
 \mathbb{E}\big[ \varphi(t^\delta , \mathbf{X}_{t^\delta },\beta_{t^\delta })| \mathbf{X}_{t}=\intens,\beta_{t }=b \big] =
 \mathbb{E}\big[ \varphi(t^\delta , \mathbf{X}_{t^\delta },\beta_{t^\delta }) | \mathcal{G}_t \big]
.\eel 
Recalling \qr{e:bt} and using the fact that $\mathbf{X}$ does not jump at time $\tau$, we have 
\beql{e:baha}
\overline{\xi}_\tau
&= 
 \mathcal{T}_\delta [\xif(\cdot,\cdot,\cdot, \beta_\tau, \Ms_{{\tau-}}, \hat{\Delta}_{\tau-})] (\tau, \mathbf{X}_{\tau }, \beta_\tau)
= \mathcal{T}_\delta [\xif(\cdot,\cdot,\cdot, \beta_\tau,\Ms_{{\tau-}}, \hat{\Delta}_{\tau-})] (\tau, \mathbf{X}_{\tau-}, \beta_{\tau}),
\eeql
where we set
$$\xif(t ,\mathbf{x},b, \beta_\tau,\Ms_{{\tau-}}, \hat{\Delta}_{\tau-})
=(1-R)\beta_\tau^{-1} b
  \Big(P(t,\mathbf{x} )+ \hat{\Delta} (t,\mathbf{x}) - \beta_\tau b^{-1} \hat{\Delta}_{\tau-}-
\Ms_{{\tau-}} \Big)^+ ,
 $$
in which $\beta_\tau$, $\Ms_{{\tau-}}$ and $\hat{\Delta}_{\tau -}$ are considered as $\cG_{\tau-}$ measurable parameters.
In view of \qr{e:baha}, we have (cf.~\citeN[Lemma 5.1]{CrepeySong15})
\beql{e:ala}
-dva_t=\gamma_t\hat{\xi}_t
=J_{t-}
 \Intens_t \mathcal{T}_\delta[\xif(\cdot,\cdot,\cdot,\beta_t, \Ms_{{t}}, \hat{\Delta}_{t-})] (t, \mathbf{X}_{t-},\beta_t) \sp \Q\times\boldsymbol\lambda \mbox{ a.e..}
\eeql
As a consequence, given an independent random variable $\zeta$ with density $p$, we can write,
using
\qr{e:ala}, the definition of $\mathcal{T}_\delta$ and \qr{e:gamtau}
to pass to the second, third and fourth line, respectively:
\bel
& -\mathbb{E}[ h_\zeta \ind_{\{\zeta\leq \tb\}} \beta_\zeta dva (\zeta,\Xt_\zeta)]=-\int_0^T \mathbb{E}\big[ h_t \beta_t \ind_{\{t< \tb\}}  dva (t,\Xt_t)\big]\, p(t)dt\\
&\qqq=
\int_0^T \mathbb{E}\Big[h_t \beta_t \ind_{\{t\leq \tau\}} \Intens _t
 \mathcal{T}_\delta [\xif(\cdot,\cdot,\cdot, \beta_t ,\Ms_{{t}}, \hat{\Delta}_{t})] (t, \mathbf{X}_{t },\beta_t) \Big]
 p(t)dt\\
&\qqq=
\int_0^T \ \mathbb{E}\left[h_t \beta_t \ind_{\{t\leq \tau\}} \Intens_t \mathbb{E}\big[ \xif(t^\delta , \mathbf{X}_{t^\delta },\beta_{t^\delta },\beta_t ,  \Ms_{{t}}, \hat{\Delta}_{t}) | \mathcal{G}_t \big]\right]\, p(t)dt\\
&\qqq=
\int_0^T \ \mathbb{E}\big[h_t \beta_t \ind_{\{t\leq \tau\}} \Intens_{\bullet}(t) \xif(t^\delta , \mathbf{X}_{t^\delta }, \beta_{t^\delta },\beta_t ,  \Ms_{{t}}, \hat{\Delta}_{t})
\big]\, p(t)dt\\
&\qqq=
\mathbb{E}\big[\ind_{\{\zeta\leq T\}} h_\zeta \beta_\zeta \ind_{\{\zeta\leq \tau\}} \Intens_{\bullet}(\zeta) %e^{-\int_\zeta^{\zeta^\delta }r(s)ds}
\xif (\zeta^\delta , \mathbf{X}_{\zeta^\delta }, \beta_{\zeta^\delta },\beta_{\zeta},\Ms_{\zeta} ,\hat{\Delta}_{\zeta })\big]. \eel

\vskip0.5cm
\paragraph{Acknowledgements}
This paper greatly benefited from regular exchanges with the quantitative research team of LCH in Paris, Quentin Archer and Julien Dosseur in particular. 
%We are grateful to the following people for stimulating
%discussions: Claudio Albanese (Global Valuation Ltd),
%Tom Bielecki (IIT Chicago),
%Samuel Drapeau (Humboldt University Berlin), 
%Christophe Perignon (HEC Paris).

%\bibliographystyle{chicago}
%\bibliography{../../ref}

\end{document}